\documentclass[preprint,aps,pre,letterpaper,superscriptaddress,longbibliography]{revtex4-1}
\usepackage{amsmath,amssymb,bm}
\usepackage{mathrsfs}
\usepackage{textcomp}
\usepackage{commath}
\usepackage{mathtools}

\usepackage{hyperref}
\hypersetup{
    colorlinks=true,
    linkcolor=blue,
}
\usepackage{natbib}
\usepackage{placeins}
\usepackage[final]{graphicx} % option final is used because graphicx inherits draft option from documentclass
\usepackage{rotating}
\usepackage{epstopdf}
\usepackage{float}
\usepackage[lofdepth=1,lotdepth]{subfig}
\usepackage[normalem]{ulem}
\usepackage[usenames]{color} %Color package is called in EngC

\usepackage{soul,xcolor}
\setstcolor{red}

\newcommand{\note}[1]{\textcolor{red}{}}%{*** #1 ***}} 
 
\newcommand{\MDGrevise}[1]{\textcolor{black}{#1}} 
\newcommand{\MDGrevisesecond}[1]{\textcolor{black}{#1}} 
\newcommand{\XZ}[1]{\textcolor{red}{}} 
\newcommand{\XZrevise}[1]{\textcolor{black}{#1}}%{blue}{#1}} 
\newcommand{\WLrevise}[1]{\textcolor{black}{#1}}%{red}{#1}}
\renewcommand{\sout}[1]{} %uncomment to make strikeouts vanish
\newcommand{\soutold}[1]{} %uncomment to make old sstrikeouts vanish
\newcommand{\Ca}{\mathrm{Ca}}
\newcommand{\gammadot}{\dot{\gamma}}

\graphicspath{{./}}  %figure location (for some reason both pairs of brackets are needed)

\makeatletter

\newcommand{\numtoRoman}[1]{\expandafter\@slowromancap\romannumeral #1@}
\makeatother

\begin{document}

\title{Flow-induced segregation and dynamics of red blood cells in sickle cell disease}%\MDGrevise{: A simulation study}}
%\title{Dynamics of deformable straight and curved prolate  capsules in simple shear flow}
\author{Xiao Zhang}
\affiliation{
Department of Chemical and Biological Engineering\\
University of Wisconsin-Madison, Madison, WI 53706-1691
}
\author{Christina Caruso}
\affiliation{
Department of Pediatrics, Division of Pediatric Hematology/Oncology, Aflac Cancer and Blood Disorders Center of Children's Healthcare of Atlanta\\
Emory University School of Medicine, Atlanta, GA 30322
}
\author{Wilbur A. Lam}
\affiliation{
Department of Pediatrics, Division of Pediatric Hematology/Oncology, Aflac Cancer and Blood Disorders Center of Children's Healthcare of Atlanta\\
Emory University School of Medicine, Atlanta, GA 30322
}
\affiliation{
Wallace H. Coulter Department of Biomedical Engineering\\
Emory University and Georgia Institute of Technology, Atlanta, GA 30332
}
\affiliation{
Winship Cancer Institute\\
Emory University, Atlanta, GA 30322
}
\affiliation{
Parker H. Petit Institute of Bioengineering and Bioscience\\
Georgia Institute of Technology, Atlanta, GA 30332}
\author{Michael D. Graham}\email{Corresponding author. E-mail: mdgraham@wisc.edu}
\affiliation{
Department of Chemical and Biological Engineering\\
University of Wisconsin-Madison, Madison, WI 53706-1691
}
\date{\today}

\begin{abstract}
Blood flow in sickle cell disease (SCD) can substantially differ from normal blood flow due to significant alterations in the physical properties of the red blood cells (RBCs). Chronic complications, such as \MDGrevise{inflammation of the endothelial cells lining blood vessel walls}, are associated with SCD, for reasons that are unclear. Here, detailed boundary integral simulations are performed to investigate an idealized model flow flow in SCD,  a binary suspension of flexible biconcave discoidal fluid-filled capsules and stiff curved prolate capsules that represent healthy and sickle RBCs, respectively, subjected to pressure-driven flow in a planar slit. The stiff component is dilute. The key observation is that, unlike healthy RBCs that concentrate around the center of the channel and form an RBC-depleted layer (i.e.~cell-free layer) next to the walls, sickle cells are largely drained from the bulk of the suspension and aggregate inside the cell-free layer, displaying strong margination. These cells are found to undergo a rigid-body-like rolling orbit near the walls. A binary suspension of flexible biconcave discoidal capsules and stiff straight (non-curved) prolate capsules is also considered for comparison, and the curvature of the stiff component is found to play a minor role in the behavior. \MDGrevise{Additionally, by considering a mixture of flexible and stiff biconcave discoids,} we reveal that rigidity difference by itself is sufficient to induce the segregation behavior in a binary suspension. Furthermore, the additional shear stress on the walls induced by the presence of cells is computed for the various cases. Compared to the small fluctuations in wall shear stress for a suspension of healthy RBCs, large local peaks in wall shear stress are observed for the binary suspensions, \MDGrevise{due to the proximity of} the marginated stiff cells to the walls. \MDGrevise{This effect is most marked for the straight prolate capsules.} \WLrevise{As endothelial cells are known to mechanotransduce physical forces such as aberrations in shear stress and convert them to physiological processes such as activation of inflammatory signals, }\MDGrevise{these results may aid in understanding mechanisms for endothelial dysfunction associated with SCD}.

\end{abstract}
\maketitle
\newpage

\section{INTRODUCTION} \label{sec:introduction}
%\input{introduction}
%\begin{itemize}
%\item A physiological overview of blood flow in both normal conditions and sickle cell disease
%\item Endothelial dysfunction in sickle cell disease and our hypothesis
%\item A review of experimental and computational studies on confined suspensions of capsules and RBCs with a discussion on the key observations: effects of deformability, size and shape differences on segregation behavior in suspensions
%\item Emphasize the significance of this work on understanding the pathological mechanism of sickle cell disease complications
%\end{itemize}
%\st{Flowing suspensions of particles in confined domains, either homogeneous or heterogeneous, are of broad interest in a large variety of areas, with extensive applications in bioengineering and pharmaceutics using microfluidics. Typical examples include rapid separation of blood cells} \cite{Hou2010,Guo2016,Shevkoplyas:2005ej}\st{ and enhanced vascular-targeted drug delivery} \cite{Geng2007,Champion2009,Charoenphol:2010hk,Namdee:2013fc,Thompson:2013dm,Kelley:2016kg}\st{ involving blood flow in microfluidic systems as a model for human microvasculature.}
Blood\sout{, a classic archetype of multicomponent suspensions,} is composed primarily of red blood cells (RBCs) suspended in plasma, with the hematocrit (i.e.~volume fraction of RBCs) ranging from $0.1 - 0.3$ in the microcirculation to $0.4 - 0.5$ in large vessels such as arteries \cite{Fung:1984up}. Two other major types of cells in blood, white blood cells (WBCs) and platelets, are both considerably outnumbered by RBCs\XZrevise{, by $\sim 500 - 1000:1$ and $12 - 32:1$, respectively \cite{Dean2005}}. In addition to the drastic difference in population density, these three cellular components exhibit distinct contrasts shape, size, and rigidity \cite{Otto:2015bx}. Typically, RBCs are fluid-filled biconcave discoids with a radius of $\sim 3.8 -  4.0$ $\mu$m, WBCs are approximately spherical and slightly larger than RBCs, while platelets are the smallest. Healthy RBCs are highly flexible, while both WBCs and platelets are much stiffer. \sout{All these cellular components in blood are suspended in plasma.} Under pathological conditions, diseased RBCs often display substantial alterations in physical properties compared to healthy RBCs. This is particularly pertinent in blood disorders such as sickle cell disease (SCD), as all SCD patients have a small percentage ($1-10\%$) of RBCs that are permanently stiffened and misshapen \cite{Caruso2019}. These alterations may lead to significant aberrance in the behavior of blood flow, and further cause complications. The insufficient understanding \sout{so far }of both the dynamics of blood flow in SCD and the mechanism for complications associated with SCD\sout{, such as endothelial inflammation,} forms the principal motivation of the present work.

A significant feature of blood flow in the microcirculation\XZrevise{, the network of microvessels ($5 - 200$ $\mu$m in diameter \cite{PAPPANO2013153}) consisting of arterioles, capillaries, and venules,} is that all types of blood cells display a non-uniform distribution across the vessels.\XZ{OK} Flexible RBCs tend to migrate away from the vessel wall and form an RBC-depleted region termed the cell-free layer next to the wall \cite{Sutera:1993un,Kumar:2012ga,NAMGUNG201419,Kumar:2013vn,Namgung:2017dz}. In contrast, the stiffer WBCs and platelets are found to reside near the vessel walls, a phenomenon known as margination \cite{Tangelder1985,Firrell1989,Shevkoplyas:2005ej,Freund:2007kx,Kumar:2012ga,Kumar:2013vn}.
%\st{, which is of physiological significance as the prerequisite for inflammatory response and hemostasis}
Many prior experimental \cite{Aarts1984,GOLDSMITH1984204,Tangelder1985,ECKSTEIN:1988vs,Firrell1989,YEH19941706,Pearson:2000wy,Jain:2009ia,Fay:2016to,Spann:2016em,Gutierrez2018}
%\st{, both in vivo and in vitro,}
 and computational  \cite{Sun:2006dp,Freund:2007kx,AlMomani2008,Crowl:2011cf,Zhao:2011do,Fedosov:2012dy,Zhao2012,Reasor:2012ey,Fedosov:2014fq,Vahidkhah:2014hy,Mehrabadi:2015ih} studies have attempted to characterize the effects of a variety of parameters, such as cell stiffness, shear rate, channel width, hematocrit, and RBC aggregation, on the strength of margination of WBCs and platelets. For example, recent experiments by Fay \emph{et al.} \cite{Fay:2016to} showed that WBC margination was substantially attenuated after treatment with dexamethasone or epinephrine, both of which were found to greatly reduce the stiffness of WBCs. Furthermore, margination-based cell separation has been realized using lab-on-a-chip devices. 
% \st{For example, motivated by WBC margination in the microcirculation, Shevkoplyas et al. }\cite{Shevkoplyas:2005ej}
% \st{ achieved efficient separation of WBCs from whole blood using a simple network of microfluidic channels. Other examples }
Examples include separation of WBCs from whole blood \cite{Shevkoplyas:2005ej}, and of stiffened RBCs such as those infected in malaria from the healthy ones \cite{Hou2010,Guo2016,Chen2017}.

\sout{The segregation behavior in blood flow, i.e., the discrepancy in cross-stream distribution propensity of different types of blood cells, primarily arises from their shape, size, and rigidity contrasts {\cite{Kumar:2011dd,Kumar:2013tu,Sinha:2016ki}}. In a more general context, these factors, among other properties of the particles, are the determinants of the segregation behavior in an arbitrary flowing multicomponent suspension.} 
A series of recent computational efforts have examined the individual roles of various \sout{physical} \MDGrevisesecond{cell} properties \sout{of the components }in the segregation behavior. Using direct simulations, Kumar \emph{et al.} \XZrevise{\cite{Kumar:2011dd,Kumar:2013tu}} determined the role of stiffness difference by considering a confined binary suspension of capsules that differ only in deformability. \XZrevise{Here the stiffness (or deformability) of a capsule is characterized by the dimensionless capillary number, which represents the ratio of viscous stresses on the particle to elastic restoring stresses.} \sout{They found that in}In both simple shear flow and pressure-driven flow, when the stiffer capsules are the dilute component, they display  substantial margination. In contrast, flexible capsules tend to enrich around the centerline of the channel when they are dilute, a behavior termed demargination. The \soutold{individual}effect of size contrast was also revealed \cite{Kumar:2013tu}: in a mixture of large and small capsules with equal \XZrevise{capillary number}\XZ{defined above}, the small capsules marginate while the large particles demarginate, in their respective dilute suspensions. \MDGrevise{Similarly, }\soutold{Later, using detailed simulations, }Sinha and Graham \cite{Sinha:2016ki} explored the shape-mediated segregation behavior in binary suspensions of spherical and ellipsoidal capsules in \sout{simple shear }flow,
%\st{. In this work, all capules have the same capillary number, while the aspect ratio $\kappa$ of the ellipsoids is varied with either the equatorial radius or the volume of the capsules held constant; here $\kappa > 1$ and $\kappa < 1$ represent prolate and oblate spheroids, respectively. They showed that when the ellipsoids have the same equatorial radius as the spheres, capsules with lower $\kappa$ marginate as the dilute component in a suspension containing primarily capsules with higher $\kappa$. When all capsules have the same volume, however, spherical capsules always marginate when they are dilute, while ellipsoidal (both oblate and prolate) capsules demarginate as the dilute component.}
and showed that the effect of shape (aspect ratio) contrast depends on whether the equatorial radius or the volume of the capsules is held constant.
%\st{ all revealed that the near-wall peak in the number density distribution profile for the marginated component shifts towards the wall as the volume fraction of the suspension increases, with a decrease in the cell-free layer thickness.}\XZcomment{I think this is irrelevant}

A number of phenomenological theories have also been developed for   the segregation behavior in flowing multicomponent suspensions{ of deformable particles} \sout{. Most of the theories} \cite{Zurita2012,Kumar:2012ie,Kumar:2013tu,Narsimhan:2013jk,Qi2017}. \sout{have been grounded on taking into account} \MDGrevisesecond{These incorporate the} two key sources of cross-stream particle fluxes in flowing suspensions under confinement: wall-induced migration flux \MDGrevisesecond{of deformable particles} away from \MDGrevisesecond{walls}\sout{ the wall, which is ubiquitous for any deformable particles in a bounded flow } \cite{Smart1991}, and shear-induced diffusive flux due to hydrodynamic pair collisions between particles.
%\st{ Incorporating the effects of both sources of particle motion, Kumar and Graham }\cite{Kumar:2012ie}\st{ presented an idealized kinetic master equation model for a binary suspension of capsules with varying rigidity, and obtained the steady-state solution using a Monte Carlo simulation technique. They found that at low volume fractions, heterogeneous pair collisions between capsules of different components drive the segregation by rigidity difference, while at higher volume fractions, wall-induced migration and heterogeneous pair collisions have comparable contributions. The authors further revealed }\cite{Kumar:2013tu}\st{ that in a binary suspension of capsules with size contrast, the segregation behavior is mainly caused by the greater wall-induced migration velocity of the larger capsules. Similar methods have been employed to understand the dynamics and particle concentration distributions in either homogeneous suspensions or more complex binary systems }\cite{Zurita2012,Narsimhan:2013jk,Qi2017}\st{. An example is a recent work by Qi and Shaqfeh }\cite{Qi2017}\st{, who predicted the distributions of RBCs and platelets in a binary suspension subjected to pressure-driven flow.} 
An example is a recent theory developed by Henr\'{\i}quez Rivera \emph{et al.} \cite{HenriquezRivera:2015fx,Rivera2016} that consists of a set of drift-diffusion equations.
%\st{Based on approximations that collisions only occur between closely adjacent particles and result in very small post-collisional displacements, a mechanistic theory was developed by Henr\'{\i}quez Rivera et al. }\cite{HenriquezRivera:2015fx}\st{ in which the master equation }\cite{Kumar:2012ie,Zurita2012,Narsimhan:2013jk}\st{ was simplified into a pair of drift-diffusion equations that have been denoted as the ``simplified drift-diffusion (SDD)" model.} 
\sout{In simple shear flow, closed-form analytical solutions for this model are attainable to predict the thickness of cell-free layer, and the } \MDGrevisesecond{Predictions of cell-free layer thickness} \sout{predictions}are found \cite{Rivera2016} to agree well with the values reported in other experimental \cite{Bugliarello1963}, numerical \cite{Kumar:2013tu} and theoretical \cite{Narsimhan:2013jk} studies. Moreover, several regimes of segregation arise in both simple shear flow and pressure-driven flow, depending on the value of a so-called ``margination parameter", which charactierizes the relative significance of the wall-induced migration and the collisional fluxes. Furthermore, a sharp ``drainage transition", characterized by complete depletion of one component from the bulk flow toward the walls, is also identified.
%\st{ This drift-diffusion model was then utilized to understand the segregation behavior in binary suspensions of spherical and ellipsoidal capsules }\cite{Sinha:2016ki}\st{, as described above, and found to well capture the key qualitative features of the shape-mediated margination and demargination. Later, this model was extended }\cite{Rivera2016}\st{ to describe suspension dynamics in pressure-driven flow, which is complicated due to the nonuniform shear rate and in particular the fact that it vanishes at the centerplane. Numerical solutions predict margination regimes and a drainage transition qualitatively similar to those in the simple shear flow case.} 

%\st{The substantial progress by the extensive numerical and theoretical studies has not only disentangled the individual effects of the physical aspects of particles on the flow-induced segregation behavior in multicomponent suspensions such as blood, but also laid the groundwork for understanding the dynamics in more complex systems.}
In light of \sout{all }these advances\sout{ in research}, in this work we investigate the dynamics of blood flow in SCD, an inherited blood disorder characterized by abnormal sickle hemoglobin. Under deoxygenated conditions, sickle hemoglobin molecules self-assemble inside RBCs\sout{ due to hydrophobic interactions}, forming long rigid polymers, a process termed sickling. These \sout{intracellular} polymers mechanically alter and damage the membrane of RBCs, leading to the classic ``sickled" (or crescent-like) shape \cite{Shen1949,Bertles1968} for sickle cells. In fact, in addition to this characteristic sickle shape, a large variation in cell morphology has been observed within the population of sample sickle cells. \sout{In a recent work, using diffraction phase microscopy, }Byun \emph{et al.} \cite{BYUN20124130} \MDGrevisesecond{recently} performed optical measurements of individual sickle RBCs, and classified them into echinocytes, discocytes, and the typical crescent-shaped irreversibly sickled cells (ISCs) based on their morphological features, similar to the observations in a prior study by Kaul \emph{et al.} \cite{Kaul1983}. Here ISCs refer to \sout{those}cells afflicted with irreversible membrane damage after repeated intracellular sickling-unsickling cycles, which have permanently lost membrane elasticity and remain sickle-shaped even when exposed to extensive oxygenation \cite{Diggs1939,Shen1949,BERTLES884}. %In addition to these experiments, computer-aided numerical simulations in the past decades have also improved the understanding of the nature of HbS polymer fiber formation and growth \cite{DOU19932068,Lei2012,LI20121130,LU20162085,Papageorgiou9473}, and corroborated experimental observations of considerable heterogeneity in cell morphologies \cite{LI20121130,Lei2012,Xu2017}.
In SCD patients, ISCs take up a small percentage ($1-10\%$) of RBCs \cite{Caruso2019} due to the much shorter life span ($\sim$ 15 days \cite{Milner1973}) than healthy RBCs ($\sim$ 115 days \cite{Cohen2008}).
%\st{Blood flow in SCD, as mentioned above, can considerably differ from the normal due to significant alterations in the physical properties of sickle cells compared to healthy RBCs. Deoxygenation of sickle hemoglobin (HbS) results in hydrophobic interactions among hemoglobin molecules, causing formation of long hemoglobin polymers. This polymer formation results in a variety of shapes for deoxygenated sickle cells }\cite{BYUN20124130}\st{, including the classic sickle shape }\cite{Bertles1968}. 
ISCs also have a much smaller volume than healthy RBCs due to dehydration \cite{ZUCKER197610,Glader1978}. In addition, a general increase in membrane rigidity has been determined for sickle cells compared to that of healthy RBCs, although the extent of \sout{membrane} stiffening \sout{ is found to vary greatly depending on the states and conditions of individual sickle cells} \MDGrevisesecond{varies greatly from cell to cell} \cite{messer1970,Nash73,Evans1443,Bunn1997,BYUN20124130,LI201734}.

%\st{Complications associated with SCD related to RBC stiffening, which cause altered microvascular blood flow and vaso-occlusion followed by consequent tissue ischemia and infarction, have long been observed clinically. These effects only partially describe the pathophysiology of SCD, however, particularly as SCD is also known to be a vasculopathic disease in which endothelial cells that line the blood vessels are dysfunctional and inflamed. SCD vasculopathy bears some resemblance to cardiovascular disease, and relatedly, stroke remains a major cause of mortality in SCD. Despite such similarities, the underlying cause of chronic SCD vasculopathy remains unknown.}
The pathophysiology of SCD has canonically been thought to lie in the acute vaso-occlusive crisis, in which diseased RBCs, upon deoxygenation, tend to obstruct microvessels due to the sickled shape and increased stiffness, resulting in restricted local blood flow and subsequent tissue ischemia accompanied by acute pain. However, this view of SCD pathophysiology overlooks a common yet important complication of SCD, chronic sickle vasculopathy, in which the endothelial cells that line the blood vessels are dysfunctional and in a pro-inflammatory state in most regions in the circulation\sout{, as in ``regular" cardiovascular disease  \cite{Aessopos2007,Kato2008,Yuditskaya2009}}. Of particular clinical importance in understanding SCD vasculopathy is its association with stroke, one of the leading yet least understood causes of mortality in SCD. Indeed, the fact that vasculopathy pervasively occurs even in the oxygenated conditions in both small and large vessels necessitates a new understanding of SCD pathophysiology that extends beyond, and is independent of, vaso-occlusion, which, occurs only under the deoxygenated conditions in microvessels. Specifically, how the abnormality in physical properties of sickle cells, such as the distorted shape and increased stiffness, alters the dynamics of blood flow, and ultimately, how this alteration may be linked to the underlying mechanism for the endothelial inflammation in SCD \MDGrevisesecond{and subsequent complications such as stroke}, are \MDGrevisesecond{important} issues to address.

\sout{Progress in the past decades in uncovering the pathophysiology of cardiovascular disease such as atherosclerosis may shed some light on the mechanism for vasculopathy in SCD. Cardiovascular bioengineering research has demonstrated that e}Endothelial cells mechanotransduce the shear forces of the hemodynamic microenvironment into cellular biological signals \cite{Harrison2006,Chien2007,Abe2014}. \sout{Additionally, pathological alterations of those forces lead to activation of pro-inflammatory signals within endothelial cells and subsequent development of atherosclerotic plaques \emph{in situ} that are prone to myocardial infarction and stroke \cite{Abe2014}.} Bao \emph{et al.} \cite{Bao1999} measured the mRNA expression of MCP-1 and PDGF-A, two genes related to endothelial inflammation, by endothelial cells subjected to laminar flows with different well-defined wall shear stress profiles, namely (1) ramp flow in which shear stress was smoothly changed from zero to a maximum value and then sustained, (2) step flow in which shear stress was abruptly applied at flow onset followed by steady shear stress for a sustained period, and (3) impulse flow in which shear stress was abruptly applied and then removed. A steady shear stress profile was also considered for comparison. They showed that rapid changes in shear stress (as in impulse flow and the onset of step flow) and steady shear stress (as in ramp flow and the steady component of step flow) stimulate and diminish the expression of both genes, respectively. In particular, the impulse flow case was found to induce the highest level endothelial expression of the pro-infammatory signals compared to the other profiles. Other studies \cite{Davies1997,DeKeulenaer1998,Chen1999Mechano,Akimoto2000,Davis2001,Dekker2002,Hsiai2003,Sorescu2004,Li2005Molecular,Harrison2006} have also elucidated a variety of shear stress-induced signal transduction pathways of the endothelium, and revealed the opposing effects of steady and unsteady shear stresses on endothelial dysfunction.

While well-studied in the context of cardiovascular disease, this issue has yet to be explored in SCD. To the best of our knowledge, there exist only a limited number of computational studies of the behavior of sickle cells in suspensions in association with mechanisms for SCD complications. Examples are \cite{Lei11326,Lei2015,DENG2019360}, which aimed to investigate the\soutold{ precise} mechanism for vaso-occlusive crisis. In these studies, simulations were performed for suspensions containing sickle cells in a cylindrical tube of diameter $D = 10$ $\mu$m, mimicking post-capillary flow, to quantify the \MDGrevise{consequences of} adhesive interactions between sickle cells and the endothelium (tube wall) during flow, and determine the effects of cell morphology and rigidity. However, given their particular focus on blood flow in post-capillaries, these studies considered suspensions with only a very small number of cells under \soutold{great}\MDGrevise{strong} confinement, so that characteristic dynamics of multicomponent suspensions such as flow-induced segregation and margination were not \soutold{described}\MDGrevise{incorporated}. 

The aim of the present study is to take a first step toward understanding \sout{the dynamics of blood flow in more complex scenarios such as}\MDGrevisesecond{these phenomena in} SCD. \soutold{Specifically, u}Using large-scale direct simulations, we investigate a number of binary suspensions containing primarily flexible healthy RBCs with a small fraction of stiff cells with different rest shapes in confined pressure-driven flow at zero Reynolds number. WBCs and platelets are not included in the suspensions\soutold{ assuming a negligible effect} due to their extremely small number fractions in blood flow. We \soutold{reveal}\MDGrevise{predict} that sickle cells display \soutold{a}strong margination toward the walls due to their increased rigidity and reduced size compared to healthy RBCs. The\soutold{ comprehensive} effects of shape, size, and rigidity differences between the components on the behavior of different suspensions are \soutold{discussed}\MDGrevise{studied}. We also characterize the orbital dynamics of single cells in the suspensions. Furthermore, the hydrodynamic effects of the cell suspensions on the walls, particularly the additional wall shear stress induced by the \soutold{suspensions}\MDGrevise{marginated cells}\soutold{ as well as the rheological properties of the suspensions} are quantified\soutold{, aiming} to establish a connection with\soutold{ addressing the medically significant} issues in SCD such as shear-induced endothelial dysfunction\soutold{ and vaso-occlusion}. 

The rest of the paper is organized as follows: in Section~\ref{sec:methods} we describe the models for the cells and suspension systems considered in this work, as well as the numerical methods and algorithms employed to compute the flow field and cell motion; in Section~\ref{sec:results} we present detailed results and discussion on the behavior of different suspensions and the dynamics of single cells, followed by a characterization of the hydrodynamic effects of the suspensions on the walls.\soutold{ The rheological properties, particular the intrinsic apparent viscosity for each suspension, are also reported.} Concluding remarks are presented in Section~\ref{sec:conclusion}.

\section{MODEL FORMULATION} \label{sec:methods}
\subsection{Model and discretization} \label{sec:model}
\begin{figure}[t]
  \centering
  \captionsetup{justification=raggedright}
  \includegraphics[width=0.7\textwidth]{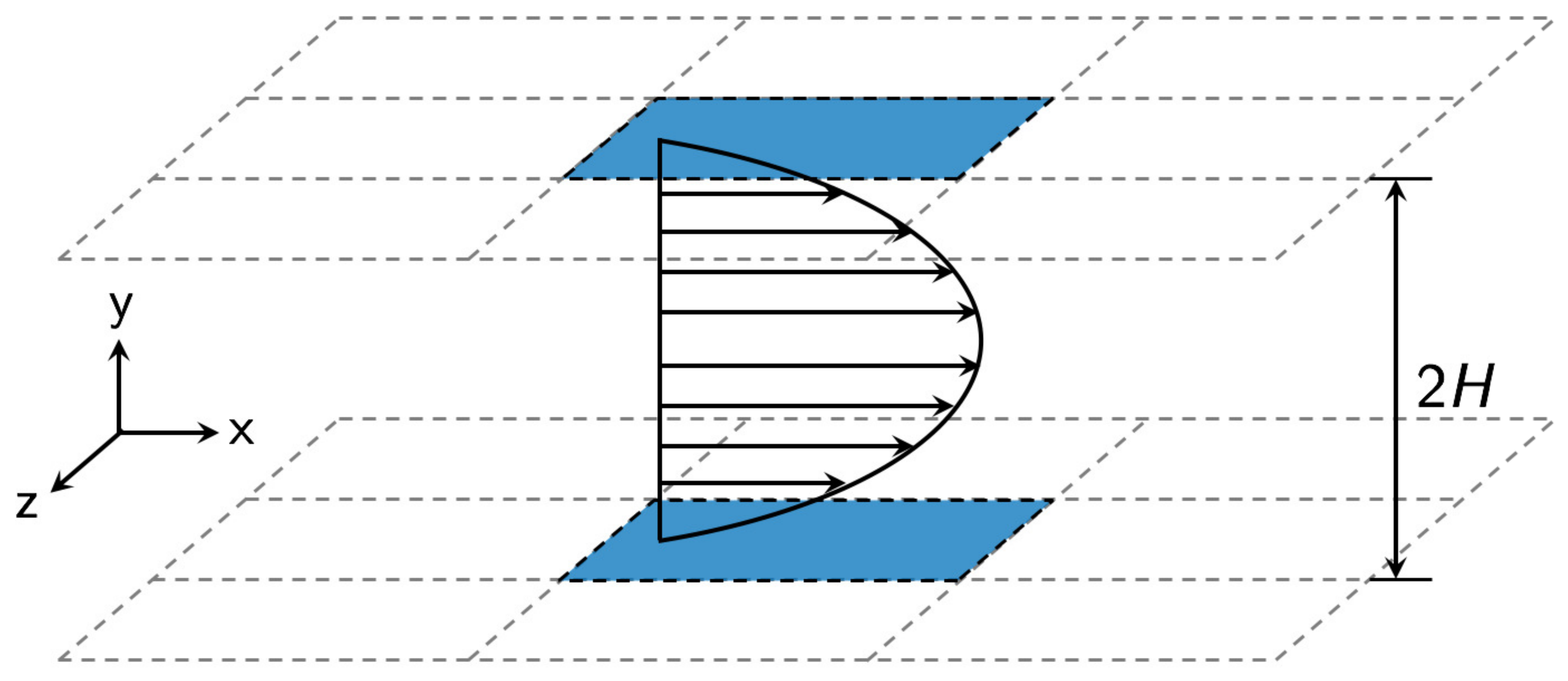}
  \caption{Schematic of the simulation domain.}
  \label{fig:simulation_domain}
  \end{figure}

We consider a flowing suspension of deformable fluid-filled elastic capsules in a planar slit bounded by two parallel rigid walls (FIG.~\ref{fig:simulation_domain}). In the wall-normal ($y$) direction, no-slip boundary conditions are applied to the two rigid walls at $y = 0$ and $y = 2H$, where $H$ is the distance from each wall to the centerplane of the slit domain. In the flow ($x$) and the vorticity ($z$) directions, periodic boundary conditions are imposed, with spatial periods $L_x$ and $L_z$, respectively. The suspension is subjected to a unidirectional pressure-driven (Poiseuille) flow, and the undisturbed flow velocity field is given by 
\begin{equation} \label{eq:Poiseuille_velocity}
  \mathbf{u}^{\infty}(x, y, z) = 2 U_0 \frac{y}{H} \big(1-\frac{y}{2 H} \big) \mathbf{e}_x,
\end{equation} 
where $U_0$ is the undisturbed flow velocity at the centerplane of the channel, and $\textbf{e}_x$ is the unit vector in the $x$ direction. \MDGrevise{In the present study a constant pressure drop is imposed, which is equivalent to fixing the mean wall shear rate at $\gammadot_w=2U_0/H$.}\soutold{ Unlike simple shear flow where the shear rate is constant, pressure-driven flow features a non-uniform shear rate profile in the wall-normal direction: the magnitude of shear rate $\dot \gamma$ is zero at the centerplane, while reaching a maximum at the walls, denoted as $\dot \gamma_w$; it is straightforward to derive that $|\dot \gamma_w| = 2U_0/H$.} We \MDGrevise{take the fluids outside and within the capsules to be}\soutold{ assume that both the suspending fluid and the fluid enclosed by the capsules are} incompressible and Newtonian with viscosity $\eta$ and $\lambda \eta$, respectively, \soutold{where}\MDGrevise{so} $\lambda$ denotes the viscosity ratio between the fluids inside and outside the capsules. 

\begin{figure}[t]
  \centering
  \captionsetup{justification=raggedright}
  \includegraphics[width=0.6\textwidth]{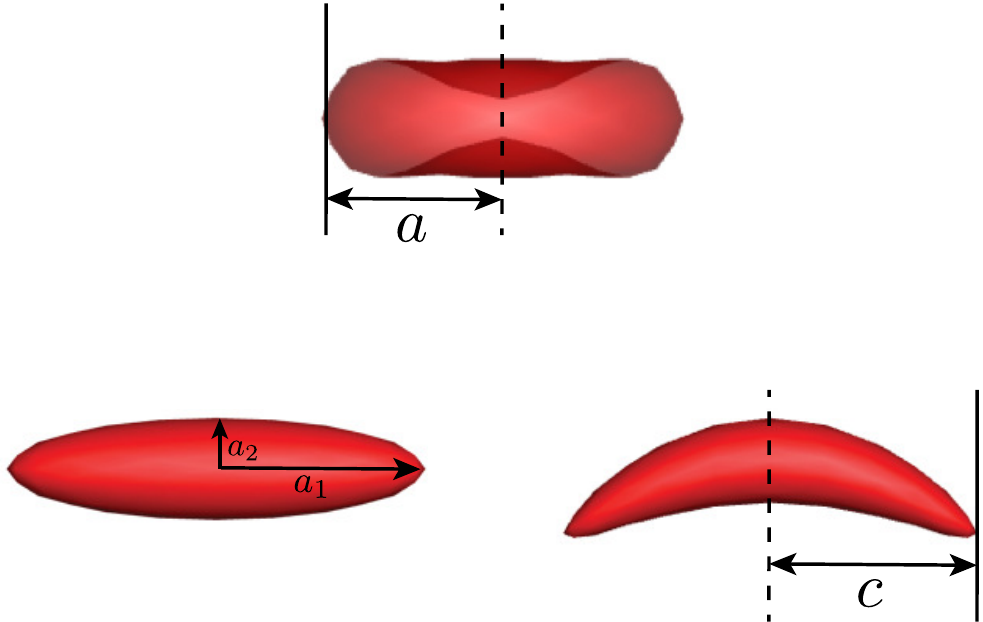}
  \caption{Rest shapes of a biconcave discoidal capsule (top) as a model for healthy RBCs, a straight prolate capsule (bottom left), and a curved prolate capsule (bottom right) as a model for typical sickle cells.}
  \label{fig:cell_schematic}
  \end{figure}

Different suspension systems comprising various types of cells are investigated in this work. Figure~\ref{fig:cell_schematic} shows the rest shapes of the cell models. A healthy RBC is modeled as a flexible capsule with a biconcave discoidal rest shape \cite{EVANS1972335,Sinha:2015wt}, with the geometry given by
\begin{equation} \label{eq:biconcave_shape}
  y = \frac{a}{2} \sqrt{1-r^2} (C_0 + C_2 r^2 + C_4 r^4),
\end{equation} 
where $r^2 = x^2+z^2 \leqslant 1$, $C_0 = 0.2072$, $C_2 = 2.0026$, and $C_4 = -1.1228$; $a$ denotes the radius of the biconcave discoid, which is $\sim 3.8 - 4.0$ $\mu$m for human RBCs. \soutold{Different from}\MDGrevise{In contrast to} a healthy RBC, a typical ISC is crescent-shaped with a characteristic length $c$ slightly greater than $a$ \cite{BYUN20124130,Park2016sickle}. In this study, an ISC is modeled as a stiff capsule with a curved prolate spheroidal rest shape. \MDGrevise{This shape is constructed by}\soutold{ model derives from} first polar stretching and equatorial compression of a spherical capsule with radius $a$, \soutold{which obtains}\MDGrevise{which results in} a slender prolate capsule with polar radius $a_1 = c = 1.2 a$ and equatorial radius $a_2 = 0.25 a$ (aspect ratio AR = $a_1/a_2$ = 4.8). The resulting straight prolate capsule is then subjected to a quadratic unidirectional displacement of membrane points perpendicular to the polar axis of the capsule, which eventually generates a curved prolate capsule. Details regarding the transformation procedure for the sickle RBC model are found in \cite{Zhang2019}. \MDGrevise{For our models,} the volume of a straight or curved prolate capsule is $\sim 20\%$ that of a biconcave discoid\soutold{ in our models}.

Having introduced the models for the morphologies of different cells, we now describe the membrane mechanics of the capsules. In this work, the capsule membrane is modeled as an isotropic and hyperelastic surface that incorporates shear elasticity, area dilatation, and bending resistance. For an arbitrary capsule, the total energy $E$ of the capsule membrane $S$ is:
\begin{equation} \label{eq:membrane_energy}
  E = \frac{K_B}{2} \int_S (2 \kappa_H + c_0)^2 dS + \overline{K_B} \int_S \kappa_G dS + \int_S W dS,
\end{equation}
where $K_B$ and $\overline{K_B}$ are the bending moduli, and $W$ is the shear strain energy density; $\kappa_H$ and $\kappa_G$ are the mean and Gaussian curvature of the membrane surface, respectively; $c_0 = -2H_0$ is the spontaneous curvature, $H_0$ being the mean curvature of the spontaneous shape. In this equation, the first two terms represent the Canham-Helfrich bending energy \cite{CANHAM197061,Helfrich1973}, and the third term corresponds to the shear strain energy. The behavior of the capsule membrane in response to the in-plane shear elastic force is described using a membrane model by Skalak \emph{et al.} \cite{SKALAK:1973tp}, in which the shear strain energy density $W$ is given by
\begin{equation} \label{eq:Skalak_model}
W_{\mathrm{SK}} = \frac{G}{4}\big[(I_1^2 + 2 I_1 - 2 I_2) + C_a I_2^2\big],
\end{equation}	
where $G$ is the in-plane shear modulus of the membrane, and $C_a$ characterizes the energy penalty for area change of the membrane. The strain invariants $I_1$ and $I_2$ are functions of the principal stretch ratios $\lambda_1$ and $\lambda_2$, defined as
\begin{equation} \label{eq:strain_invariants}
I_1 = \lambda_1^2 + \lambda_2^2 - 2, \quad I_2 = \lambda_1^2 \lambda_2^2 - 1.
\end{equation} 
Barth\`{e}s-Biesel \emph{et al.} \cite{DBB-diaz-dhenin-2002} showed that for $C_a \gtrsim 10$, the tension of a Skalak membrane becomes nearly independent of $C_a$ under a simple uniaxial deformation, so $C_a$ is set to $10$ for all capsules in our simulations. Sinha and Graham \cite{Sinha:2015wt} have validated this model for the membrane mechanics of a healthy RBC, showing that the strain hardening behavior of the membrane predicted by this model agrees very well with the experimentally determined response of an RBC membrane to optical tweezer stretching by Mills \emph{et al.} \cite{Mills2004}. 

The deformability of a capsule in pressure driven flow is characterized by the dimensionless wall capillary number $\Ca$ =$\eta \dot{\gamma}_w l/G$, where $l$ is the characteristic length of the capsule. The bending modulus of a capsule is expressed nondimensionally by $\hat{\kappa}_B = K_B/l^2 G$, which is $\sim O(10^{-4} - 10^{-2})$ in the physiological context \cite{Evans2008,Betz15320}; here we set $\hat{\kappa}_B = 0.04$ for all capsules, which also prevents the cell membranes from any buckling instabilities. Taking the first variation of the total membrane energy $E$ in Eq.~\ref{eq:membrane_energy} gives the total membrane strain force density:
\begin{equation} \label{eq:membrane_force}
\mathbf{f^m} = \mathbf{f^b} + \mathbf{f^s}, 
\end{equation}
where $\mathbf{f^b}$ and $\mathbf{f^s}$ are bending and shear elastic force densities, respectively. For any capsule in this work, unless noted otherwise, the natural shape for shear and the spontaneous shape for bending elasticities of the capsule membrane are both chosen to be the same as the rest shape of the capsule, so that any in-plane or out-of-plane deformation would lead to an increase in the membrane energy and thus to an elastic restoring force. The capsule membrane is discretized into $N_{\triangle}$ piecewise flat triangular elements; in this work $N_{\triangle} = 320$, resulting in 162 nodes. We have verified that increasing the number of nodes makes no difference to the cell dynamics, and this mesh resolution keeps the computational cost manageable given that very long simulation times are required for the dynamics of the suspensions to reach steady state. Note that in this work, steady state refers to the state in which the suspension dynamics are statistically invariant or stationary. Based on this discretization, the calculation of the total membrane force density $\mathbf{f^m}$ follows the work of Kumar and Graham \cite{Kumar:2012ev} and Sinha and Graham \cite{Sinha:2015wt} using approaches given by Charrier \emph{et al.} \cite{charrier1989free} for the in-plane shear force density $\mathbf{f^s}$ and Meyer \emph{et al.} \cite{Meyer:2002vh} for the out-of-plane bending force density $\mathbf{f^b}$, respectively. Details regarding these calculations are found in \cite{Kumar:2012ev} and \cite{Sinha:2015wt}.   

Now we describe the suspension systems considered in this work. A base case is a homogeneous suspension of flexible biconcave discoidal capsules representing purely healthy RBCs. In our simulations $\Ca$ is always set to $1.6$ for flexible biconcave discoids in all suspensions, which is in accordance with the biomechanical properties of the membrane of a healthy RBC within the physiological ranges, assuming the shear rate of blood flow in the microcirculation $\dot{\gamma}_w \sim O(10^2 - 10^3)$ s$^{-1}$ \cite{Lipowsky2013}, the viscosity of plasma $\eta \sim$ $1.2$ mPa s \cite{Harkness1970}, and the in-plane shear modulus of a healthy RBC membrane $G \sim 7.1 \pm 1.6$ $\mu$N/m \cite{BYUN20124130}. 

One of the major aims of this study is to understand the effects of deformability, shape, and size contrast on the dynamics of a flowing suspension of capsules. To this end, different cases of binary suspensions are considered. The ``primary" component, i.e., the component with a larger population, is always the flexible biconcave discoids (healthy RBCs), and denoted as `$p$'. The other component is \XZrevise{stiff, }termed ``trace" and denoted as `$t$'. The number fractions for the capsules of these two components in all binary suspensions considered in this work are \XZrevise{set to} $X_p = 0.9$ and $X_t = 0.1$, respectively, with a number density ratio $n_p/n_t = 9$. These values are within the physiological ranges in the context of SCD \cite{Caruso2019}. In the first binary suspension, \XZrevise{the `$t$' component is stiff biconcave discoidal capsules with the same rest shape as the flexible (healthy) ones to investigate the isolated role of rigidity difference in the dynamics of the binary suspension. In the second case, we study a binary suspension with the `$t$' component being stiff curved prolate capsules that represent typical sickle cells, and this system serves as an idealized model for blood flow in SCD. For comparison, we also consider a suspension containing stiff straight prolate capsules with no curvature to illustrate the effect of curvature on the behavior of the suspension.}\soutold{In addition, we also consider the case with the `$t$' component being stiff biconcave discoidal capsules with the same rest shape as the flexible ones to investigate the isolated role of deformability contrast.} \XZrevise{In all cases of the binary suspensions, $\Ca_t$ is always set to $0.4$ for the `$t$' component to control variables. Indeed, this value is physiological given that the membrane shear modulus of a typical sickle cell is experimentally determined to be roughly four times that of a healthy RBC \cite{BYUN20124130}.} 

In this study, the total volume fraction of the capsules is $\phi \approx 0.15$ for all suspensions, with $\phi$ being slightly lower for the cases containing curved and straight prolate capsules due to their smaller volume than a biconcave discoid, as noted above. This volume fraction is consistent with the physiological hematocrit in small vessels ($\sim$ 40 $\mu$m across) in the microcirculation \cite{Sarelius1982,LIPOWSKY1980297}. The domain is periodic in $x$ and $z$ with lengths $L_x = L_z = 2H = 10 a$, where $a$ again is the radius of a biconcave discoid as described above. The domain height $H=5a$ so the confinement ratio $C = H/a = 5$. We have verified that changing the domain period in the $x$ direction from $L_x = 10a$ to $L_x = 20a$ makes negligible difference in the cross-stream distribution and dynamics of capsules (not shown). Furthermore, the suspending fluid and the fluid inside the capsules are always assumed to have the same viscosity, i.e., $\lambda = 1$ for all suspensions, to keep the computational cost manageable\soutold{ for the large-scale simulations}. \MDGrevise{This is different than the physiological viscosity ratio of $\sim 5$. However, our previous investigations have revealed that the dynamics of a single sickle cell \cite{Zhang2019} or straight prolate capsule \cite{zhang2019multiplicity} remain qualitatively unchanged as the viscosity ratio $\lambda$ is increased from 1 to 5. Furthermore, as shown by the simulation results below (Section~\ref{sec:pure_healthy}), the dynamics of single healthy RBCs in the suspensions assuming $\lambda = 1$ are consistent with the orbital behavior observed in a number of prior experimental \cite{Goldsmith351,Bitbol1986,Lanotte13289} and numerical \cite{Cordasco:2013hb,Mendez2018} studies with $\lambda = 5$.} 

    \subsection{Fluid motion} \label{sec:BIM}
%\input{BIM.tex}
%In this work, the particle Reynolds number, defined as $\textnormal{Re}_p = \rho \gammadot_w l^2/\eta$, is $\sim O(10^{-3} - 10^{-2})$ assuming that the characteristic wall shear rate $\gammadot_w$ of blood flow in the microcirculation is $\sim 10^2 - 10^3$ s$^{-1}$ \cite{Lipowsky2013}, the length scale $l$ of an arbitrary type of cells considered here $\sim 4 - 6$ $\mu$m \cite{BYUN20124130}, the viscosity of plasma $\eta \sim 1.20 - 1.71$ mPa s \cite{Harkness1970,Laogun1980}, and its density $\rho \sim 10^3$ kg/$\textrm{m}^3$. \MDG{aren't you repeating the values of these quantities? Rewrite this paragraph} The \soutold{magnitude of the} particle Reynolds number is sufficiently small \MDG{what is it -- you just provided all the numbers needed to compute it!  Also, is it sufficient that the particle Reynolds number is small? What about the channel Reynolds number?}  
\XZrevise{In our simulations, the particle and channel Reynolds numbers, defined as $\textnormal{Re}_p = \rho \gammadot_w l^2/\eta$ and $\textnormal{Re}_c = \rho \gammadot_w H^2/\eta$, respectively, both have magnitude of $\sim O(10^{-3} - 10^{-2})$ based on the physiological range of the parameter values for blood flow in the microcirculation as noted in Section~\ref{sec:model}. The Reynolds numbers are assumed to be sufficiently small} so that the fluid motion is governed by the Stokes equation. To determine the velocity field at each time instant, we use an accelerated boundary integral method of Kumar and Graham \cite{Kumar:2012ev} for simulations of capsule suspensions. The fluid velocity $\mathbf{u}$ at any point $\mathbf{x_0}$ in the simulation domain can be written as:
\begin{equation} \label{eq:BI_equation}
  u_j(\mathbf{x_0}) = u_j^{\infty}(\mathbf{x_0}) + \sum_{m=1}^{N_p} \int_{S^m} q_i(\mathbf{x}) G_{ji}(\mathbf{x_0}, \mathbf{x}) dS(\mathbf{x}),
\end{equation}
where the single layer density $\mathbf{q(x_0)}$ satisfies 
\begin{equation} \label{eq:single_layer_density}
  q_j(\mathbf{x_0}) + \frac{\lambda - 1}{4 \pi (\lambda + 1)} n_k(\mathbf{x_0}) \sum_{m=1}^{N_p} \int_{S^m} q_i(\mathbf{x}) T_{jik}(\mathbf{x_0}, \mathbf{x}) dS(\mathbf{x}) = - \frac{1}{4 \pi \mu} \big(\frac{\Delta f_j(\mathbf{x_0})}{\lambda + 1} + \frac{\lambda - 1}{\lambda + 1} f_j^{\infty}(\mathbf{x_0})\big). 
\end{equation} 
Here $\mathbf{u}^{\infty}(\mathbf{x_0})$ is the undisturbed fluid velocity at a given point $\mathbf{x_0}$, $S^m$ denotes the surface of capsule $m$; $\mathbf{f}^{\infty}(\mathbf{x_0})$ is the traction at $\mathbf{x_0}$ due to the stress generated in the fluid corresponding to the undisturbed flow $\mathbf{u}^{\infty}(\mathbf{x_0})$; $\Delta \mathbf{f}(\mathbf{x_0})$ is the hydrodynamic traction jump across the membrane interface, which relates to the total membrane force density by $\Delta \mathbf{f}(\mathbf{x_0}) = -\mathbf{f^m}$ assuming the membrane equilibrium condition; $\mathbf{G}$ and $\mathbf{T}$ are the Green's function and its associated stress tensor. 

The acceleration in this implementation is achieved by the use of the General Geometry Ewald-like Method (GGEM) by Hernandez-Ortiz \emph{et al.} \cite{HernandezOrtiz:2007p1314}. The key idea is to decompose the overall problem into a local and a global problem, specifically by splitting the Green's function for the Stokes equation into a singular but exponentially-decaying (short-ranged) part and a smooth but long-ranged part. The solution associated with the local problem is obtained assuming free-space boundary conditions, i.e., it is independent of the geometry of interest.
%\st{ An important feature of the local solution is that it is short-ranged and therefore can be neglected beyond a length scale $\sim \alpha^{-1}$ from the pole of the local density. In this work, a cutoff length is taken as $r_{\textnormal{cut}} = 4/\alpha$ with good accuracy, and a cell-linked algorithm }\cite{allen_1989}\st{ is applied to generate the near-neighbor list required for the computation of the local solution.} 
The solution associated with the global problem, on the other hand, is nonsingular and ensures that the boundary conditions for the overall problem are satisfied. In this work, we compute the global solution by employing the discrete Fourier series approximation in the periodic $x$ (flow) and $z$ (vorticity) directions and the discrete Chebyshev polynomial approximation in the $y$ (wall-normal) direction.
%\st{ Note that in a traditional Ewald-sum-based method, the domain is triply periodic and a discrete Fourier representation is used in all three directions. In this approach, the pressure drop associated with the flow field induced by the capsules is always zero over the spatial period of the simulation domain, which ensures that the pressure drop obtained from this boundary integral implementation for a flowing suspension always equals the pressure drop in the flow in the absence of the capsules. In other words, these simulations are performed at constant pressure drop.} 
This approach ensures that the simulations in this work are performed at constant pressure drop. The overall solution for the velocity or stress field is the sum of the local and global solutions.
%\st{ Based on extensive numerical tests }\cite{Kumar:2012ev}\st{, we set $\alpha h = 0.5$ to guarantee the convergence of the GGEM solution, where $h$ is the characteristic mesh spacing in the global problem. Once the overall solution is obtained on the mesh points, the velocity and stress at the discrete nodes on the capsule membrane are determined by 4th-order Lagrange interpolation.} 
For a flowing suspension in a slit geometry as considered in the present study, the computational cost of the algorithm scales as $O(N \textnormal{log}N)$, where $N$ is proportional to the product of the total number of capsules $N_p$ and the number of triangular elements $N_{\triangle}$ upon surface discretization. Details of the numerical method and algorithm are found in \cite{Kumar:2012ev}. 

Once the flow field is determined, the positions of the element nodes on the discretized capsule membrane are \soutold{then}advanced in time using the second-order explicit Adams-Bashforth method with adaptive time step $\Delta t = 0.02 \Ca d$, where $d$ is the minimum node-to-node distance. Time is nondimensionalized with the wall shear rate $\dot{\gamma}_w$, and in this work, $t$ always represents dimensionless time.

\section{RESULTS AND DISCUSSION} \label{sec:results}
\XZrevise{In this section, we first present and discuss simulation results for the dynamics of different suspension systems as described above (Section~\ref{sec:cell_distribution_and_dynamics}). Specifically, we focus on understanding the flow-induced segregation behavior as well as determining the orbital dynamics of single cells, especially the stiff cells, in each suspension. Substantial margination of the stiff cells is observed in all binary suspensions. Following these results, we further characterize the hydrodynamic effects of the suspensions on the walls, particularly the additional wall shear stress induced by the marginated stiff cells (Section~\ref{sec:wall_shear_stress}). We reveal that compared to the small fluctuations in wall shear stress observed in the case of a homogeneous suspension of healthy RBCs, the marginated cells in the binary suspensions induce intermittent local wall shear stress peaks.} We further discuss the implications of these findings on the mechanism for endothelial inflammation in SCD. 

\subsection{Cross-stream distribution and dynamics of cells in different suspensions} \label{sec:cell_distribution_and_dynamics}  
      \subsubsection{Homogeneous suspension of healthy (flexible) RBCs} 
      \label{sec:pure_healthy}
\begin{figure}[t]
\centering
\captionsetup{justification=raggedright}
\subfloat[]% caption for subfigure a
{
    \includegraphics[width=0.5\textwidth]{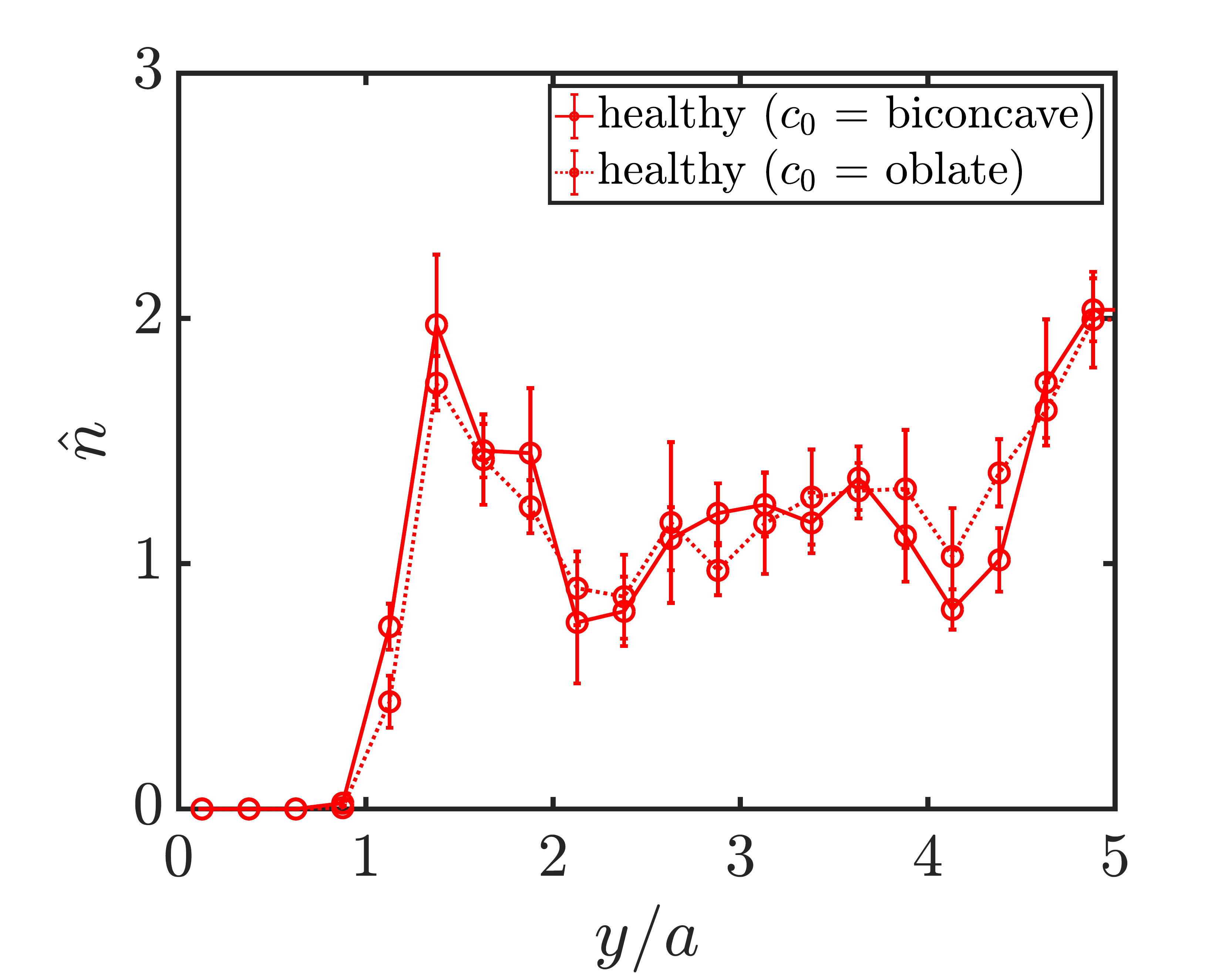}
    \label{fig:pure_healthy_profile_different_spontaneous}
}
%\subfloat[]% caption for subfigure a
%{
%    \includegraphics[width=0.30\textwidth]{./healthy_RBC_orientation.pdf}
%    \label{fig:healthy_RBC_orientation_schematic}
%}
\subfloat[]% caption for subfigure a
{
    \includegraphics[width=0.5\textwidth]{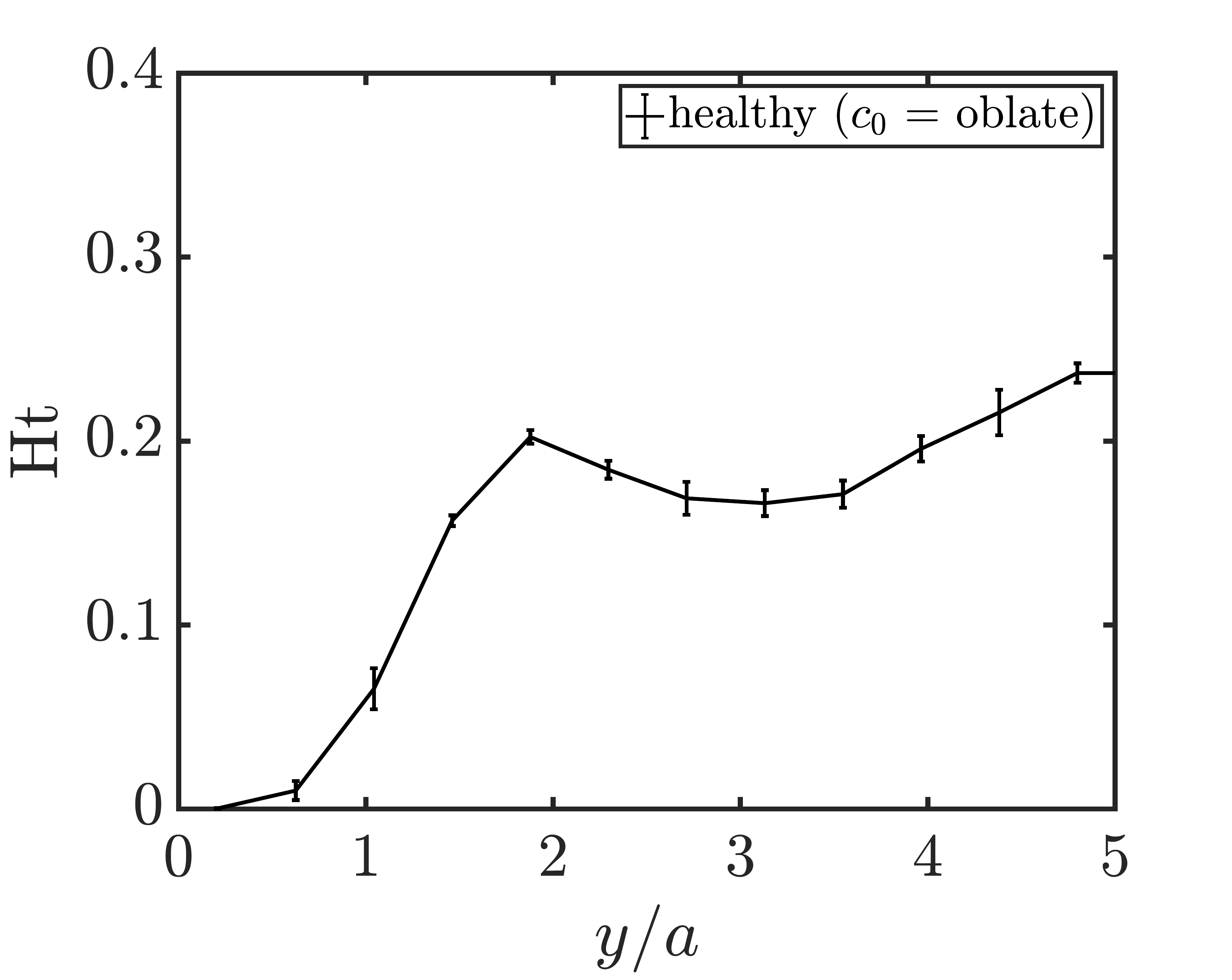}
    \label{fig:healthy_hematocrit_profile}
}
\caption{(a) Wall-normal number density profile $\hat n$ for healthy RBCs assuming different spontaneous shapes ($c_0$) in a homogeneous suspension at steady state. (b) Hematocrit profile for the healthy RBC suspension. The two positions at $y/a = 0$ and $y/a = 5$ correspond to the wall and centerplane of the channel, respectively.}
  \label{fig:healthy_profiles}
  \end{figure}

We first consider the base case with a homogeneous suspension of flexible biconcave discoidal capsules representing purely healthy RBCs, and assume that the spontaneous shape of the capsules is the same as their rest shape, i.e., a biconcave discoid. The effect of spontaneous shape will be discussed shortly.
%\st{ Again, the capillary number of healthy RBCs is $\Ca = 1.6$. Figure~}\ref{fig:two_regions_healthy}\st{ shows a simulation snapshot for this suspension at steady state. Two periods of the simulation domain are shown in both $x$ and $z$ directions in this snapshot as well as all future snapshots for other suspensions. One observation is that RBCs are depleted in the near-wall regions.} 
To obtain the cross-stream distribution for the cells of component $\alpha$ in a general suspension, we compute the normalized number density profile of the cells based on their wall-normal center-of-mass positions, given by $\hat n_{\alpha}(y) = n_{\alpha}(y)/n_{\alpha}^0$, where $n_{\alpha}^0$ is the mean number density of the cells of component $\alpha$ in the suspension; for a uniform distribution, $\hat n_{\alpha}(y) = 1$ at all $y$ positions. Figure~\ref{fig:pure_healthy_profile_different_spontaneous} shows the time-averaged wall-normal number density distribution profile for healthy RBCs in a homogeneous suspension. The error bars represent estimated error using the ``blocking" method \cite{Flyvbjerg1989}. A cell-free layer is evidently observed next to the wall ($y/a = 0$), which is a consequence of the competition between the effects of wall-induced migration \cite{Smart1991} and hydrodynamic pair collisions.
%\st{, with a thickness of $\sim$ 1 radius of an RBC. The formation of this layer is attributed to the hydrodynamic interactions between the cells and the wall that lead to migration of the cells away from the wall towards the center of the channel} \cite{Smart1991,Goldsmith19711578}. \st{In addition, t}
The two substantial peaks in the profile, which have been observed in prior experiments \cite{Tatsumi2019}, simulations \cite{FEDOSOV2010,Kumar:2013tu,KATANOV201557,Sinha:2016ki,Mehrabadi2016}, and theoretical predictions \cite{Rivera2016,Tatsumi2019} for suspensions of RBCs and other deformable capsules, indicate that the healthy RBCs accumulate both around the centerplane of the channel and near the wall right beyond the cell-free layer. In addition, we also determine the hematocrit profile by computing the local volume fraction of the RBCs. The time-averaged hematocrit profile is shown in FIG.~\ref{fig:healthy_hematocrit_profile}. We note that this profile closely resembles those in a previous numerical study by Zhao, Shaqfeh and Narsimhan \cite{Zhao2012} for an RBC suspension with similar parameters.
%\st{ The near-wall peak is a consequence of the competition between the hydrodynamic migration away from the wall and the diffusive motion due to the interparticle collisions in a confined suspension. This peak has also been observed in simulations for suspensions of deformable spherical and ellipsoidal capsules }\cite{Kumar:2013tu,Sinha:2016ki}\st{. The peak at the centerplane results from the parabolic profile of the undisturbed velocity field in pressure-driven flow, which has also been observed in prior experiments }\cite{Tatsumi2019}\st{, simulations }\cite{FEDOSOV2010,Zhao2012,KATANOV201557,Mehrabadi2016}\st{, and theoretical predictions }\cite{Rivera2016,Tatsumi2019}\st{ for suspensions of RBCs with varying volume fraction and cell deformability.}

\begin{figure}[t]
\centering
\captionsetup{justification=raggedright}
 \subfloat[]% caption for subfigure a
{
    \includegraphics[width=0.5\textwidth]{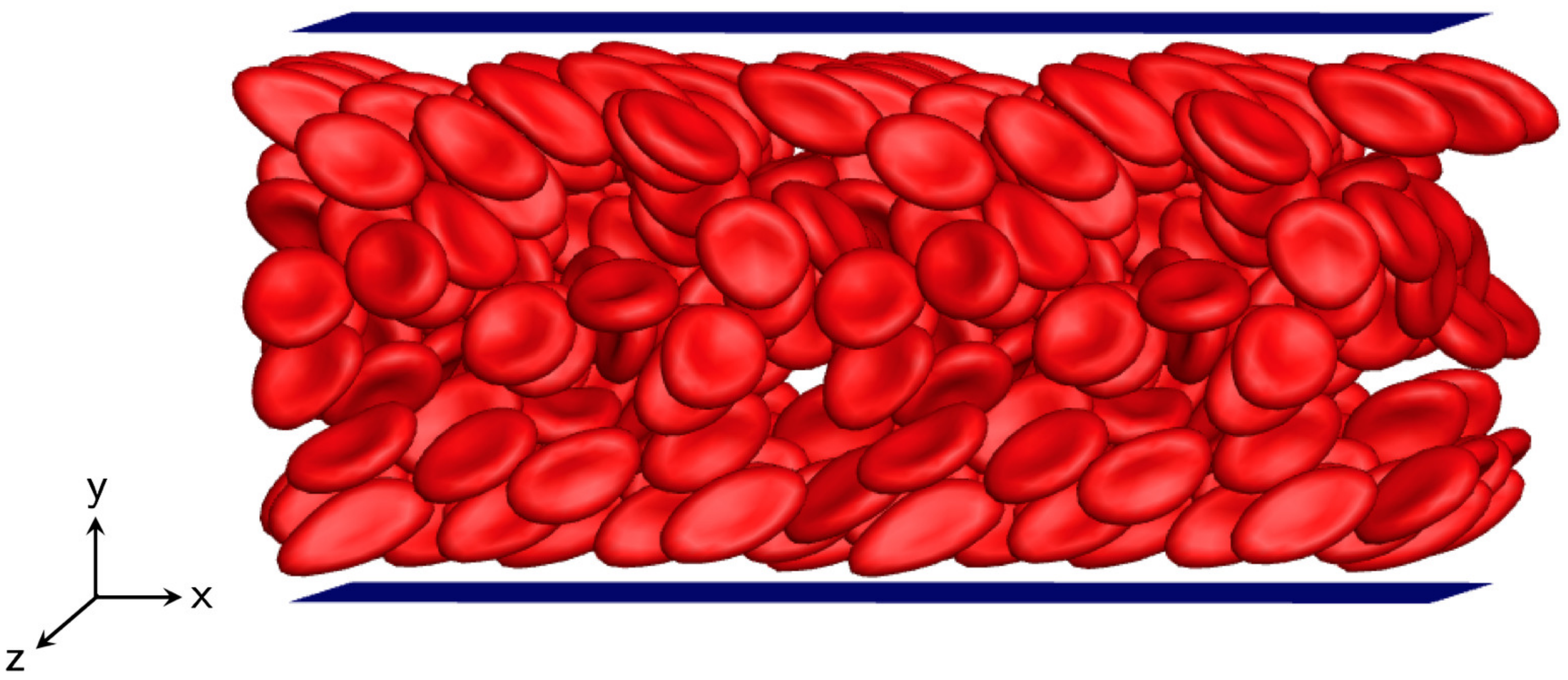}
    \label{fig:healthy_RBC_orientation_schematic}
}
 \subfloat[]% caption for subfigure a
{
    \includegraphics[width=0.49\textwidth]{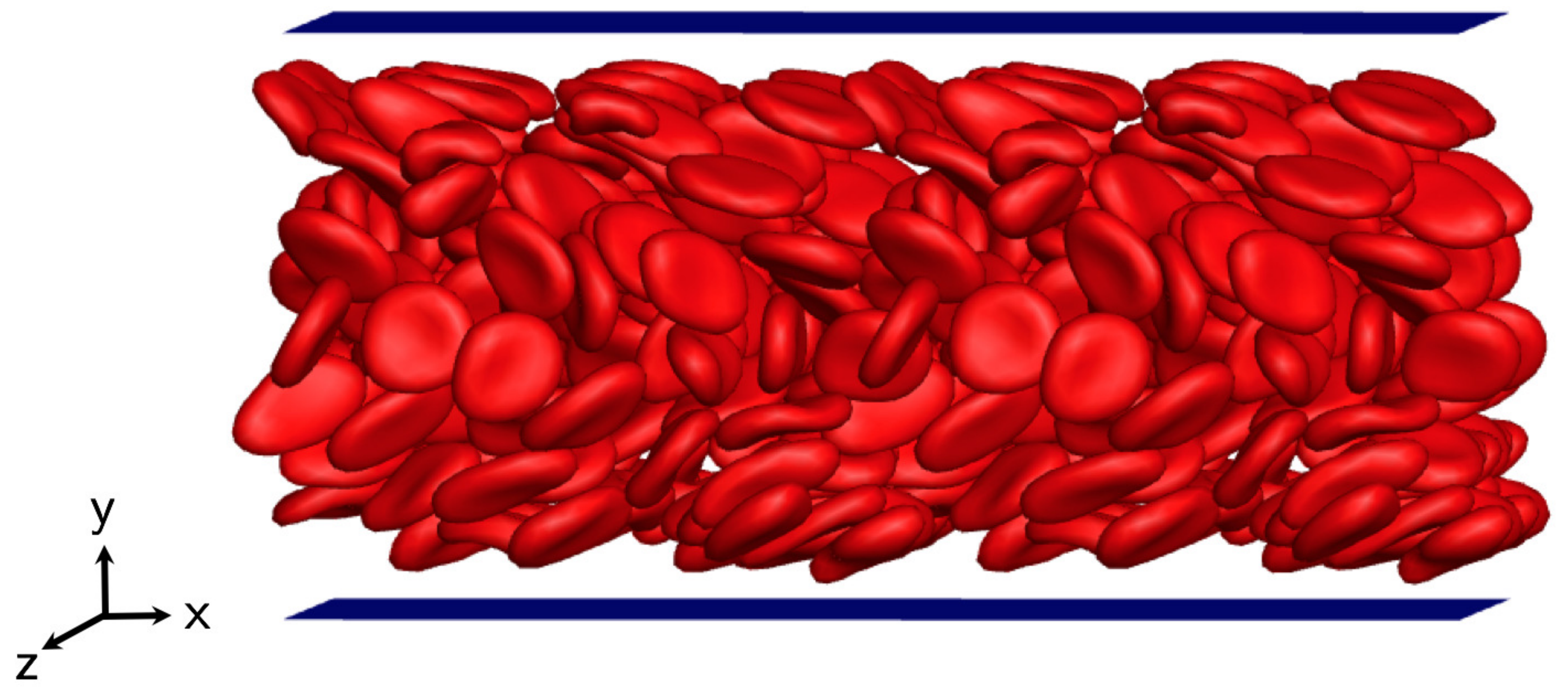}
    \label{fig:healthy_RBC_orientation_oblate_schematic}
}
\caption{Simulation snapshots for a homogeneous suspension at steady state of flexible biconcave discoidal capsules (healthy RBCs) with $\Ca = 1.6$ assuming either a biconcave discoidal (a) or an oblate spheroidal (b) spontaneous shape.}
  \label{fig:healthy_snapshots_different_spontaneous_shape}
  \end{figure}

Furthermore, in the near-wall region where the local shear rate is high, healthy RBCs approximate a ``rolling" orientational motion,  where the axis of symmetry (the short axis in the case of an oblate object) orients in the $z$-direction and the cell rolls like a wheel, as observed in the simulation snapshot (FIG.~\ref{fig:healthy_RBC_orientation_schematic}). Two periods of the simulation domain are shown in both $x$ and $z$ directions in this snapshot as well as all future snapshots for other suspensions. This rolling orbit assumed by an RBC at high shear rate, or a transition of the cell dynamics toward the rolling orbit upon increasing shear rate, has also been revealed in both experimental \cite{Goldsmith351,Bitbol1986,Lanotte13289} and numerical \cite{Cordasco:2013hb,Cordasco:2014go,Mendez2018} studies for single RBCs under physiological conditions in which the viscosity ratio between the fluids inside and outside the cell is typically $\lambda \approx 5$, although the viscosity ratio is set to $\lambda = 1$ in this work, again, to make the computational cost manageable. We note, though, that the spontaneous shape ($c_0$) has a nontrivial effect on the dynamics of single RBCs. For example, as shown in FIG.~\ref{fig:healthy_snapshots_different_spontaneous_shape}, the near-wall healthy RBCs reorient from a rolling orbit into a ``tank-treading" configuration when their spontaneous shape is changed from a biconcave discoid to an oblate spheroid, consistent with the findings in a prior numerical investigation by Sinha and Graham \cite{Sinha:2015wt}. (In tank-treading, points on the surface of the cell move but the cell shape itself remains nearly time-independent.)  In spite of the change in orientational dynamics when the spontaneous shape is changed, no substantial change is observed in the wall-normal number density profile for healthy RBCs (FIG.~\ref{fig:pure_healthy_profile_different_spontaneous}). Indeed, we have further verified that for the binary suspensions that will be presented next, the segregation behavior and dynamics of stiff cells are only weakly affected by the orientational dynamics of healthy (flexible) RBCs.
%\st{ In the near-centerplane region where the local shear rate is lower, on the other hand, the cell orientations exhibit relatively larger inhomogeneity, which is likely caused by the high number density of cells near the centerplane coupled with the interparticle interactions.}

      \subsubsection{Binary suspension of flexible and stiff RBCs: isolated effect of rigidity difference} 
      \label{sec:healthy_stiff_RBCs}
\begin{figure}[t]
\centering
\captionsetup{justification=raggedright}
 \subfloat[]% caption for subfigure a
{
    \includegraphics[width=0.5\textwidth]{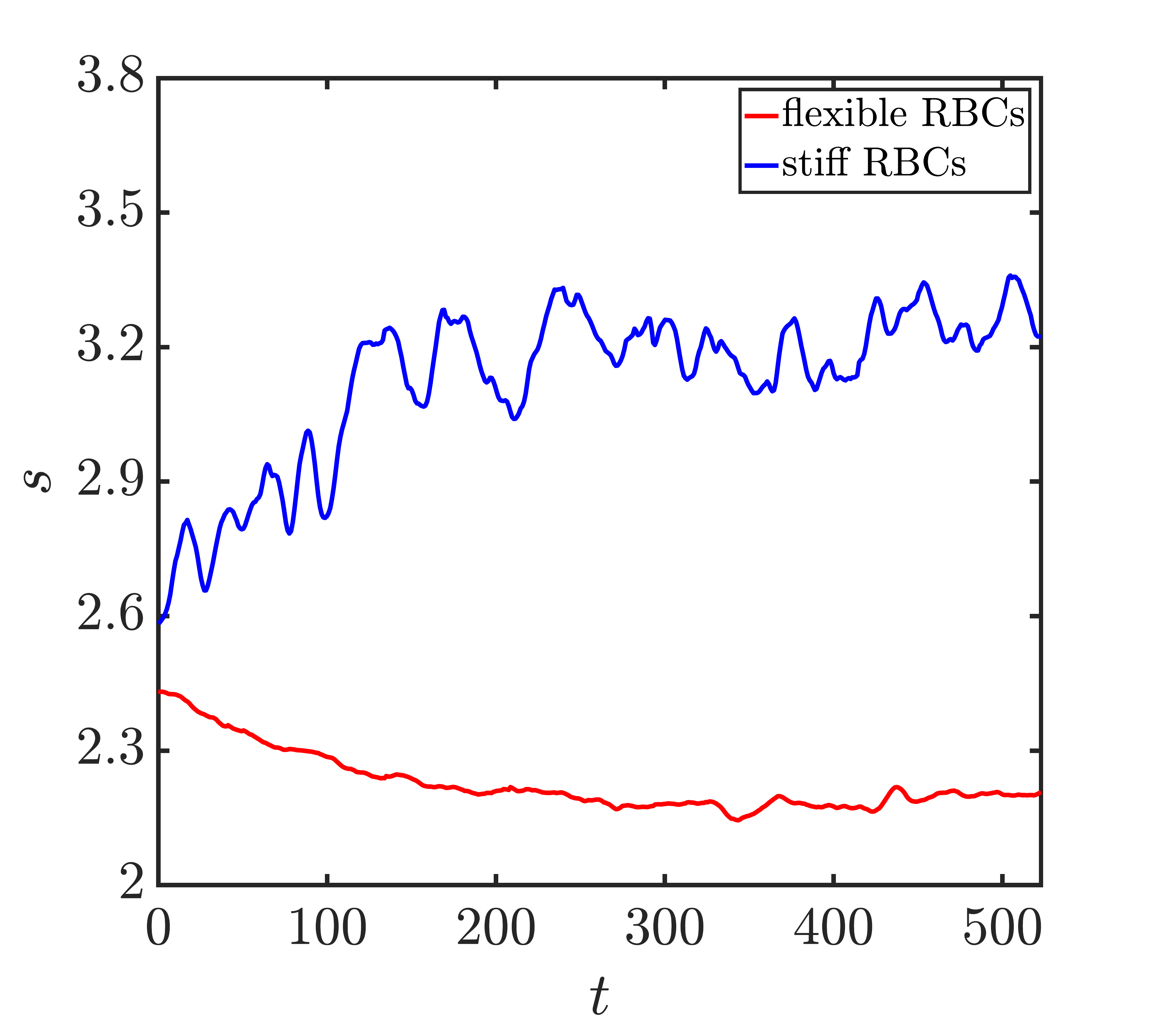}
    \label{fig:segregation_healthy_stiff}
}
 \subfloat[]% caption for subfigure a
{
    \includegraphics[width=0.5\textwidth]{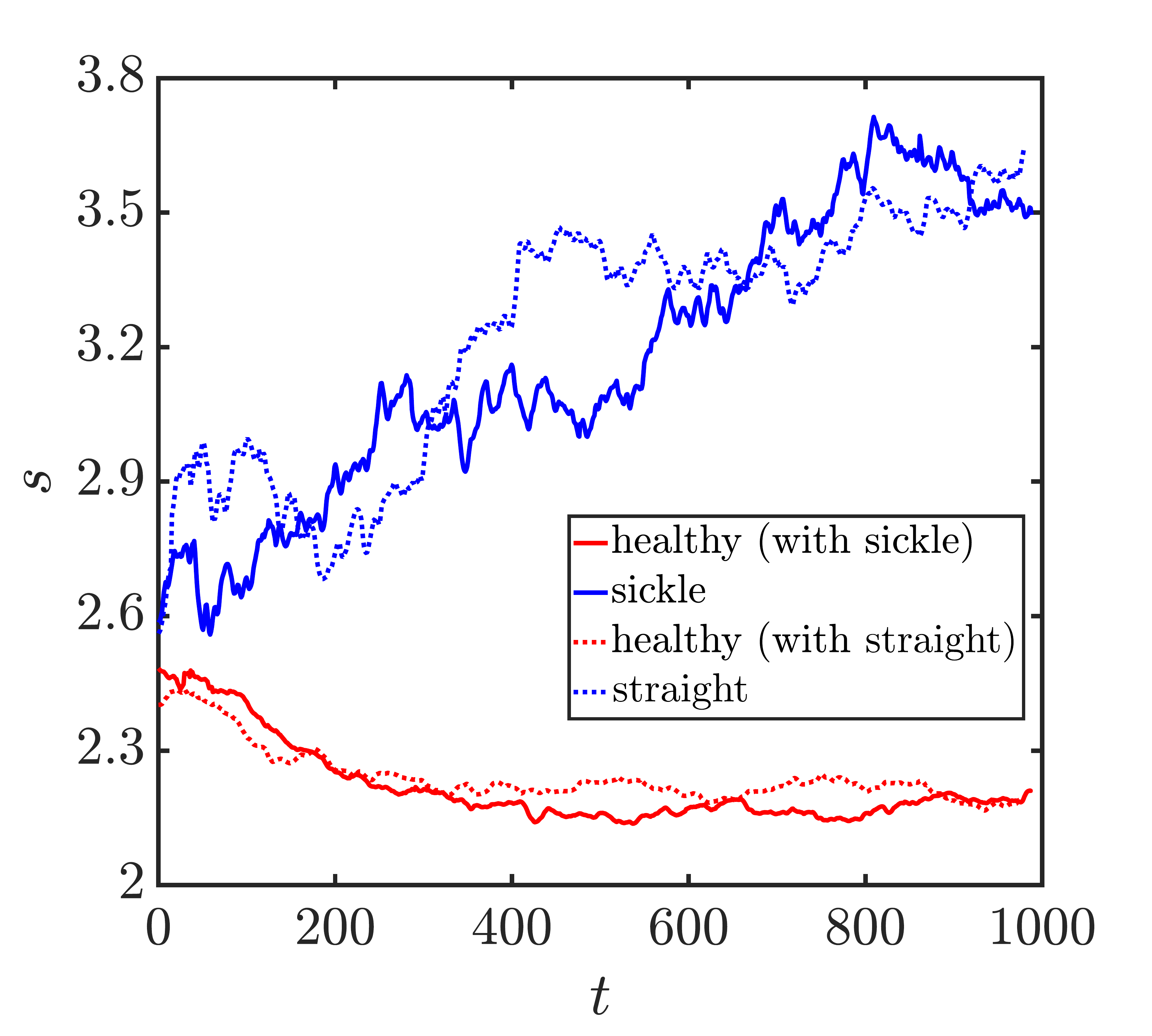}
    \label{fig:segregation_healthy_sickle_prolate}
}
\caption{Time evolution of the parameter $s = \big \langle (y_{cm} - H)^2 \big \rangle^{1/2}/a$ for (a) flexible (i.e.~healthy) RBCs in suspension with stiff RBCs, and for (b) healthy RBCs in suspension with sickle cells and straight prolate capsules, respectively.}
  \label{fig:segregation_parameter_all_cases}
  \end{figure}

We now consider a binary suspension of flexible ($\Ca_p = 1.6$) and stiff ($\Ca_t = 0.4$) RBCs with a number density ratio $n_p/n_t = 9$. To characterize the collective dynamics of the cells of each component, we define a\soutold{ so-called segregation} parameter $s = \big \langle (y_{cm} - H)^2 \big \rangle^{1/2}/a$, which \soutold{is essentially}\XZrevise{measures} the \soutold{averaged}\XZrevise{root-mean-square (RMS)} distance of the cells from the centerplane of the channel; here $y_{cm}$ is the center-of-mass position of an arbitrary cell in the wall-normal direction. The time evolution of $s$ for the cells of each component is plotted in FIG.~\ref{fig:segregation_healthy_stiff}. It is obvious that though starting with comparable values for both components, $s$ decreases for flexible RBCs while increasing for stiff RBCs, before reaching a plateau (steady state) at $s_p \approx 2.2$ and $s_t \approx 3.2$, respectively, indicating a segregation behavior in the binary suspension.

\begin{figure}[t]
  \centering
  \captionsetup{justification=raggedright}
  \includegraphics[width=0.7\textwidth]{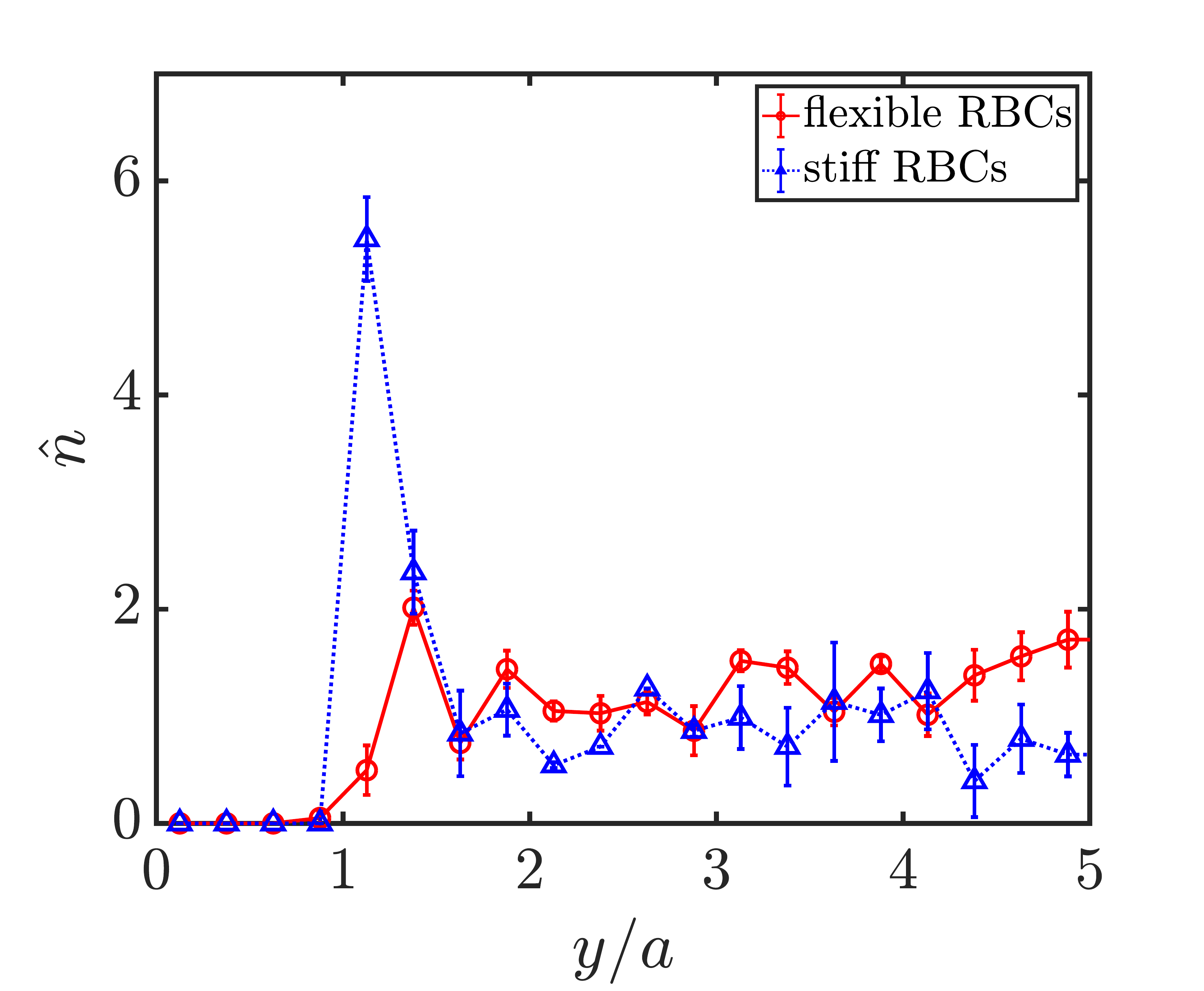}
  \caption{Wall-normal number density profiles $\hat n$ for flexible (solid red) and stiff (dotted blue) RBCs in a binary suspension at steady state.}
  \label{fig:healthy_stiff_profile_comparison}
  \end{figure}

\begin{figure}[h]
\centering
\captionsetup{justification=raggedright}
 \subfloat[]% caption for subfigure a
{
    \includegraphics[width=0.48\textwidth]{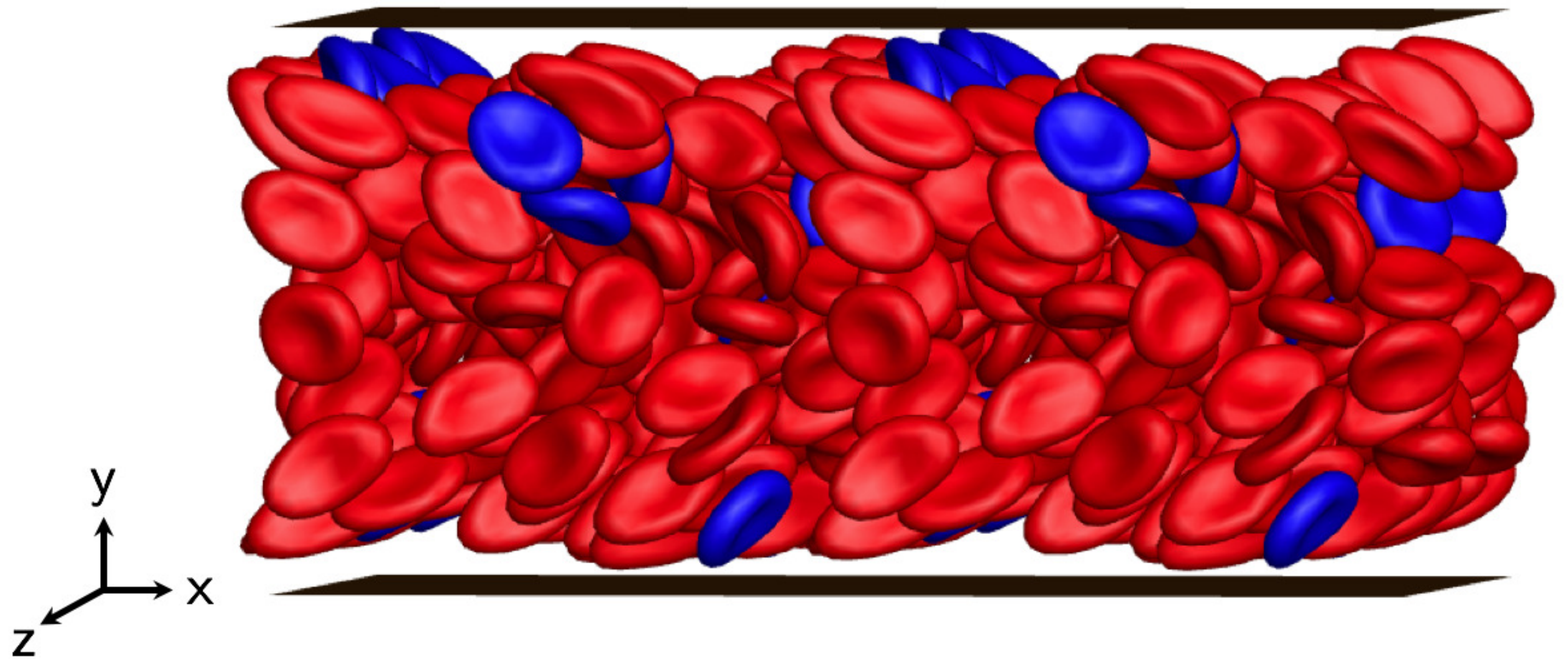}
    \label{fig:healthy_stiff_side}
}
 \subfloat[]% caption for subfigure a
{
    \includegraphics[width=0.48\textwidth]{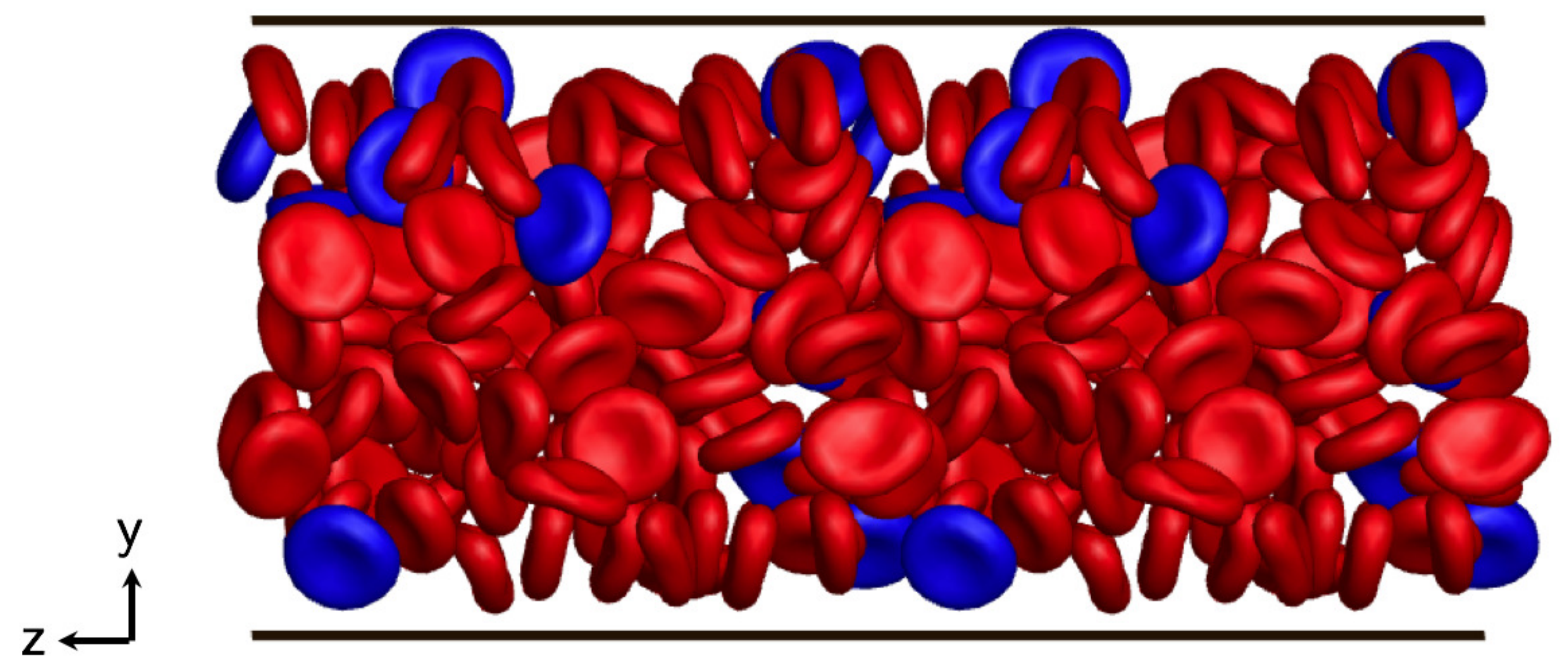}
    \label{fig:healthy_stiff_front}
}
\caption{Simulation snapshots ((a) side view, (b) front view) for a binary suspension of flexible (red) and stiff (blue) RBCs at steady state.}
  \label{fig:healthy_stiff_snapshots}
  \end{figure}

Furthermore, to quantify the cell distributions, the steady-state wall-normal number density profiles $\hat n(y)$ are computed for both components and shown in FIG.~\ref{fig:healthy_stiff_profile_comparison}. In this case where stiff RBCs are dilute, the key observation is that, in contrast to flexible RBCs that exhibit a profile similar to that in their pure suspension, a majority of stiff RBCs are drained from the center of the channel and aggregate in the near-wall region, undergoing substantial margination. This margination and segregation behavior is also evident in snapshots from simulation for this binary suspension (FIG.~\ref{fig:healthy_stiff_snapshots}). These results demonstrate that rigidity difference by itself is sufficient to induce the segregation behavior in a binary suspension, which is consistent with the findings in prior studies \cite{Kumar:2011dd,Kumar:2013tu,Rivera2016} considering a suspension of spherical capsules that differ only in rigidity. The marginated stiff RBCs are found to take an approximate in-plane ``tumbling" orbit near the walls: i.e. the axis of symmetry undergoes as an approximately circular trajectory in the $x$-$y$ plane. In addition, we have verified that the existence of a small number fraction of stiff RBCs makes little difference to the orientational dynamics of flexible RBCs (not shown).

%\st{The results above demonstrate that rigidity difference by itself is sufficient to induce the segregation behavior in a binary suspension, which has also been found in simulations for a suspension of spherical capsules with inhomogeneous rigidity }\cite{Kumar:2013tu,Rivera2016}\st{. As an application in biotechnology, rapid separation of diseased RBCs with increased stiffness (e.g. malaria-infected RBCs) has been achieved based on cell margination or lateral displacement using microfluidics }\cite{Hou2010,Guo2016}.

      \subsubsection{Binary suspension of healthy and sickle RBCs: combined effect of rigidity, shape, and size differences} 
      \label{sec:healthy_sickle_RBCs}
In this section we present simulation results for a binary suspension of healthy and sickle RBCs (specifically ISCs), which is the focus of this work. Again, healthy RBCs are modeled as flexible biconcave discoidal capsules ($\Ca_p = 1.6$), while sickle cells as stiff curved prolate capsules ($\Ca_t = 0.4$); the number density ratio between these two components is $n_p/n_t = 9$. In the meantime, results for the case in which sickle cells, the trace component, are replaced by stiff straight prolate capsules ($\Ca_t = 0.4$ and $X_t = 0.1$) are also presented for comparison to illustrate the effect of curvature for the trace component. 

\begin{figure}[b]
\centering
\captionsetup{justification=raggedright}
 \subfloat[]% caption for subfigure a
{
    \includegraphics[width=0.5\textwidth]{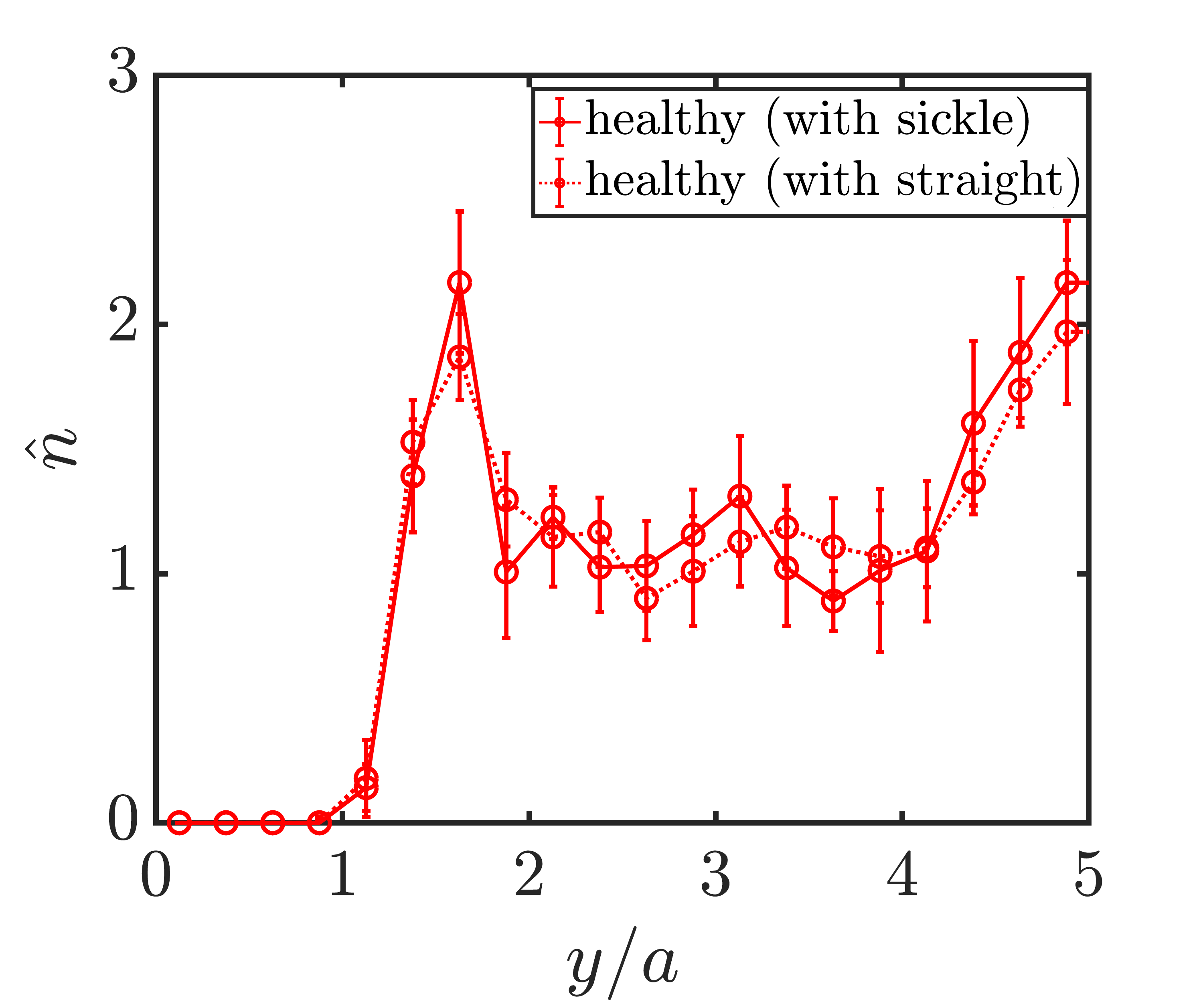}
    \label{fig:healthy_profile_comparison}
}
 \subfloat[]% caption for subfigure a
{
    \includegraphics[width=0.5\textwidth]{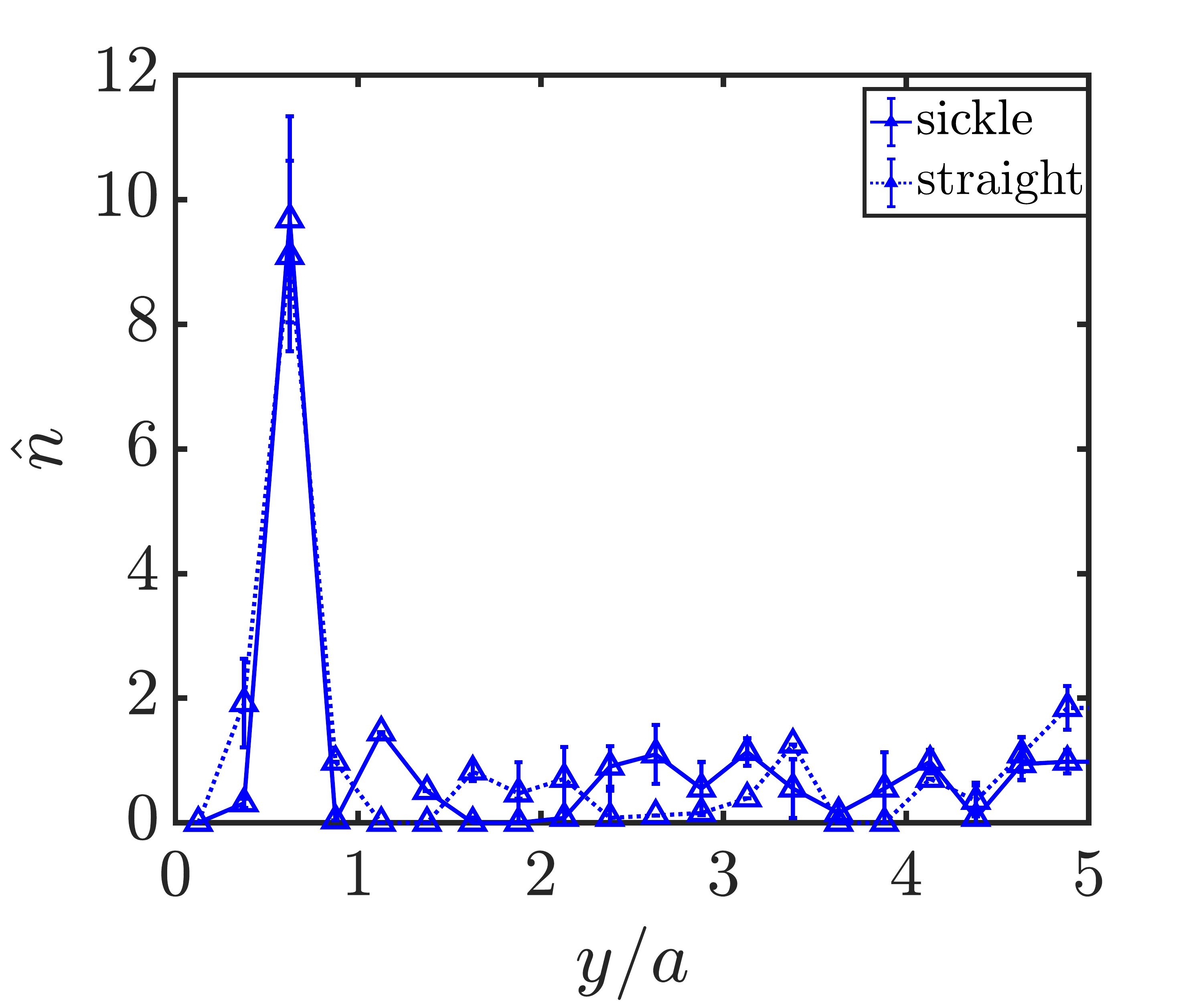}
    \label{fig:sickle_prolate_profile_comparison}
}
\caption{Wall-normal number density profiles $\hat n$ for (a) healthy RBCs in suspension with sickle cells (solid red) and straight prolate capsules (dotted red), and (b) for sickle cells (solid blue) and straight prolate capsules (dotted blue) in suspension with healthy RBCs, respectively.}
  \label{fig:curvature_effects_on_profile}
  \end{figure}

\begin{figure}[b]
\centering
\captionsetup{justification=raggedright}
 \subfloat[]% caption for subfigure a
{
    \includegraphics[width=0.48\textwidth]{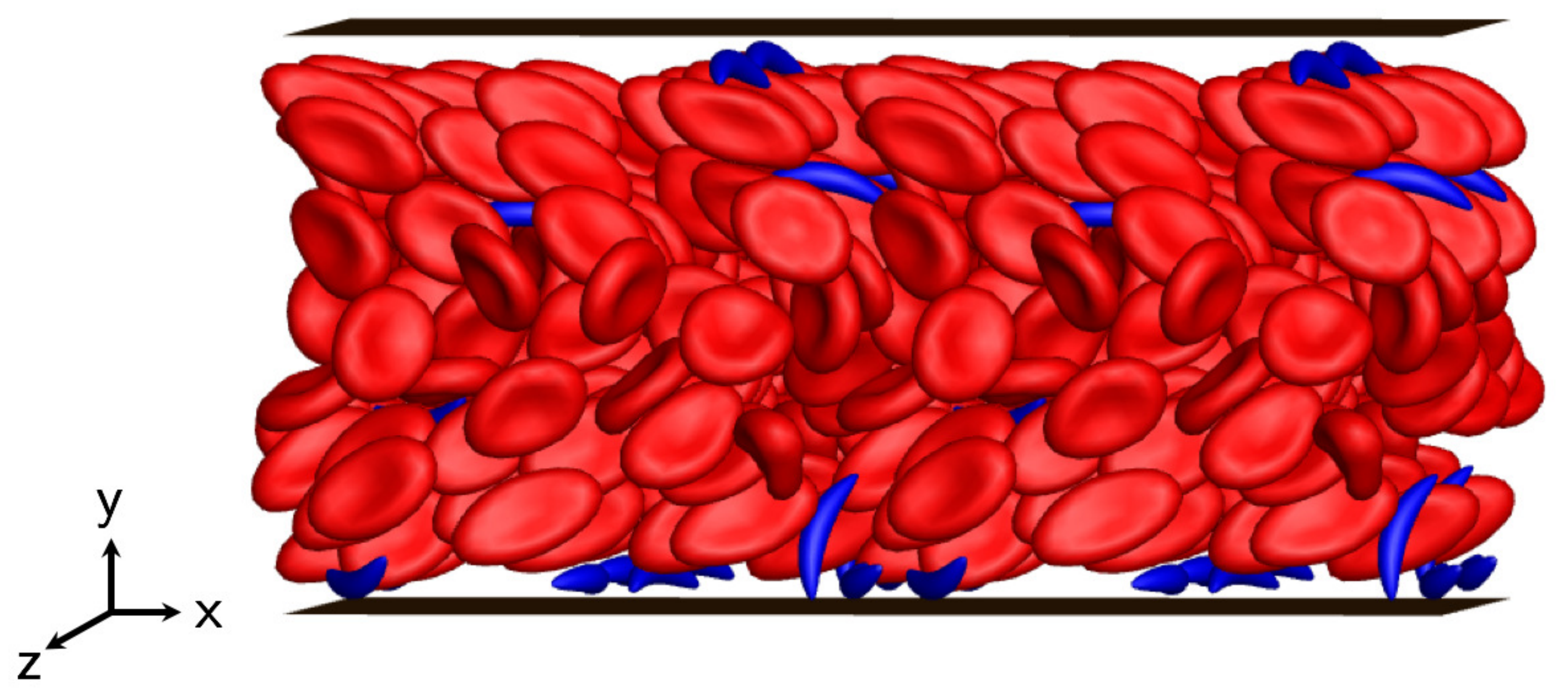}
    \label{fig:healthy_sickle_side_view}
}
 \subfloat[]% caption for subfigure a
{
    \includegraphics[width=0.48\textwidth]{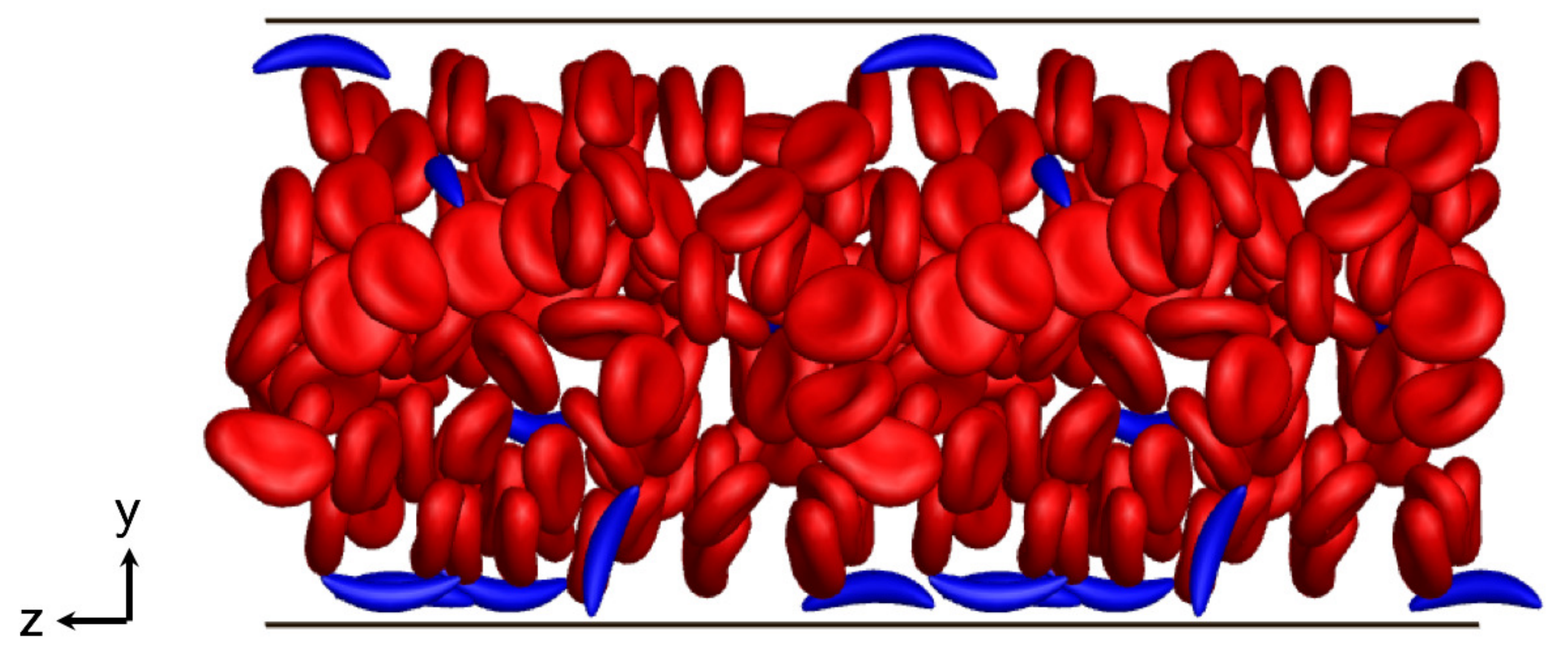}
    \label{fig:healthy_sickle_front_view}
}
\\
 \subfloat[]% caption for subfigure a
{
    \includegraphics[width=0.48\textwidth]{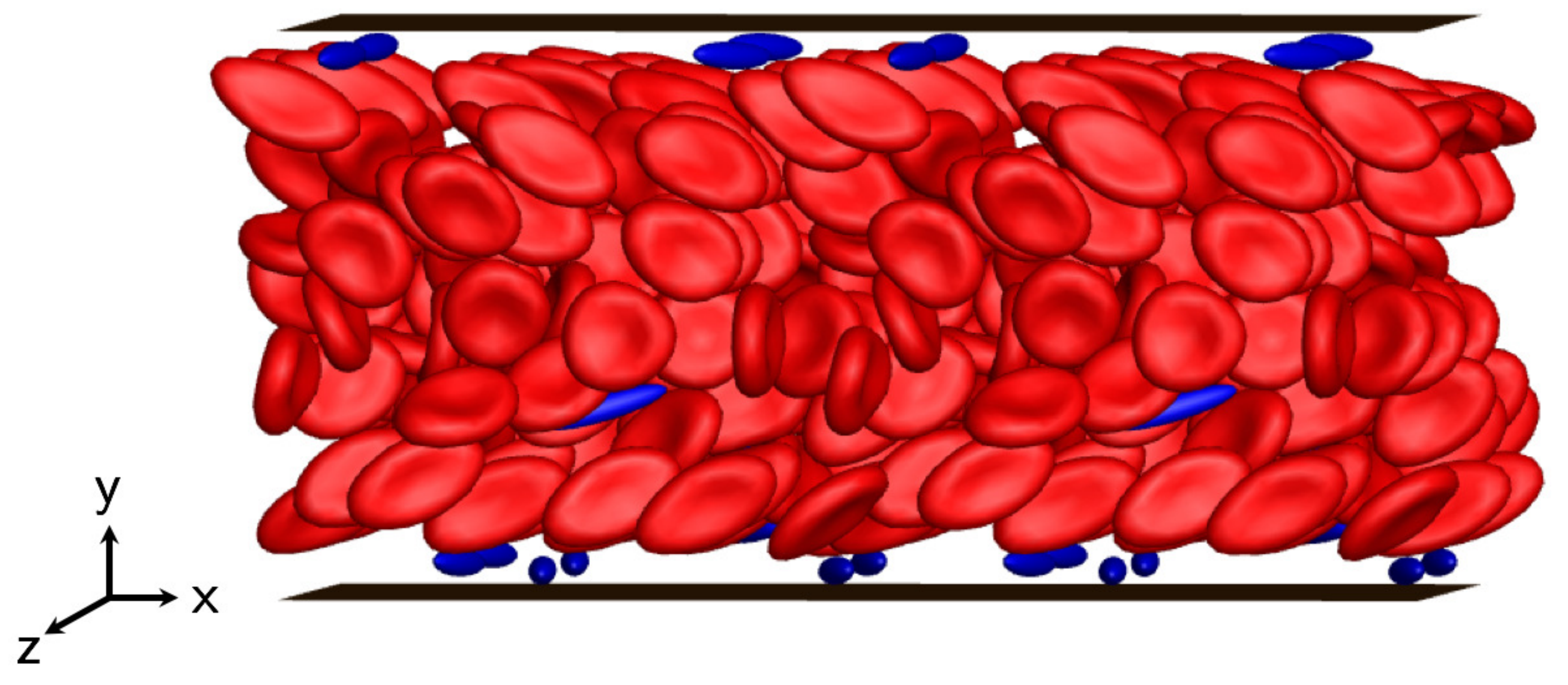}
    \label{fig:healthy_prolate_side_view}
}
 \subfloat[]% caption for subfigure a
{
    \includegraphics[width=0.48\textwidth]{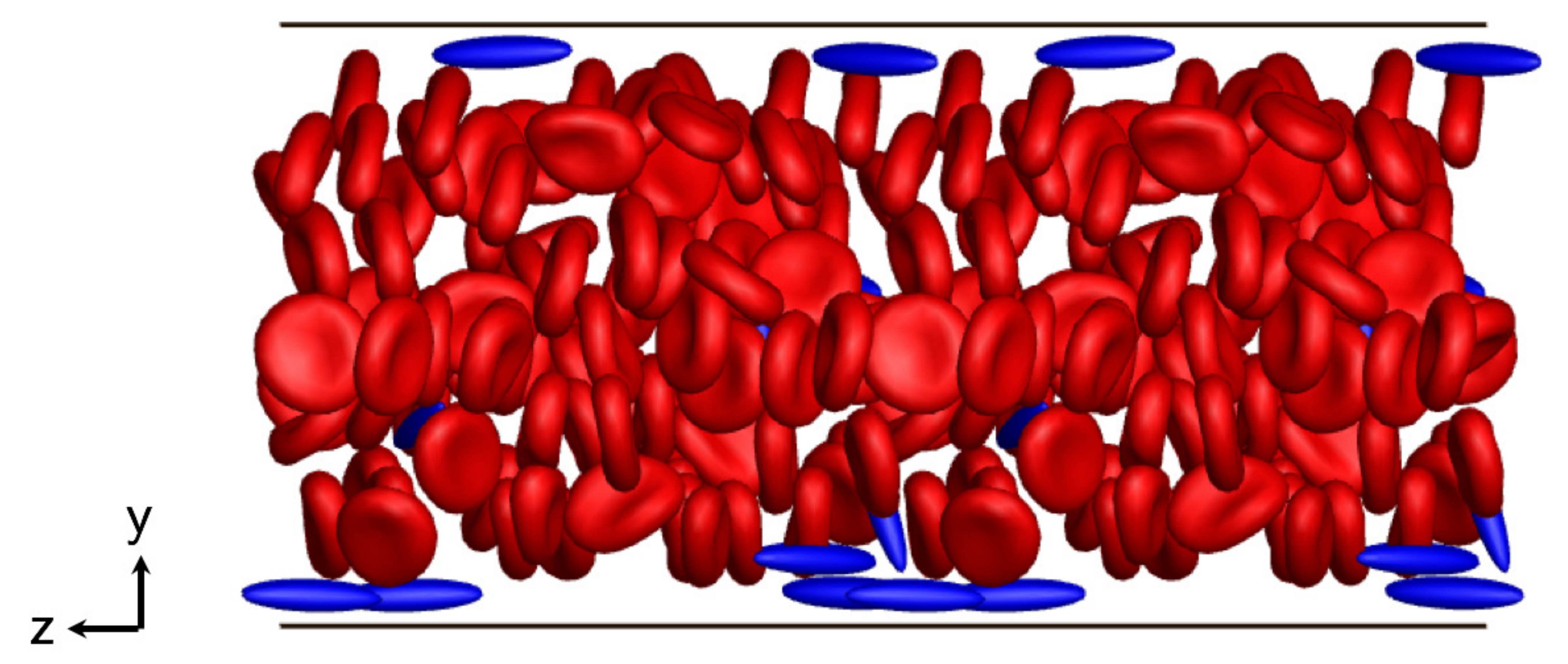}
    \label{fig:healthy_prolate_front_view}
}
\caption{Simulation snapshots (left: side view; right: front view) for binary suspensions of healthy RBCs with sickle cells (a,b) and straight prolate capsules (c,d), respectively, at steady state.}
  \label{fig:healthy_sickle_prolate_snapshots}
  \end{figure}

%  \begin{figure}[t]
%  \centering
%  \captionsetup{justification=raggedright}
%  \includegraphics[width=0.7\textwidth]{./segregation_comparison.pdf}
%  \caption{Time evolution of $s$ for healthy RBCs in suspension with sickle cells (solid red) and straight prolate capsules (dotted red), and for sickle cells (solid blue) and straight prolate capsules (dotted blue) in suspension with healthy RBCs, respectively.}
%  \label{fig:segregation_healthy_sickle_prolate}
%  \end{figure}
%
Figure~\ref{fig:segregation_healthy_sickle_prolate} shows the \XZrevise{time} evolution of the\soutold{ segregation} parameter $s = \big \langle (y_{cm} - H)^2 \big \rangle^{1/2}/a$\soutold{ with time} for each component in both suspensions.
%\st{Similar to the previous case with flexible and stiff RBCs, here the}
Segregation behavior is observed for both cases. 
%\st{It shows from direct comparison that the evolution of $s$ for healthy RBCs is not greatly affected by the curvature of the trace component. }
The trends of increase in $s$ display minor differences for sickle cells and straight prolate capsules, both reaching a plateau (steady state) at $s_t \approx 3.6$. This steady-state value is greater than that for stiff RBCs ($s_t \approx 3.2$). We also note that compared to the case with stiff RBCs, the cases with sickle cells and straight prolate capsules take a longer time to reach steady state.

\begin{figure}[t]
\centering
\captionsetup{justification=raggedright}
 \subfloat[]% caption for subfigure a
{
    \includegraphics[width=0.48\textwidth]{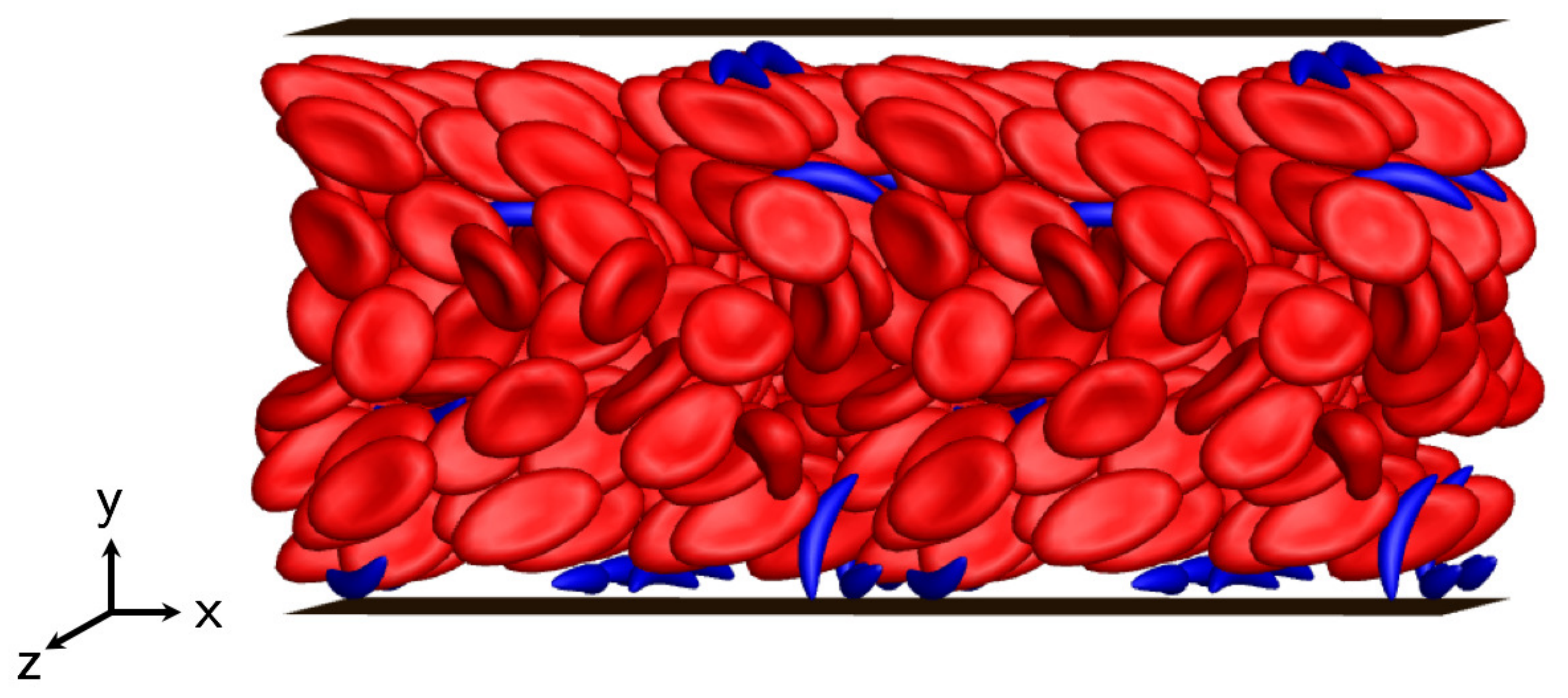}
    \label{fig:healthy_sickle_side_view_2}
}
 \subfloat[]% caption for subfigure a
{
    \includegraphics[width=0.48\textwidth]{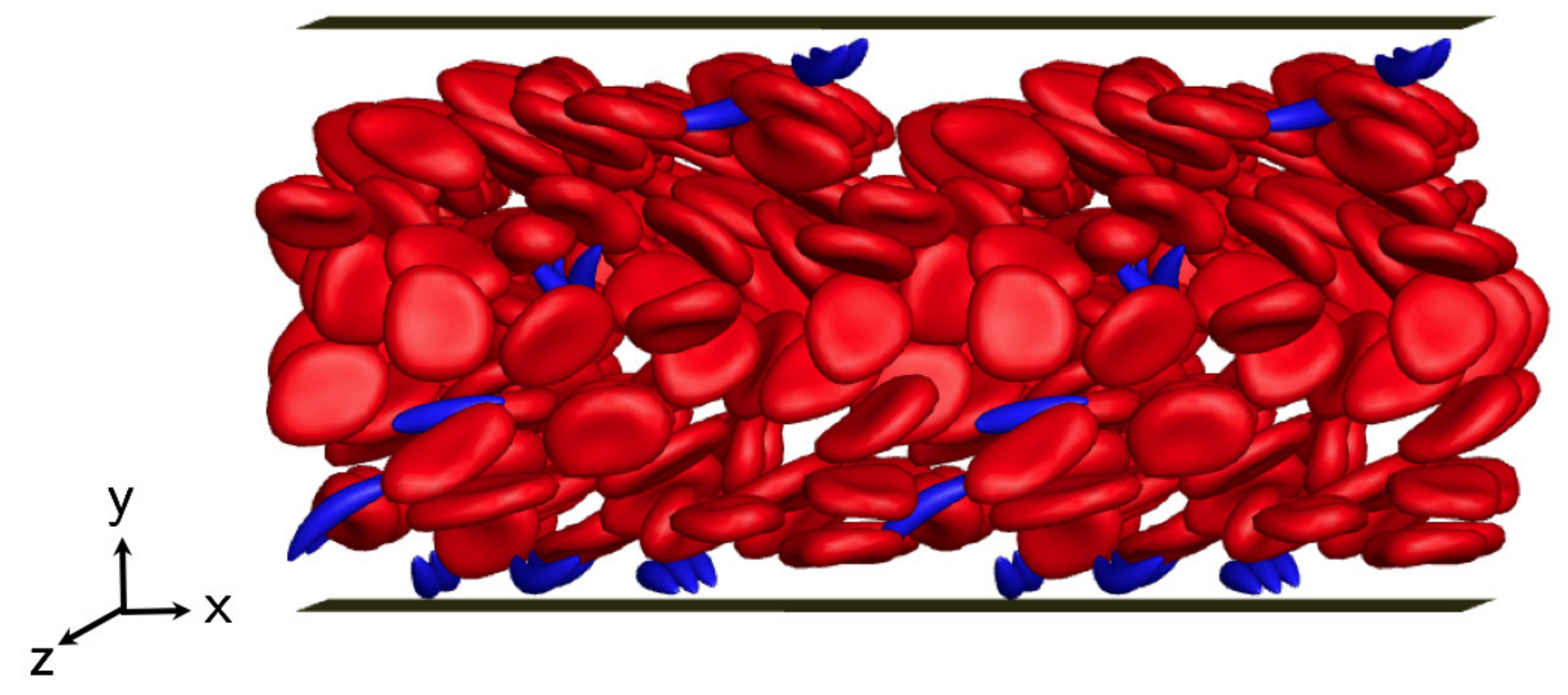}
    \label{fig:healthy_sickle_side_view_oblate_spontaneous}
}
\\
 \subfloat[]% caption for subfigure a
{
    \includegraphics[width=0.5\textwidth]{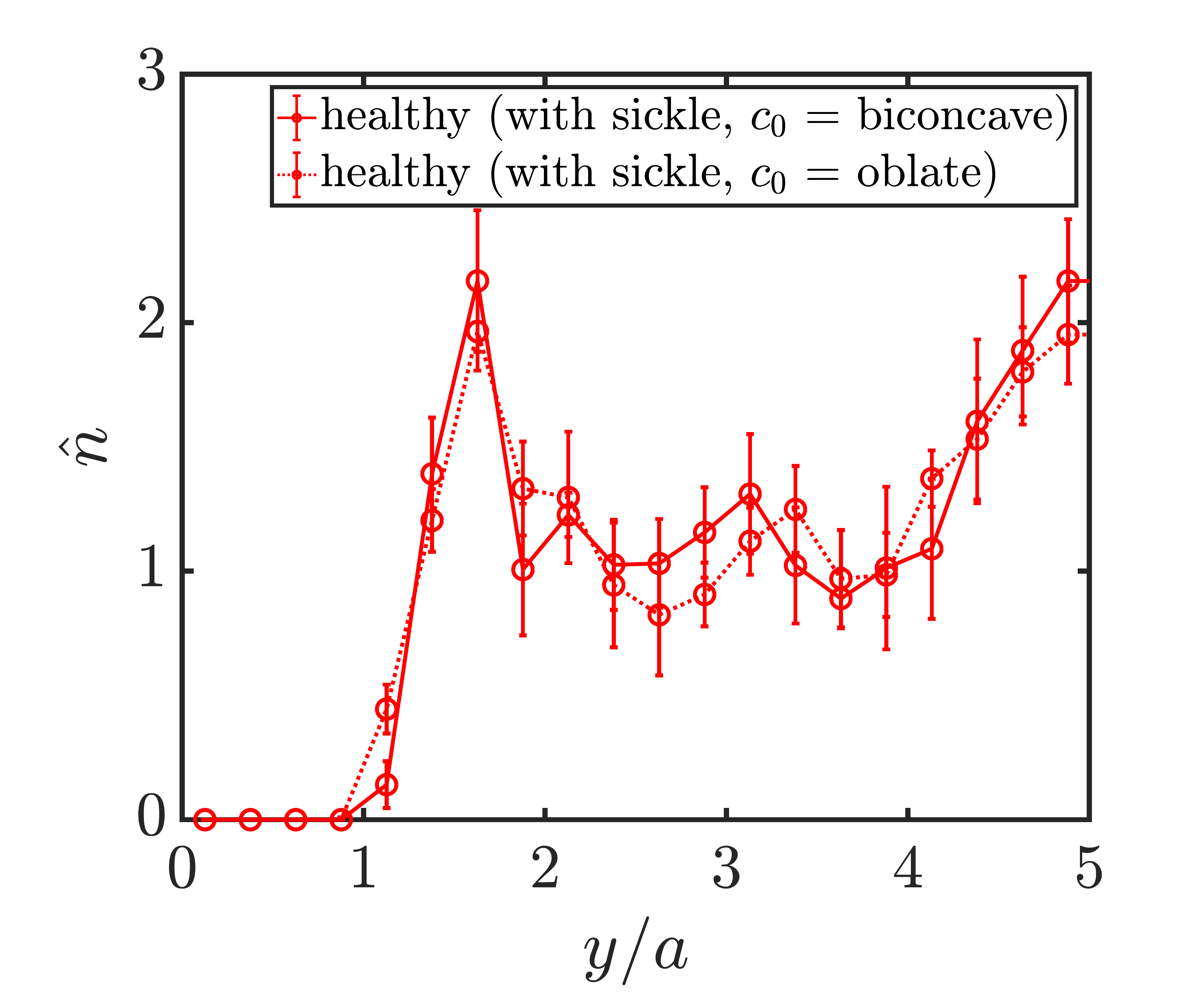}
    \label{fig:healthy_profile_comparison_different_spontaneous}
}
 \subfloat[]% caption for subfigure a
{
    \includegraphics[width=0.5\textwidth]{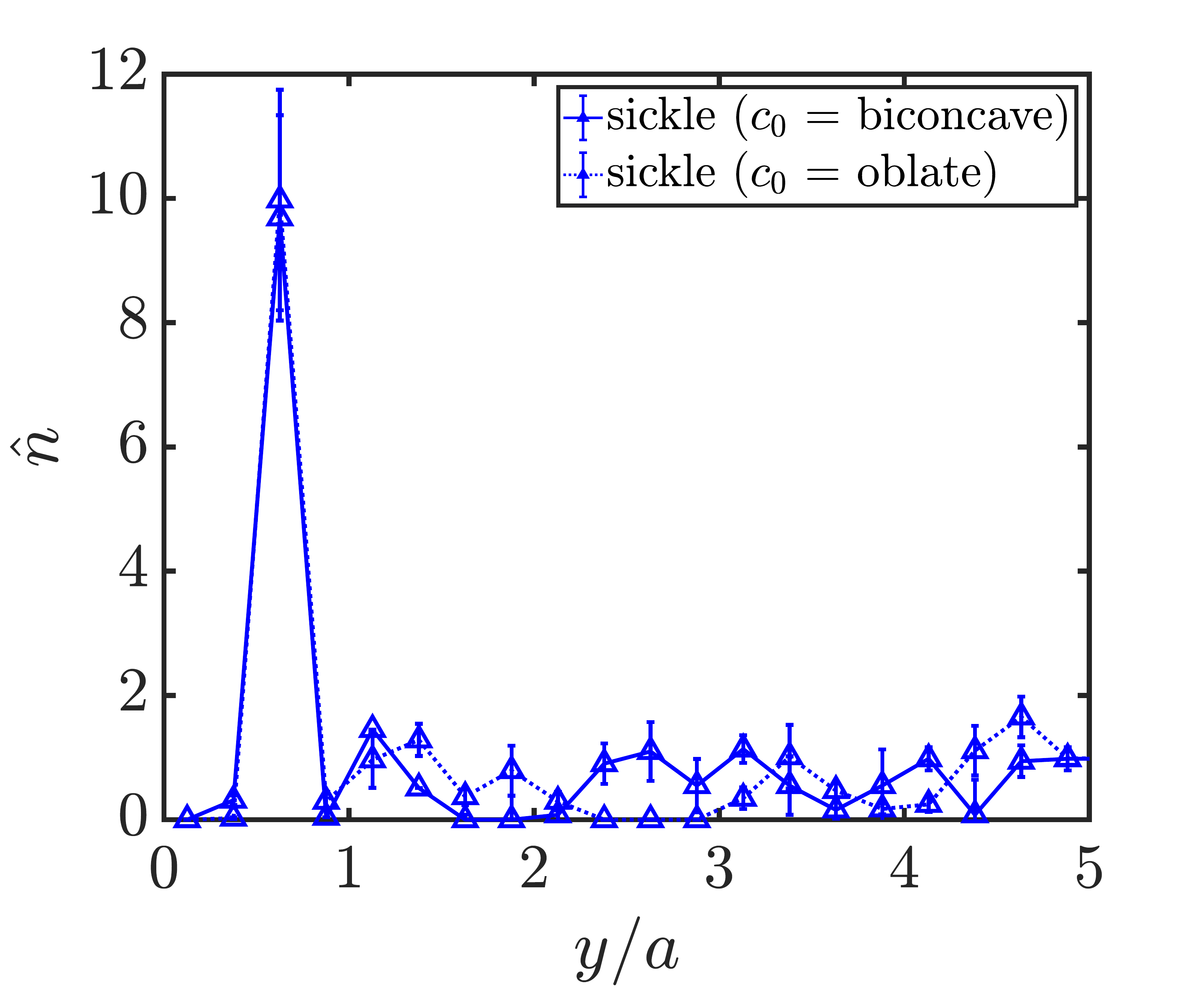}
    \label{fig:sickle_profile_comparison_different_spontaneous}
}
\caption{Simulation snapshots (a,b) for a steady-state binary suspension of healthy RBCs with sickle cells, as well as wall-normal number density profiles $\hat n$ (c,d) for both components, assuming a biconcave discoidal (a) and an oblate spheroidal (b) spontaneous shape ($c_0$) for healthy RBCs, respectively.}
  \label{fig:healthy_sickle_profile_different_spontaneous}
  \end{figure}

More details can be revealed by computing the cross-stream distribution profiles for each component in both suspensions. In FIG.~\ref{fig:healthy_profile_comparison}, a cell-free layer is observed in the wall-normal number density profiles $\hat n$ for healthy RBCs in both cases, as well as the two peaks both around the centerplane of the channel and right beyond the cell-free layer. In general, the $\hat n$ profiles for healthy RBCs in these two suspensions are similar to those in previous cases. For sickle cells and straight prolate capsules, the profiles both show a high near-wall peak inside the cell-free layer (FIG.~\ref{fig:sickle_prolate_profile_comparison}), which suggests that the trace component in both cases is largely drained from the bulk of the suspension, displaying strong margination as seen in the simulation snapshots in FIG.~\ref{fig:healthy_sickle_prolate_snapshots}. No substantial differences are observed in the $\hat n$ profiles for both cases. We also show in FIG.~\ref{fig:healthy_sickle_profile_different_spontaneous} that when the spontaneous shape ($c_0$) of healthy RBCs is changed from a biconcave discoid to an oblate spheroid, although an orientational transformation is observed for the dynamics of healthy RBCs (FIGs.~\ref{fig:healthy_sickle_side_view_2} and~\ref{fig:healthy_sickle_side_view_oblate_spontaneous}), insignificant differences are found in the distribution profiles for both healthy RBCs and sickle cells in a binary suspension (FIGs.~\ref{fig:healthy_profile_comparison_different_spontaneous} and~\ref{fig:sickle_profile_comparison_different_spontaneous}).

%\begin{figure}[h]
%\centering
%\captionsetup{justification=raggedright}
% \subfloat[]% caption for subfigure a
%{
%    \includegraphics[width=0.5\textwidth]{./healthy_different_Lx.pdf}
%    \label{fig:healthy_different_Lx}
%}
% \subfloat[]% caption for subfigure a
%{
%    \includegraphics[width=0.5\textwidth]{./sickle_different_Lx.pdf}
%    \label{fig:sickle_different_Lx}
%}
%\caption{Wall-normal number density profiles $\hat n$ for healthy RBCs (a) and sickle cells (b), respectively, in a binary suspension with different spatial periods of the simulation domain: $L_x = 10a$ (solid lines) and $L_x = 20a$ (dotted lines).\MDG{how long did you run in the longer domain before averaging?}\MDG{I don't think we need to include a plot of this in the paper -- we can just say that we did simulations in a domain twice as long and found quantitatively similar results. Put the plot in your thesis.}}
%  \label{fig:healthy_sickle_profiles_different_Lx}
%  \end{figure}  
  \begin{figure}[h]
  \centering
  \captionsetup{justification=raggedright}
  \includegraphics[width=0.8\textwidth]{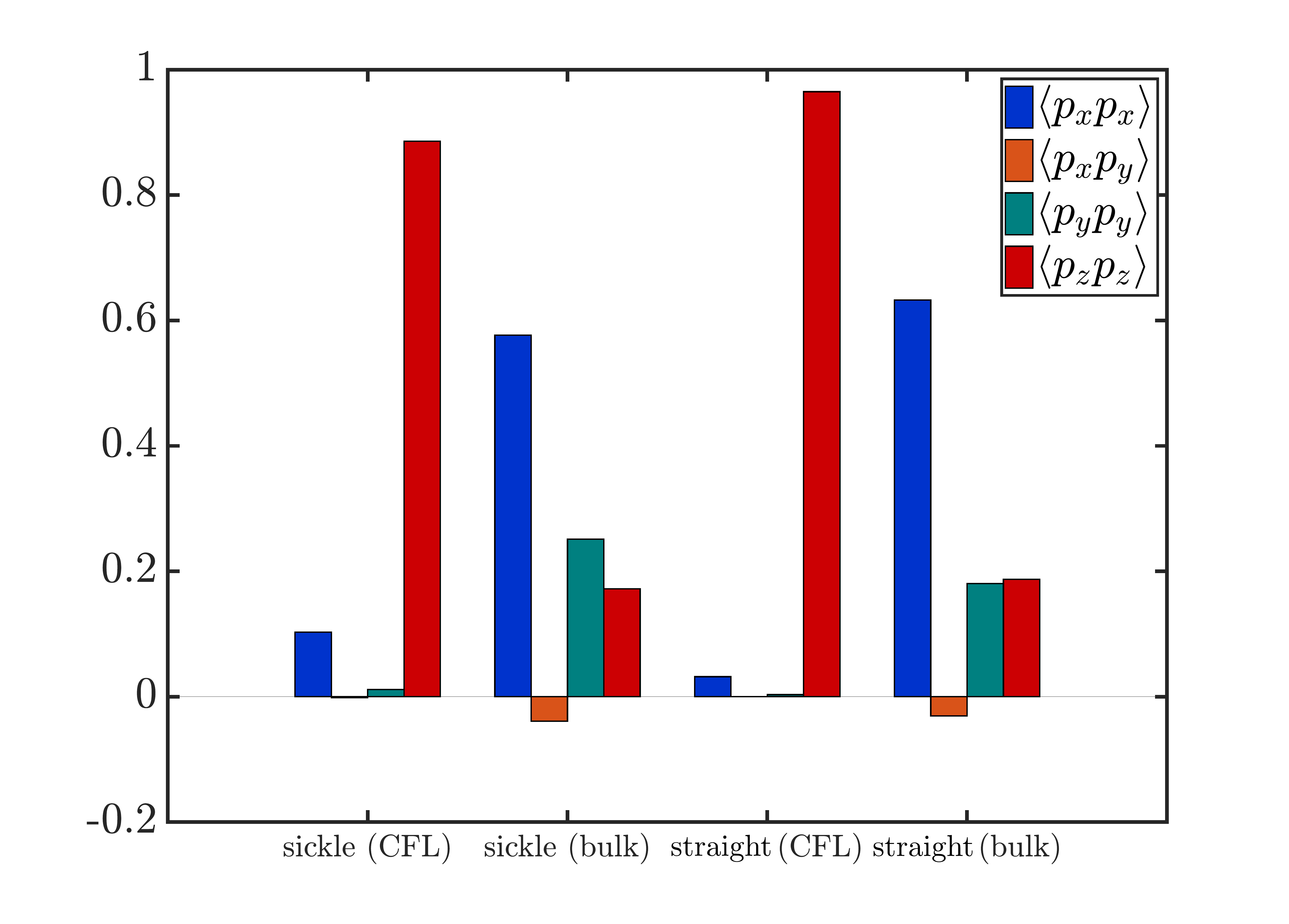}
  \caption{Magnitude of the $x$-$x$, $x$-$y$, $y$-$y$, and $z$-$z$ components of the tensor $\textbf{S} = \langle \textbf{pp} \rangle$ for sickle cells and straight prolate capsules inside the cell-free layer (labelled ``CFL") and in the bulk of the suspension (labelled ``bulk"), respectively.}
  \label{fig:sickle_prolate_orbital_dynamics}
  \end{figure}

It is noteworthy that the near-wall peaks in the $\hat n$ profiles for sickle cells and straight prolate capsules (FIG.~\ref{fig:sickle_prolate_profile_comparison}) are both of greater magnitude and closer to the wall compared to that in the profile for stiff RBCs (FIG.~\ref{fig:healthy_stiff_profile_comparison}). To explain this, we first characterize the orbital dynamics of sickle cells and straight prolate capsules in the respective suspension. Here the cells (capsules) are classified into two groups based on their wall-normal center-of-mass positions: one for the cells inside the cell-free layer, and the other for those in the bulk of the suspension. The instantaneous orientation of a single sickle cell or straight prolate capsule in flow is given by a unit end-to-end vector $\textbf{p}$ that connects the two tips of the cell (capsule). We then define a tensor $\textbf{S}$ to quantify the ensemble-averaged orientation of the cells in each group, given as
\begin{equation} \label{eq:orientation_tensor}
\mathbf{S} = \langle \mathbf{pp} \rangle= 
\begin{pmatrix} 
  \langle p_x p_x \rangle & \langle p_x p_y \rangle & \langle p_x p_z \rangle \\ 
  \langle p_y p_x \rangle & \langle p_y p_y \rangle & \langle p_y p_z \rangle \\
  \langle p_z p_x \rangle & \langle p_z p_y \rangle & \langle p_z p_z \rangle
\end{pmatrix}
.
\end{equation}  
In particular, we care about the magnitude of the components $\langle p_x p_x \rangle$, $\langle p_x p_y \rangle (=\langle p_y p_x \rangle)$, $\langle p_y p_y \rangle$, and $\langle p_z p_z \rangle$, assuming that the values for the other components are essentially zero given the equal probability of $p_z$ being positive or negative for an arbitrary cell. The steady-state magnitude of each of these components is presented using a bar graph in FIG.~\ref{fig:sickle_prolate_orbital_dynamics} for sickle cells and straight prolate capsules inside the cell-free layer and in the bulk of the suspension, respectively. We observe that for the cells inside the cell-free layer, the component $\langle p_z p_z \rangle$ dominates with the magnitude close to 1, while the other components are vanishingly small. This suggests that the marginated sickle cells and straight prolate capsules tend to approximate a log-rolling orbital motion inside the cell-free layer, with the end-to-end vector $\textbf{p}$ nearly aligned with the $z$ axis, which is in agreement with the observations from the simulation snapshots in FIG.~\ref{fig:healthy_sickle_prolate_snapshots}. Compared to the marginated stiff RBCs that approximate a tumbling motion, the near-wall sickle cells or straight prolate capsules are able to approach closer to the walls owing to a minimal volume exclusion effect because of their near-rolling orbits, smaller volume, and slenderness in shape, which leads to a near-wall peak in the $\hat n$ profile that is both greater in magnitude and closer to the wall.
  
%\begin{figure}[h]
%\centering
%\captionsetup{justification=raggedright}
% \subfloat[]% caption for subfigure a
%{
%    \includegraphics[width=0.45\textwidth]{./sickle_near_wall_pp.pdf}
%    \label{fig:near_wall_sickle_orientation}
%}
%\subfloat[]% caption for subfigure a
%{
%    \includegraphics[width=0.45\textwidth]{./prolate_near_wall_pp.pdf}
%    \label{fig:near_wall_prolate_orientation}
%}
%\\
% \subfloat[]% caption for subfigure a
%{
%    \includegraphics[width=0.45\textwidth]{./sickle_bulk_pp.pdf}
%    \label{fig:sickles_outside_CFL}
%}
% \subfloat[]% caption for subfigure a
%{
%    \includegraphics[width=0.45\textwidth]{./prolate_bulk_pp.pdf}
%    \label{fig:prolate_outside_CFL}
%}
%\caption{Magnitude of the $x$-$x$, $x$-$y$, $y$-$y$, and $z$-$z$ components of the tensor $\textbf{S} = \langle \textbf{pp} \rangle$ for sickle cells (left) and prolate capsules (right) inside the cell-free layer (a,b) and in the bulk of the suspension (c,d), respectively.\MDG{These could be combined into a single bar graph}}
%  \label{fig:sickle_prolate_orbital_dynamics}
%  \end{figure}
For sickle cells and straight prolate capsules in the bulk of the suspension, in contrast, $\langle p_x p_x \rangle$ dominates, with $\langle p_y p_y \rangle$ and $\langle p_z p_z \rangle$ being smaller but non-zero, indicating that the cells in the bulk generally take a ``kayaking" orbit (in which the long axis of the cell lies out of the $x$-$y$ plane but is not fully aligned along the $z$-axis). Here the long axis is only slightly out of the shear ($x$-$y$) plane, consistent with our previous findings on the dynamics of a single sickle cell in unbounded simple shear flow \cite{Zhang2019}. During the evolution of the suspension dynamics, this orbit becomes favorable for these slender cells with a high aspect ratio, given limited space between cells (primarily healthy RBCs) in suspension. The dynamics of the healthy RBCs do not show much difference compared to previous cases.      
      
   \subsection{Effect of marginated cells on the walls in binary suspensions}      \label{sec:wall_shear_stress}

\begin{figure}[t]
\centering
\captionsetup{justification=raggedright}
% \subfloat[]% caption for subfigure a
%{
%    \includegraphics[width=0.3\textwidth]{./purely_healthy_wall_shear_stress_600.png}
%    \label{fig:purely_healthy_wall_shear_stress_600}
%}
%\subfloat[]% caption for subfigure a
%{
%    \includegraphics[width=0.3\textwidth]{./purely_healthy_wall_shear_stress_700.png}
%    \label{fig:purely_healthy_wall_shear_stress_700}
%}
\subfloat[]% caption for subfigure a
{
    \includegraphics[width=0.38\textwidth]{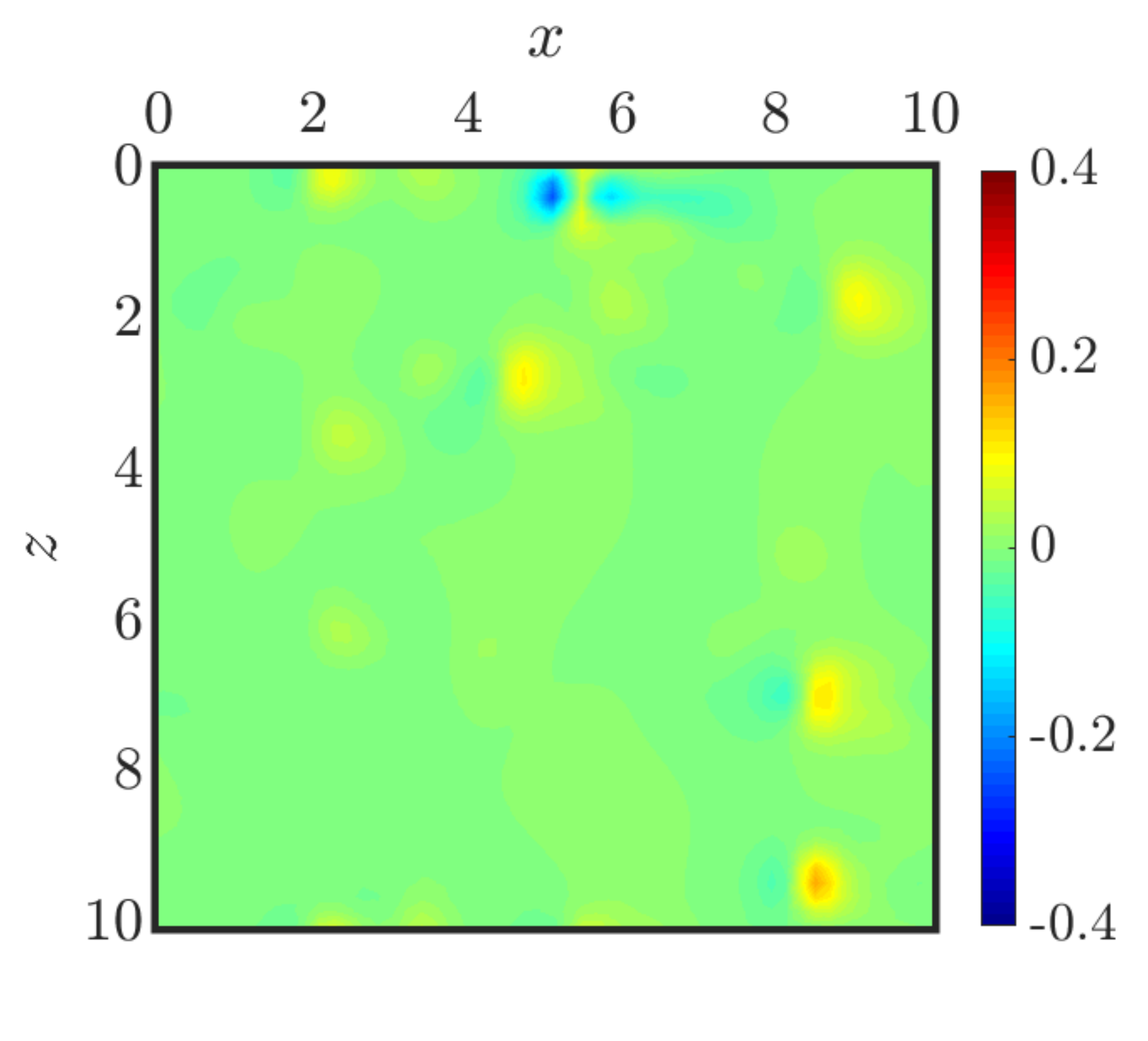}
    \label{fig:purely_healthy_wall_shear_stress_800}
}
\subfloat[]% caption for subfigure a
{
    \includegraphics[width=0.38\textwidth]{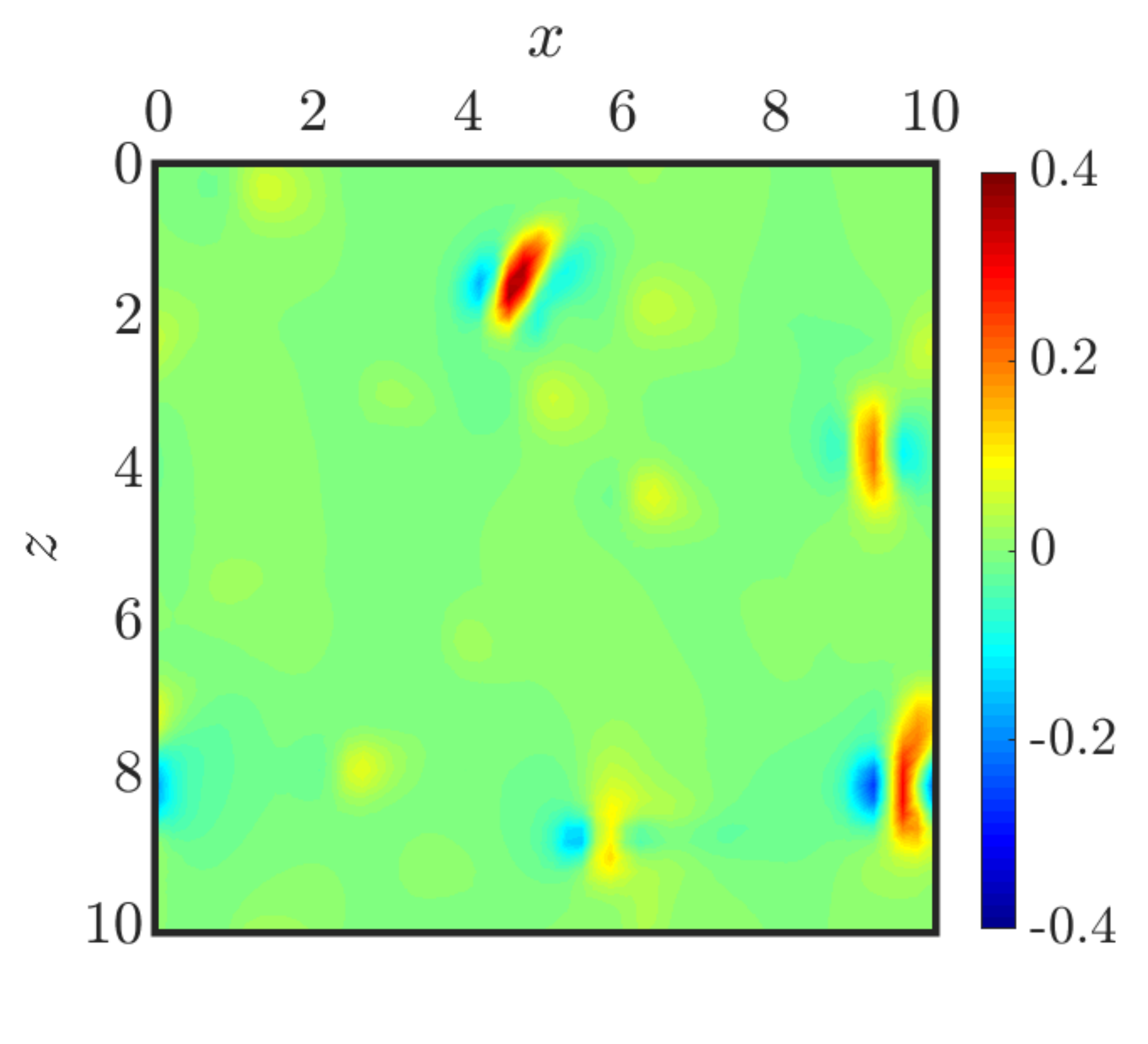}
    \label{fig:purely_healthy_wall_shear_stress_800_2}
}
\\
% \subfloat[]% caption for subfigure a
%{
%    \includegraphics[width=0.3\textwidth]{./healthy_sickle_wall_shear_stress_600.png}
%    \label{fig:healthy_sickle_wall_shear_stress_600}
%}
 \subfloat[]% caption for subfigure a
{
    \includegraphics[width=0.38\textwidth]{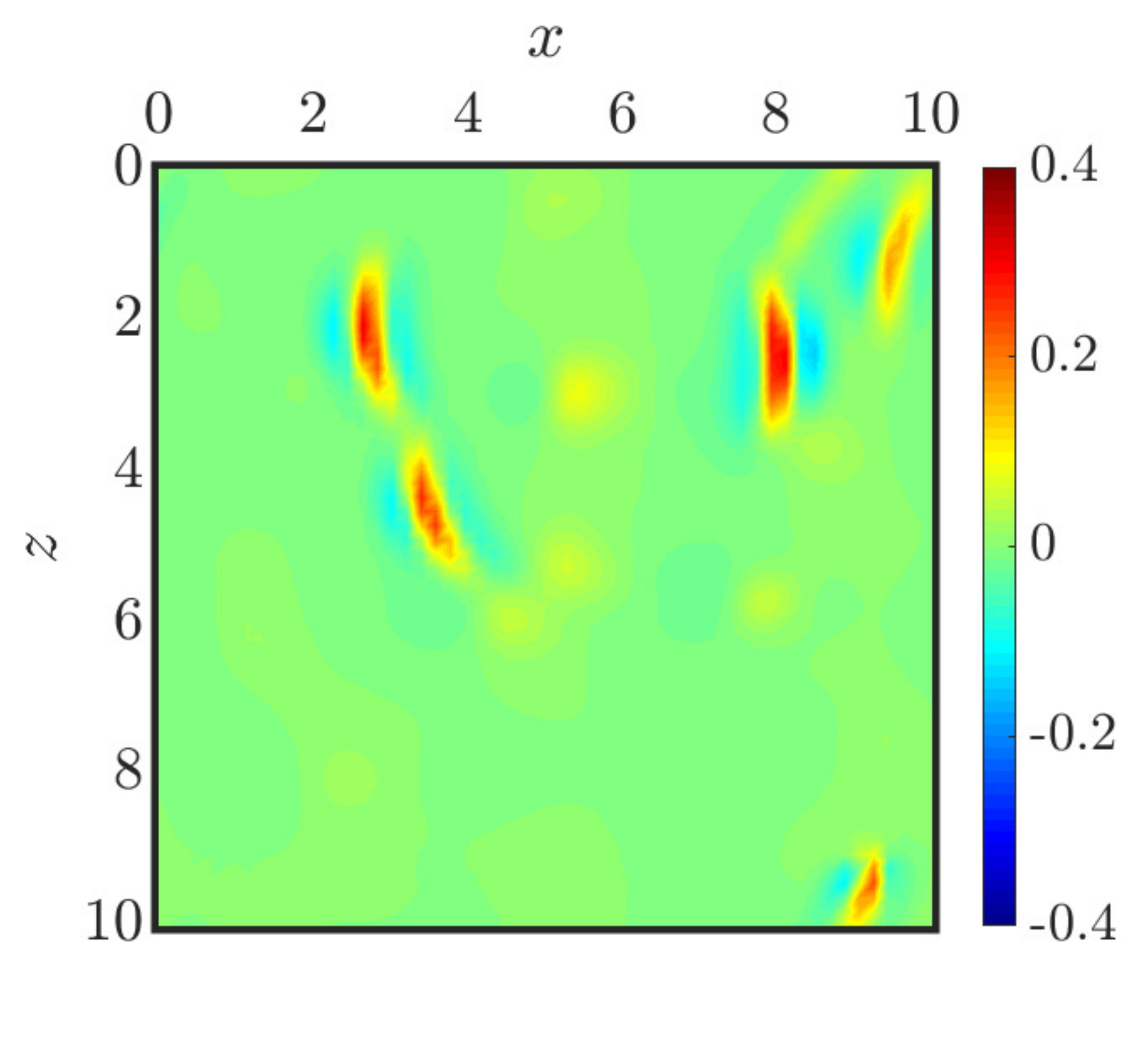}
    \label{fig:healthy_sickle_wall_shear_stress_700}
}
%\subfloat[]% caption for subfigure a
%{
%    \includegraphics[width=0.3\textwidth]{./healthy_sickle_wall_shear_stress_800.png}
%    \label{fig:healthy_sickle_wall_shear_stress_800}
%}
%\\
% \subfloat[]% caption for subfigure a
%{
%    \includegraphics[width=0.3\textwidth]{./healthy_prolate_wall_shear_stress_600.png}
%    \label{fig:healthy_prolate_wall_shear_stress_600}
%}
% \subfloat[]% caption for subfigure a
%{
%    \includegraphics[width=0.3\textwidth]{./healthy_prolate_wall_shear_stress_700.png}
%    \label{fig:healthy_prolate_wall_shear_stress_700}
%}
\subfloat[]% caption for subfigure a
{
    \includegraphics[width=0.38\textwidth]{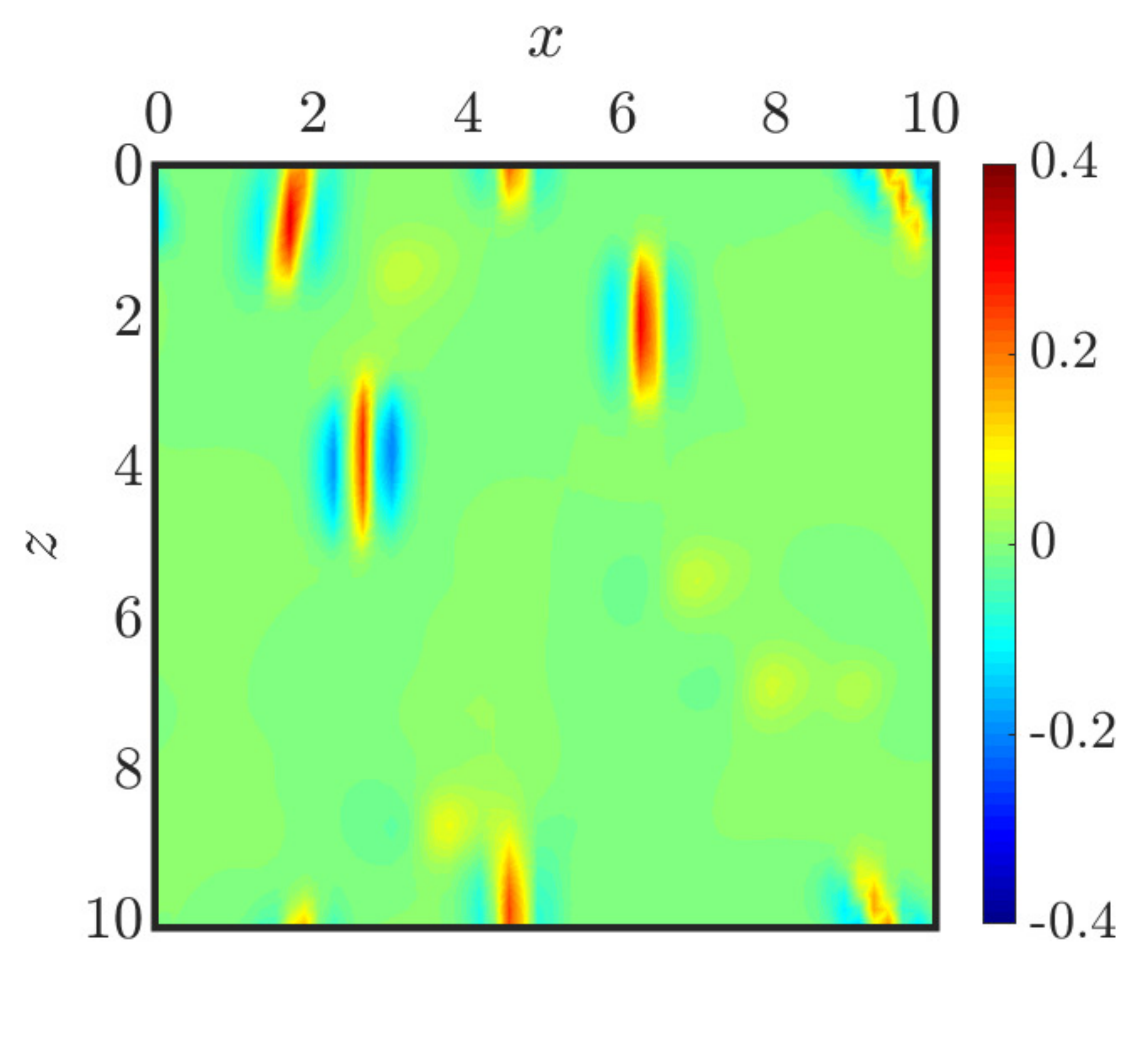}
    \label{fig:healthy_prolate_wall_shear_stress_800}
}
\caption{Examples of the spatial distribution of the additional wall shear stress $\hat \tau_w$ on the bottom wall ($y = 0$) induced by a homogeneous suspension of healthy RBCs (a) and binary suspensions of healthy (flexible) RBCs with stiff RBCs (b), sickle cells (c), and straight prolate capsules (d), respectively, at steady state.}
\label{fig:wall_shear_stress_distribution_comparison}
\end{figure}

Having determined the cross-stream distribution and cell dynamics in different suspensions, in this section we characterize the hydrodynamic effects of the suspensions on the walls, and compare the results for the binary suspensions with the case containing purely healthy RBCs to illustrate the impact of the marginated stiff cells in the binary suspensions. Particularly, we compute the shear stress at the walls associated with different suspensions. For an undisturbed planar pressure-driven flow in the absence of the capsules confined by two walls at $y = 0$ and $y = 2H$, the mean wall shear stress is given by $\tau_w = 2\eta U_0/H$.  
In the case of a flowing suspension of capsules, however, additional wall shear stress $ \tau_w'$ can be induced by the presence of the capsules. In this work $\tau_w'$ is computed numerically using an accelerated boundary integral method \cite{Kumar:2012ev}, as introduced in Section~\ref{sec:BIM}. It is worth emphasizing, though, that since the simulations are performed at constant pressure drop, the mean wall shear stress is the same for all suspensions considered here, and is equal to the undisturbed wall shear stress $\tau_w$, which is nondimensionalized to 2. Figure~\ref{fig:wall_shear_stress_distribution_comparison} shows examples of the spatial distribution of $\hat \tau_w=\tau_w'/\tau_w$ on the bottom wall ($y = 0$) induced by different suspensions at steady state. Note that this quantity and all the wall shear stress quantities reported below have been normalized by the undisturbed wall shear stress $\tau_w$ unless noted otherwise. It is clear that except for the case with purely healthy RBCs in which barely any fluctuations in wall shear stress are observed, the binary suspensions all induce local shear stress peaks on the wall, which is certainly accredited to the marginated cells in each suspension. 
%To quantify the average additional wall shear stress, we further take the root mean square of $\overline{\sigma_{xy}^w}$ over the wall mesh grid to obtain $\overline{\sigma_{xy}^w}_{\textnormal{RMS}}$, and normalize it by the undisturbed wall shear stress, which gives $\widehat{\sigma_{xy}^w} = \overline{\sigma_{xy}^w}_{\textnormal{RMS}}/|{\sigma_{xy}^{w,\infty}}|$. Here we report the time average of $\widehat{\sigma_{xy}^w}$ for different suspensions at steady state: 0.015 for a homogeneous suspension of healthy RBCs, 0.029 for a binary suspension of healthy and sickle RBCs, and 0.040 for a binary suspension of healthy RBCs and prolate capsules. Prolate capsules in a binary suspension induce a higher averaged wall shear stress than do sickle RBCs, which can be accounted for by the steady near-wall rolling motion with smaller fluctuations of their cross-stream positions. 

\begin{figure}[t]
\centering
\captionsetup{justification=raggedright}
 \subfloat[]% caption for subfigure a
{
    \includegraphics[width=0.5\textwidth]{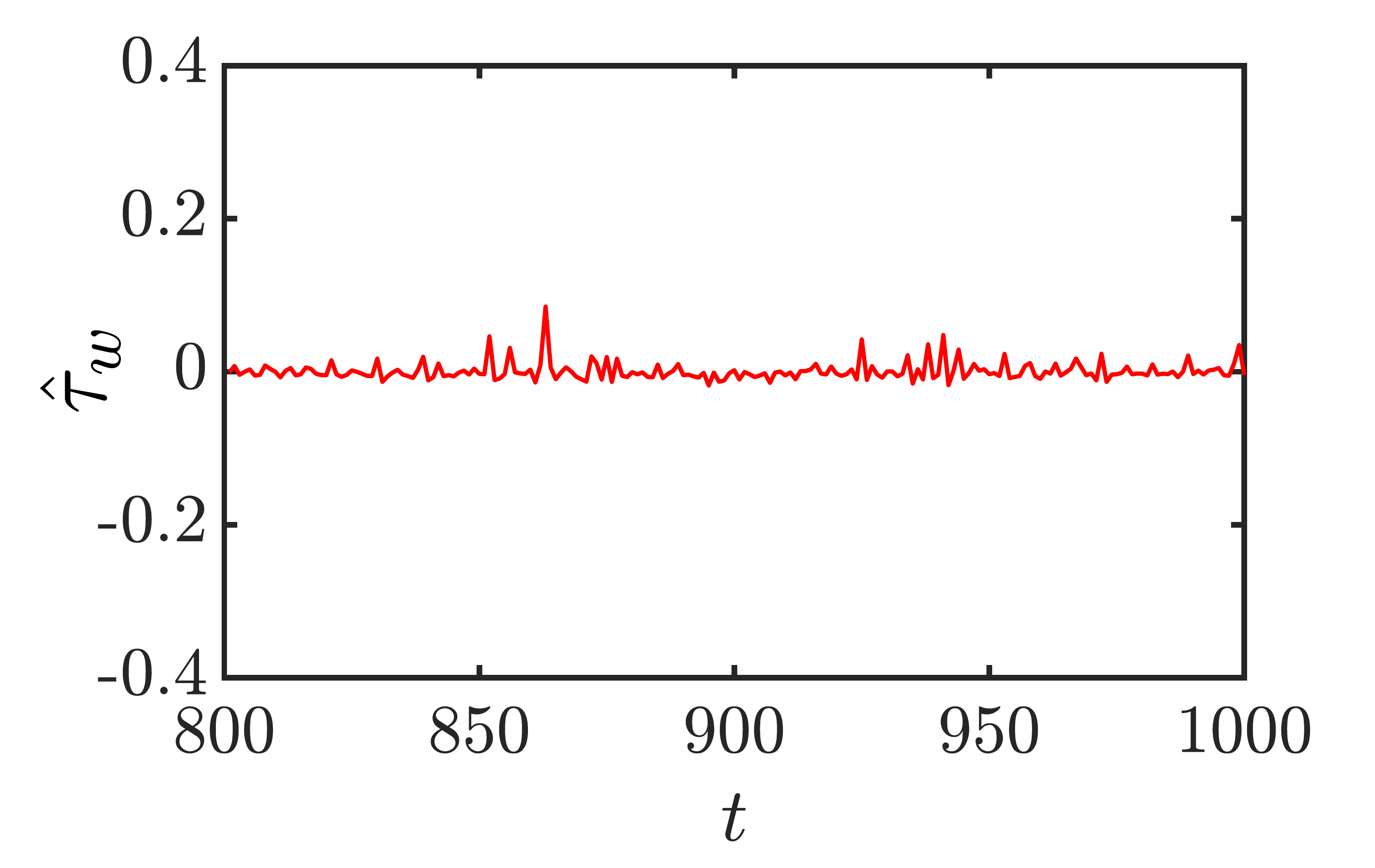}
    \label{fig:fluctuation_healthy}
}
 \subfloat[]% caption for subfigure a
{
    \includegraphics[width=0.5\textwidth]{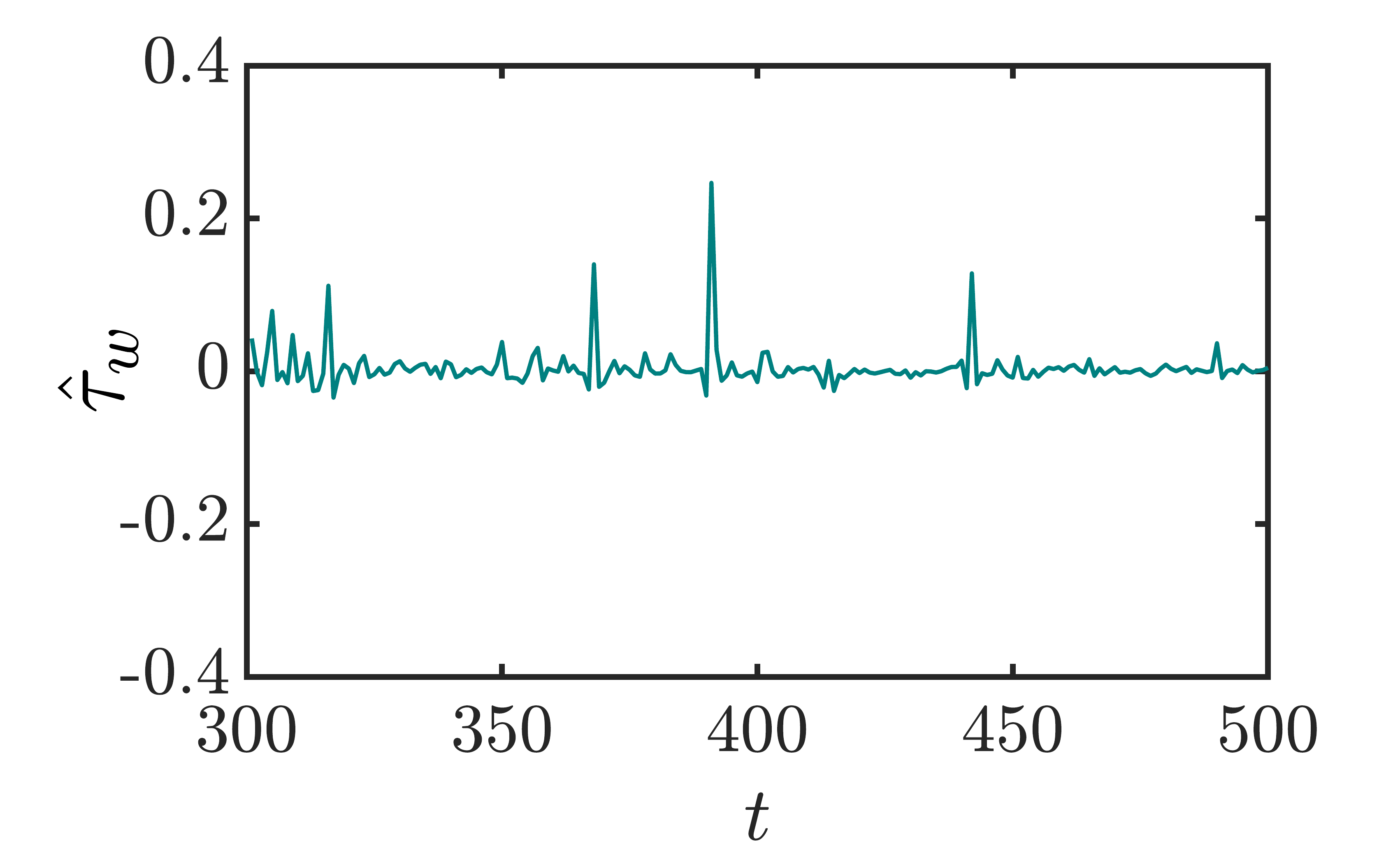}
    \label{fig:fluctuation_stiff}
}
\\
 \subfloat[]% caption for subfigure a
{
    \includegraphics[width=0.5\textwidth]{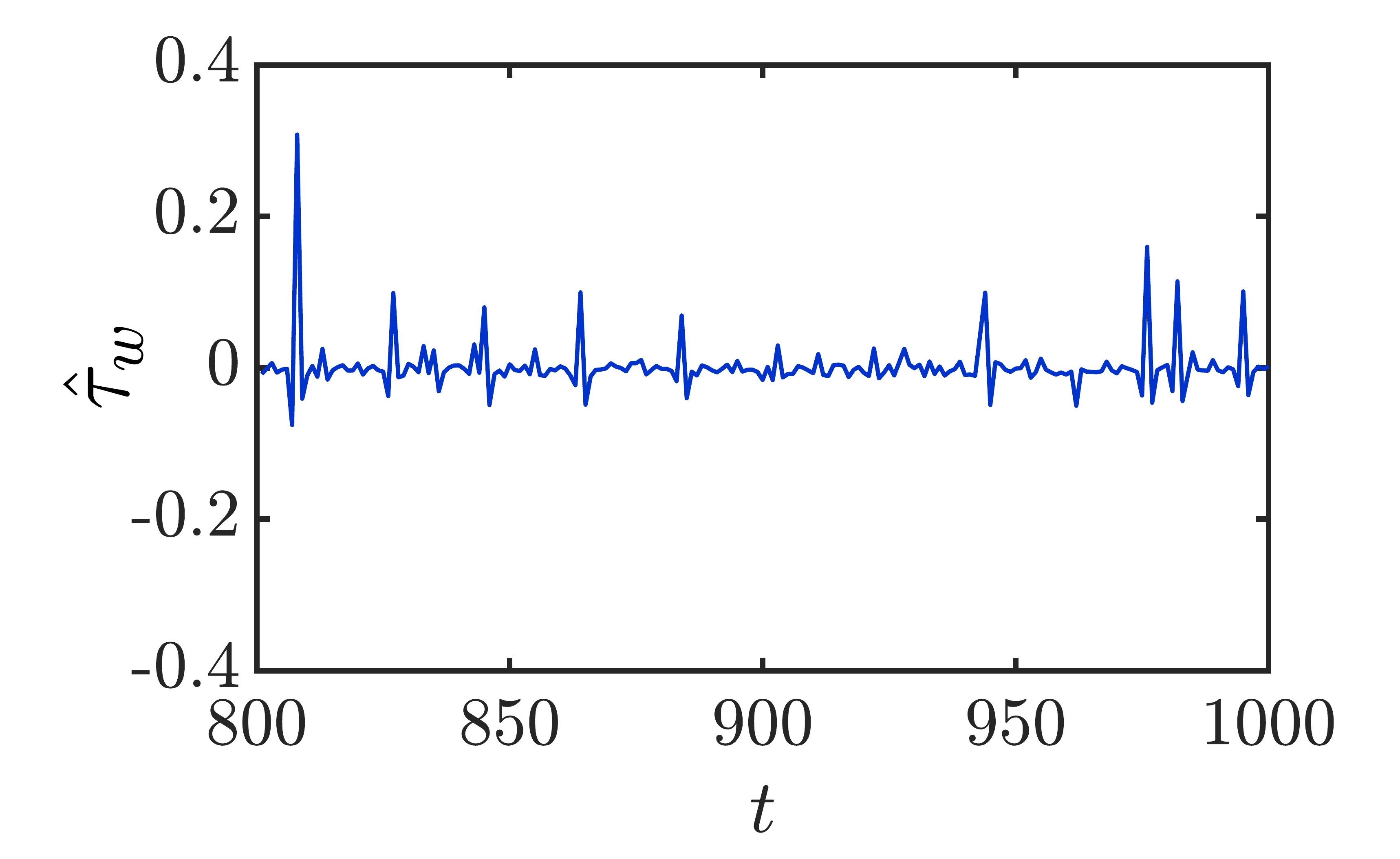}
    \label{fig:fluctuation_sickle}
}
 \subfloat[]% caption for subfigure a
{
    \includegraphics[width=0.5\textwidth]{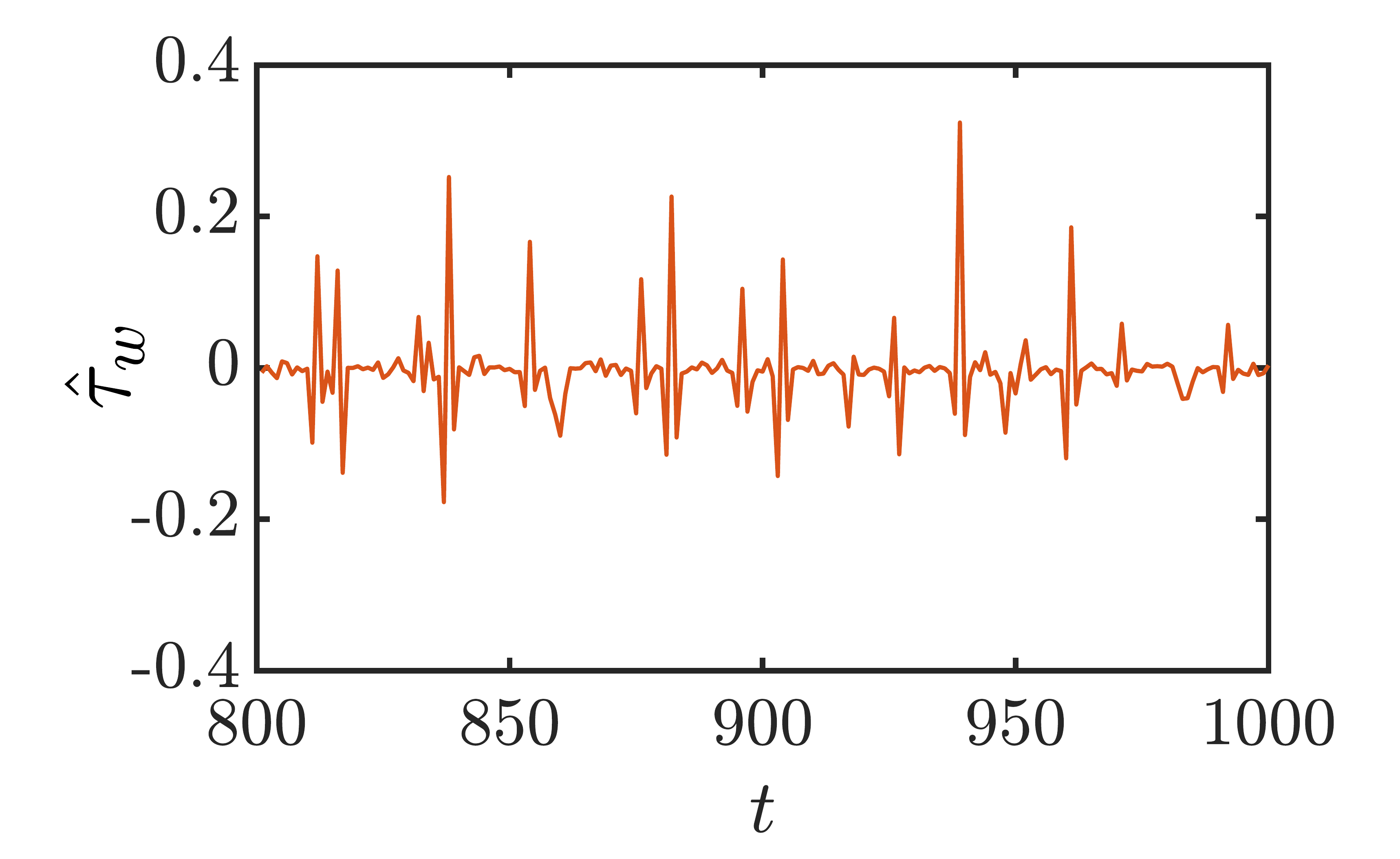}
    \label{fig:fluctuation_prolate}
}
\caption{Time evolution of the additional wall shear stress $\hat \tau_w$ at a fixed wall position $(x,y,z) = (0.74a,0,1.67a)$ for the cases of a homogeneous suspension of healthy RBCs (a) and binary suspensions of healthy (flexible) RBCs with stiff RBCs (b), sickle cells (c), and straight prolate capsules (d), respectively, at steady state.}
\label{fig:wall_shear_stress_one_position_comparison}
\end{figure}

\begin{figure}[t]
\centering
\captionsetup{justification=raggedright}
 \subfloat[]% caption for subfigure a
{
    \includegraphics[width=0.45\textwidth]{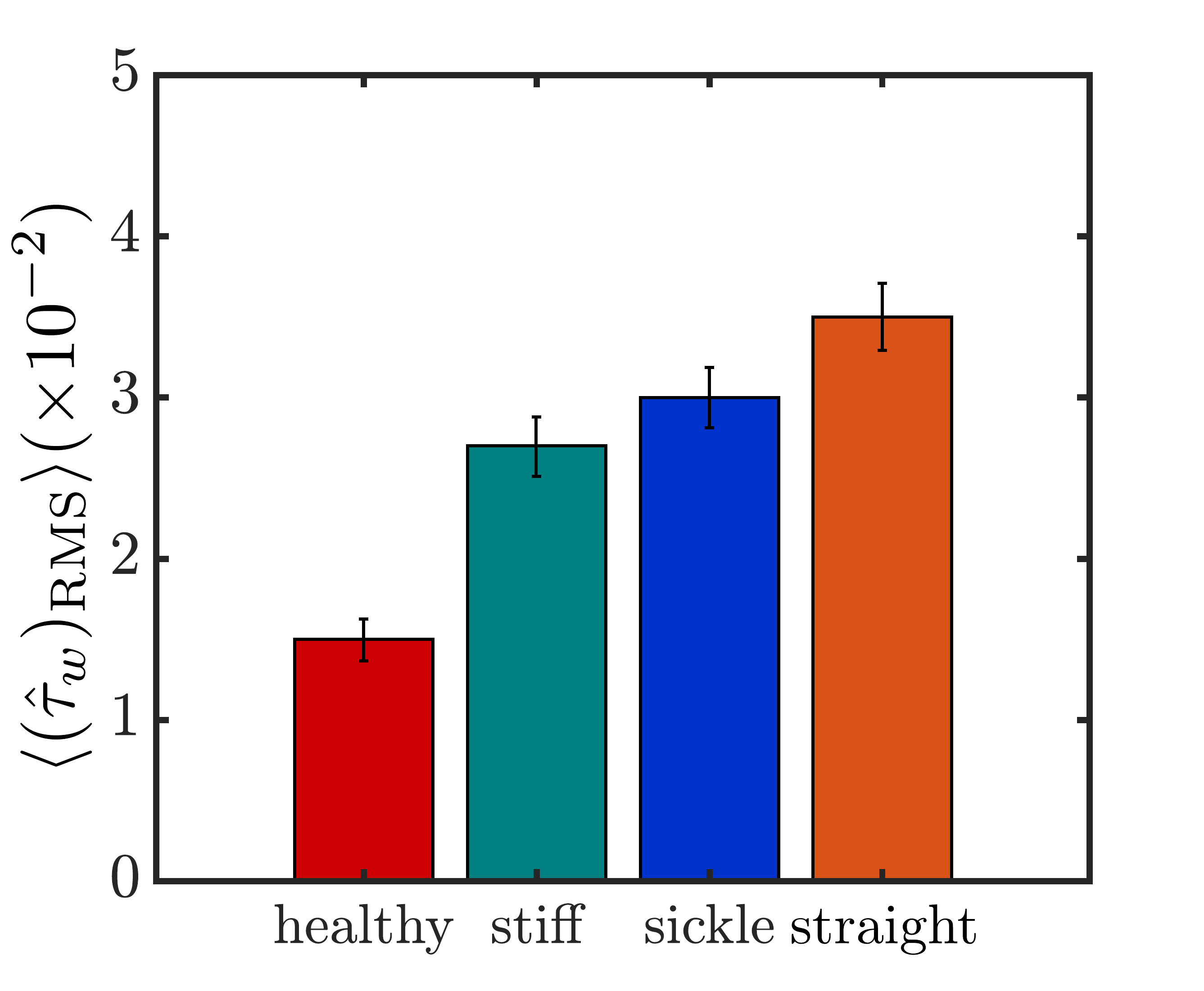}
    \label{fig:sigma_rms}
}
 \subfloat[]% caption for subfigure a
{
    \includegraphics[width=0.45\textwidth]{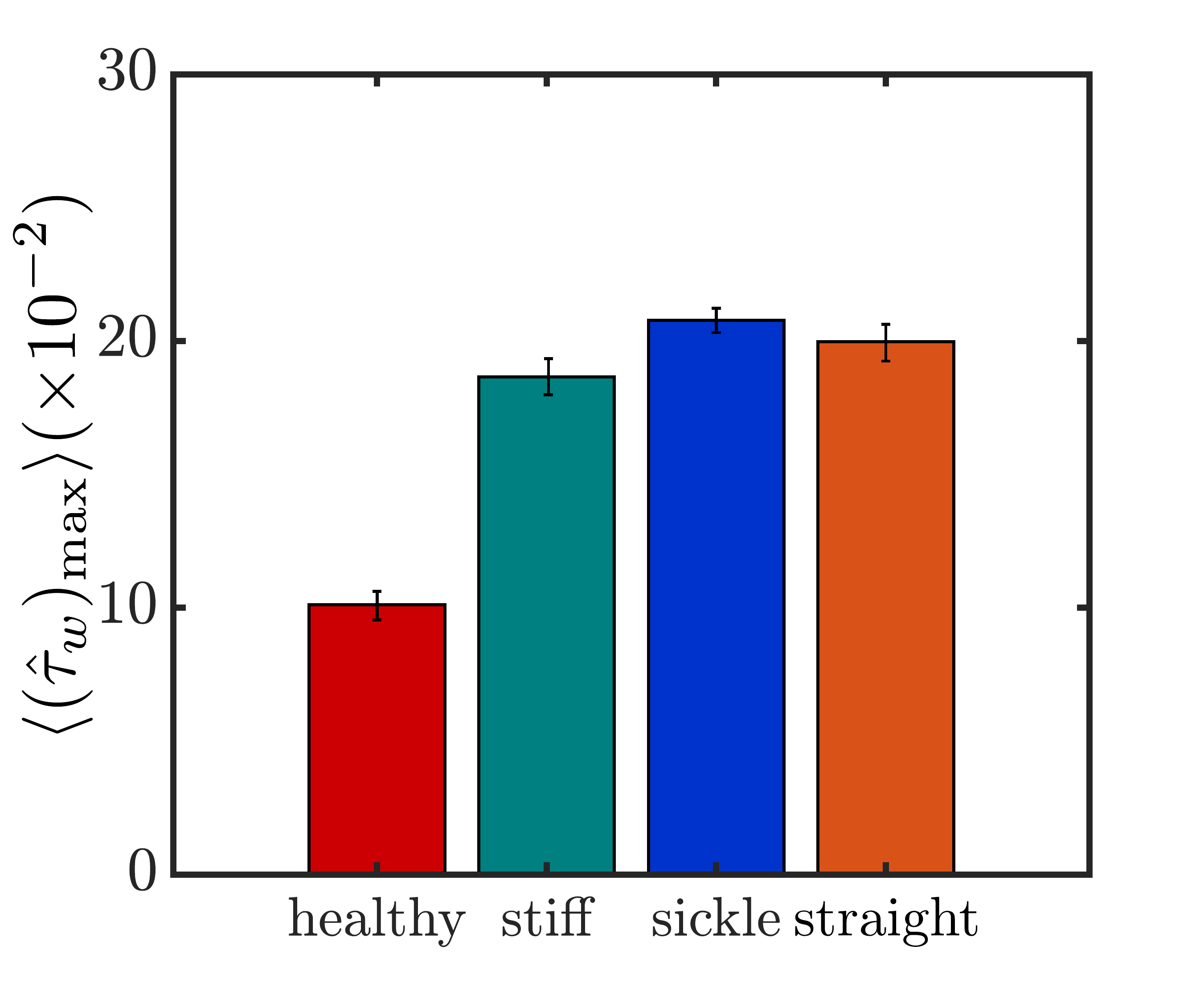}
    \label{fig:sigma_max}
}
\caption{The\soutold{ ensemble-averaged root mean square $\langle (\sigma_{xy}^w)_{\textnormal{rms}} \rangle$} \XZrevise{spatially-averaged RMS $\langle (\hat \tau_w)_{\textnormal{RMS}} \rangle$} (a) and maximum $\langle (\hat \tau_w)_{\textnormal{max}} \rangle$ (b) of the additional wall shear stress for different suspensions at steady state. For each quantity, the error bars represent standard error.}
  \label{fig:sigma_rms_max}
  \end{figure}

In general, an arbitrary wall position can experience dramatic fluctuations in $\hat \tau_w$ due to the dynamical motion of the near-wall cells in a flowing suspension. Examples are shown in FIG.~\ref{fig:wall_shear_stress_one_position_comparison} for the evolution of $\hat \tau_w$ experienced at a fixed wall position $(x,y,z) = (0.74a,0,1.67a)$ for different suspensions \XZrevise{at steady state}. A number of peaks with large magnitude are observed in each case of the binary suspensions, suggesting that this wall position experiences high additional shear stress intermittently.\soutold{ We first compute the averaged wall shear stress at each wall position by taking the root mean square of $\sigma_{xy}^w$ over a total sampling time $T$, which gives $(\sigma_{xy}^w)_{\textnormal{rms}}$. The spatial (ensemble) average of $(\sigma_{xy}^w)_{\textnormal{rms}}$ over all wall positions, denoted as $\langle (\sigma_{xy}^w)_{\textnormal{rms}} \rangle$, is then obtained and plotted in FIG.~\ref{fig:sigma_rms} for each case of the suspensions.} \XZrevise{We now detailedly characterize the hydrodynamic effect of each suspension on the walls. First, we compute the RMS of the additional wall shear stress $\hat \tau_w$ for each wall position, and the results for the spatially-averaged RMS wall shear stress, $\langle (\hat \tau_w)_{\textnormal{RMS}} \rangle$, are plotted in FIG.~\ref{fig:sigma_rms} for different suspensions.}\XZ{good point -- no we don't need to do it separately. I just need a better way to describe it in text.} \XZrevise{It is observed that all}\soutold{All} three types of cells as the trace component, i.e., stiff RBCs, sickle cells, and straight prolate capsules, cause an increase (of approximately two folds) in the averaged RMS wall shear stress in the case of the corresponding binary suspension compared to the case with purely healthy RBCs. The magnitude of the increase is the greatest for the case with straight prolate capsules, though the differences are small among the binary cases. 
%which may be explained by both the steady rolling motion of the marginated prolate capsules (with negligible fluctuations in positions due to the absence of curvature) and their slender shape that may allow them to ``sweep" over a greater area . 
In addition, the spatially-averaged maximum additional wall shear stress, $\langle (\hat \tau_w)_{\textnormal{max}} \rangle$, is also computed for different cases (FIG.~\ref{fig:sigma_max}). Again, no substantial differences are observed among the binary cases, although the magnitude of this quantity is slightly greater for the case with sickle cells than with stiff RBCs or straight prolate capsules.
%, which can be explained by the findings in Section~\ref{sec:tip_wall_interaction} regarding tip-wall interactions: during the rolling motion of the cell bodies, the tips of marginated sickle cells are likely to approach closer to the walls than those of prolate capsule due to the curvature, thus causing a higher maximal wall shear stress. 

\begin{figure}[t]
\centering
\captionsetup{justification=raggedright}
 \subfloat[]% caption for subfigure a
{
    \includegraphics[width=0.45\textwidth]{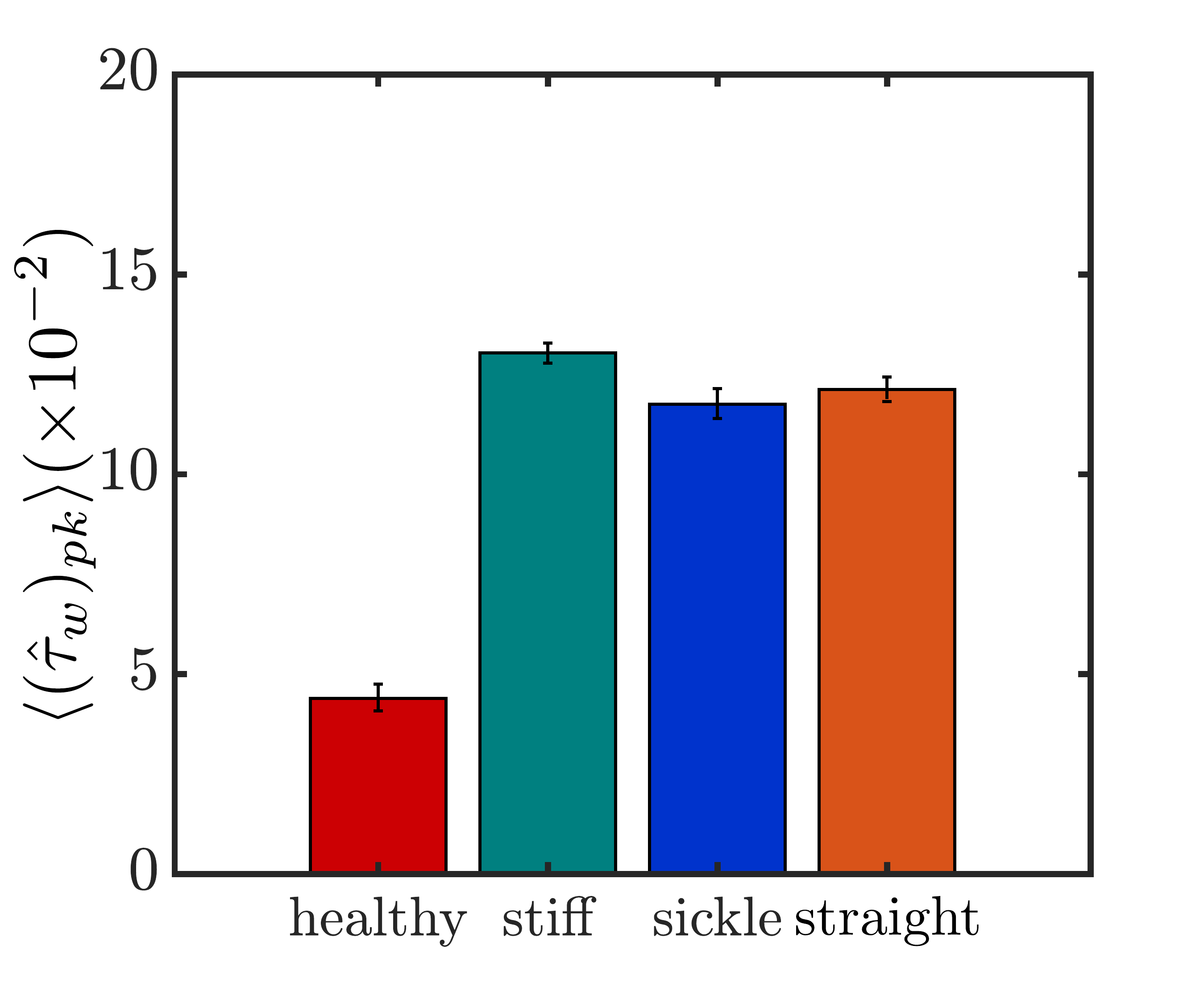}
    \label{fig:sigma_p}
}
 \subfloat[]% caption for subfigure a
{
    \includegraphics[width=0.45\textwidth]{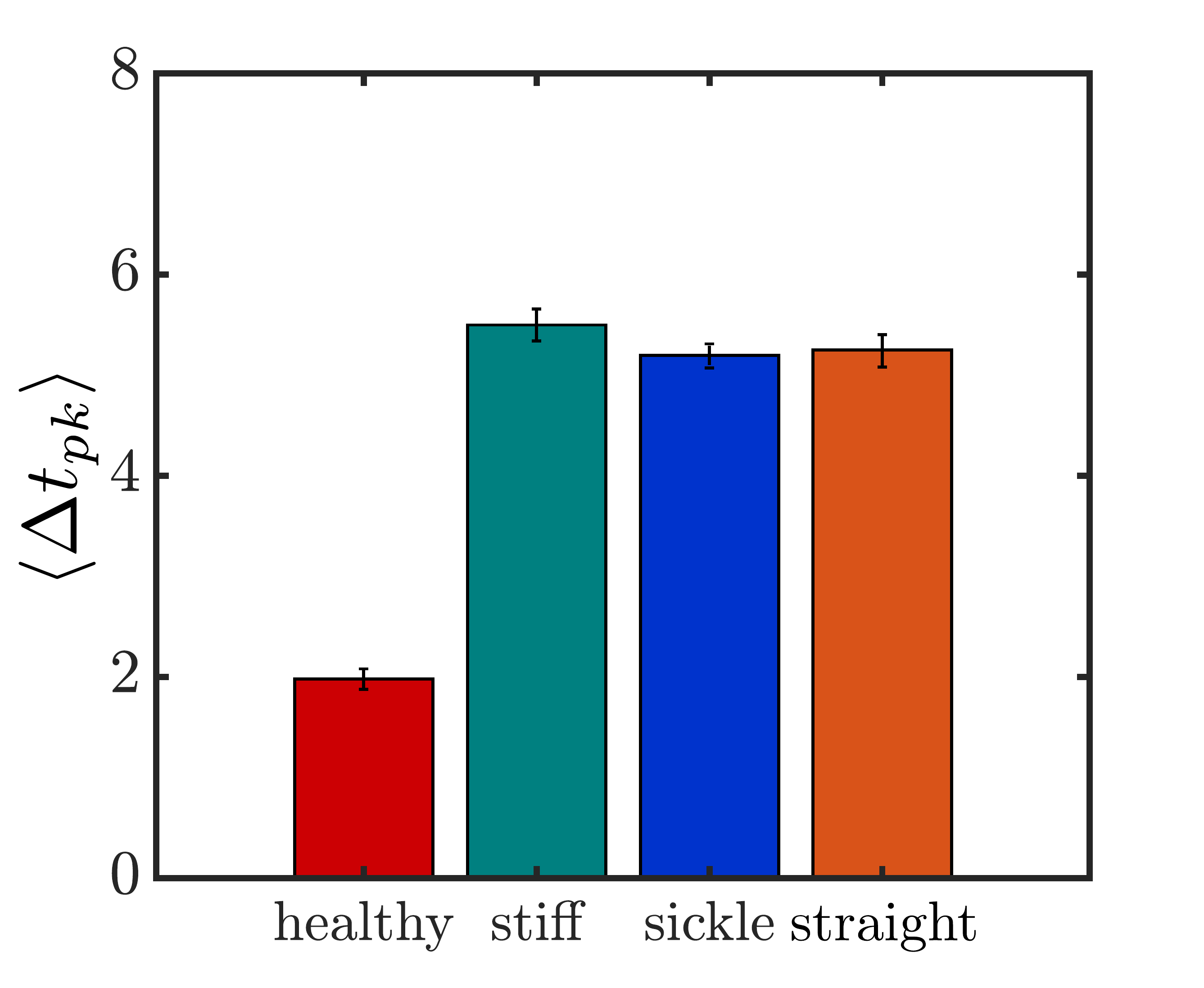}
    \label{fig:duration_p}
}
\\
 \subfloat[]% caption for subfigure a
{
    \includegraphics[width=0.45\textwidth]{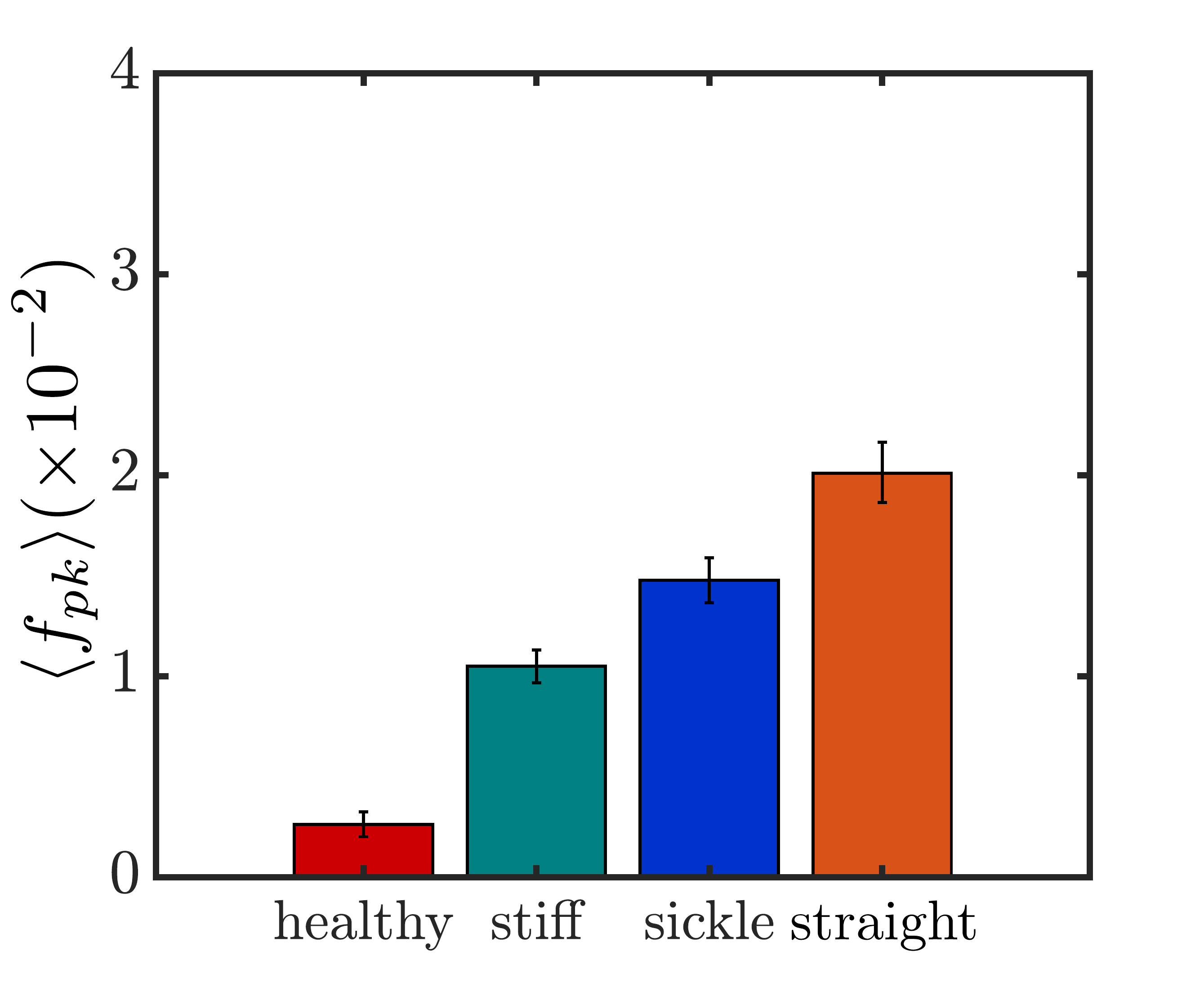}
    \label{fig:f_p}
}
 \subfloat[]% caption for subfigure a
{
    \includegraphics[width=0.45\textwidth]{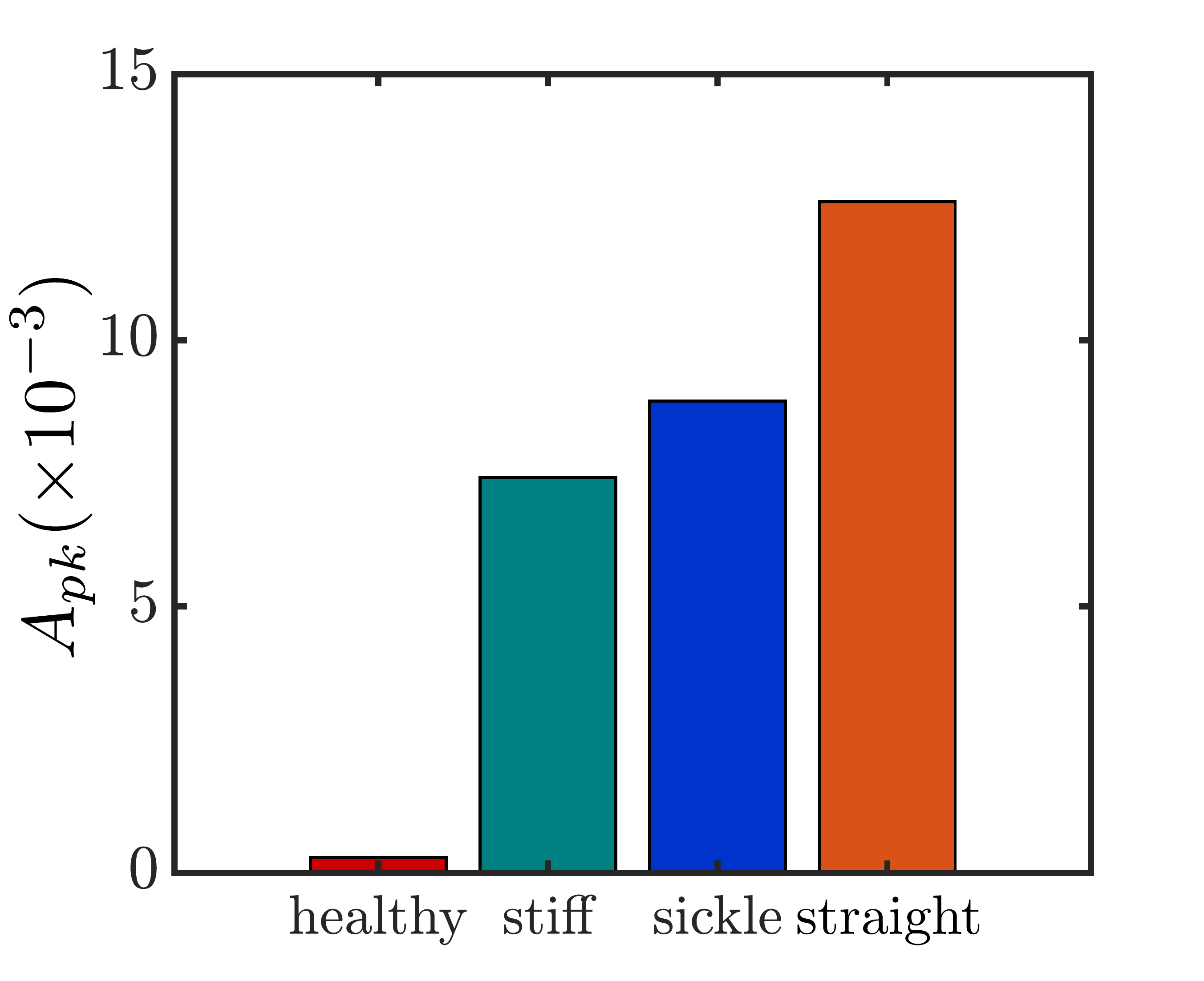}
    \label{fig:A_p}
}
\caption{The ensemble-averaged magnitude $\langle (\hat \tau_w)_{pk} \rangle$ (a), duration\soutold{ $\langle \Delta (\sigma_{xy}^w)_{pk} \rangle$} \XZrevise{$\langle \Delta t_{pk} \rangle$} (b), and frequency $\langle f_{pk} \rangle$ (c), respectively, of the wall shear stress peaks experienced by all wall positions for different suspensions at steady state, and the product $A_{pk}$ of these three quantities (d). The error bars in (a), (b), and (c) represent standard error for each quantity.}
\label{fig:real_peaks_quantification}
\end{figure}

However, despite the general indications, neither the RMS nor the maximum is able to reveal the influence of the marginated cells in a binary suspension on the walls in detail. Indeed, it is mainly these near-wall cells that generate the intermittent peaks in the additional wall shear stress, as shown in the examples in FIG.~\ref{fig:wall_shear_stress_one_position_comparison}. To this end, we now compute three quantities for each of the suspensions: the ensemble-averaged magnitude, duration, and frequency of the peaks in the additional wall shear stress. To do this, we\soutold{ first set a critical peak amplitude to the value of $\langle (\sigma_{xy}^w)_{\textnormal{max}} \rangle$ for the case with purely healthy RBCs, so that those local maxima with very small amplitude (data noises) during the time evolution of $\sigma_{xy}^w$ are eliminated. The remaining ones are considered ``real" peaks} \XZrevise{consider only those wall shear stress peaks in each case that are greater than the expected maximum value for the healthy case}, and the associated quantities are denoted using the subscript ``$pk$"\note{I still think it's better to keep the subscript ``$pk$" (I just realized the subscript ``p" has been used for the ``primary component in binary suspensions) - it tells the audience these quantities characterize the wall shear stress PEAKS as an effect of the marginated stiff cells. I'm not sure how we should name the magnitude otherwise, and to be consistent, all other quantities should have the same subscript. Also, I think adding this subscript doesn't make the nomenclature more nasty.}. The ensemble-averaged magnitude, duration, and frequency of these peaks\soutold{ over a total sampling time $T$} are then computed\soutold{ by averaging over the peaks in shear stress experienced by all wall positions}, which gives $\langle (\hat \tau_w)_{pk} \rangle$,\soutold{ $\langle \Delta (\sigma_{xy}^w)_p \rangle$} \XZrevise{$\langle \Delta t_{pk} \rangle$}, and $\langle f_{pk} \rangle$, respectively. \XZrevise{Note that an arbitrary peak in $\hat \tau_w$, as defined above, is always characterized by a spike and two dips right before and after it (FIG.~\ref{fig:wall_shear_stress_one_position_comparison}). Accordingly, the duration of a peak is defined as the time difference between the onset of the dip before the spike and the end of the one after.}\soutold{ Here the frequency of the peaks in $\sigma_{xy}^w$ at an arbitrary wall position is simply defined as
%\begin{equation} \label{eq:peak_frequency}
%f_p = n_p/T,
%\end{equation} 
where $n_p$ is the number of peaks}\soutold{ in $\sigma_{xy}^w$ experienced at this wall position over a total sampling time $T$.} Finally, we take the product $A_{pk}$ of these three quantities, defined as 
\begin{equation} \label{eq:peak_quantity_product}
A_{pk} = \langle (\hat \tau_w)_{pk} \rangle \times \langle \Delta t_{pk} \rangle \times \langle f_{pk} \rangle,
\end{equation}
which quantifies the overall effect of each suspension on the wall shear stress. 

The results are summarized in FIG.~\ref{fig:real_peaks_quantification}. It is obvious that compared to the case with purely healthy RBCs where $A_{pk}$ is vanishingly small, all three binary suspensions exert a substantial overall effect on the wall shear stress (FIG.~\ref{fig:A_p}). The averaged magnitude and duration of the wall shear stress peaks, $\langle (\hat \tau_w)_{pk} \rangle$ and $\langle \Delta t_{pk} \rangle$, show minor differences among the cases of binary suspensions, although both are slightly greater for the case with stiff RBCs than with sickle cells or straight prolate capsules. Significant differences, however, are observed in the averaged frequency of the wall shear stress peaks, with $\langle f_{pk} \rangle$ being the greatest when the trace component is straight prolate capsules. As a result, the product $A_{pk}$ also yields the greatest value for the case with straight prolate capsules, while being slightly greater for the case with sickle cells than with stiff RBCs.
 
The differences in $\langle f_{pk} \rangle$ among the binary suspension cases, as shown in FIG.~\ref{fig:f_p}, may be explained by the orbital dynamics of the marginated cells. As determined in Section~\ref{sec:healthy_stiff_RBCs}, marginated stiff RBCs tend to approximate an in-plane tumbling orbit. Theoretically, substantial additional wall shear stress is induced when the cell is oriented perpendicular to the wall with a minimal distance between the edge of the cell and the wall. However, during the periodic orbital motion of the cell, this orientation is transient before the cell is quickly flipped by the shear effect of the flow and spend longer time at the edge of the cell-free layer becoming parallel with the wall, which conversely leads to negligible additional wall shear stress and, as a consequence, the lowest frequency of the wall shear stress peaks among the binary suspension cases. The difference in $\langle f_{pk} \rangle$ between the cases with sickle cells and straight prolate capsules, on the other hand, is majorly an effect of curvature: the steady rolling motion of marginated straight prolate capsules inside the cell-free layer, with negligible fluctuations in the wall-normal position of the cell body compared to the larger fluctuations for near-wall rolling sickle cells caused by curvature, ensures that at any wall position, a peak in wall shear stress is induced whenever a straight prolate capsule goes by. 

As mentioned in Introduction, in this work WBCs and platelets are not considered in the suspensions assuming negligible effects due to their extremely small number fractions in blood. To justify this assumption, we have tested the case in which one stiff spherical capsule slightly larger than a biconcave discoid, which serves as a model for the WBC, is placed near the wall in suspension with healthy RBCs ($X_p = 0.99$, $X_t = 0.01$). We reveal that the overall effect of this suspension on wall shear stress, characterized by $A_p$ as defined above, is indeed negligible compared to that for the case containing sickle cells; note that this number fraction of WBCs is even higher than the physiological range \cite{Dean2005}. Therefore, it is reasonable to exclude WBCs in our simulations. Similarly, we expect that platelets should also have a negligible effect on wall shear stress due to both the small number fraction and the tumbling motion taken near the walls \cite{Zhao2012} as noted above for stiff RBCs.

The findings in the present work may aid in understanding the pathophysiology of chronic endothelial inflammation in SCD from a biophysical perspective. Indeed, the intermittent peaks in wall shear stress induced by the marginated stiff sickle cells, as observed in this work, resemble the spiky wall shear stress profile considered in Bao \emph{et al.} \cite{Bao1999} on flow-mediated endothelial mechanotransduction. This spiky profile, characterized by abrupt onset and termination of shear stress, was found to induce the greatest upregulation of pro-inflammatory signals compared to the other shear stress profiles. More generally, rapid changes in shear stress, as opposed to steady shear stress, have been demonstrated to contribute to endothelial inflammation and subsequent atherogenesis via specific mechanotransduction pathways of the endothelium \cite{DeKeulenaer1998,Dekker2002,Hsiai2003,Sorescu2004,Harrison2006}. In this sense, our results provide a possible mechanism for chronic SCD vasculopathy. 

%\XZrevisesecond{Additionally, prior experimental \cite{Kaul3356,Kaul1994,Frenette2002} and computational \cite{Lei11326,Lei2015} studies have demonstrated the significant role of adhesive interactions between sickle cells and the endothelium in initiating the acute vaso-occlusion. Our present work probes the mechanism for the chronic endothelial inflammation, particularly seeking to address the question of whether non-adhesive interactions between sickle cells and endothelial cells are  the though cell-wall adhesion is not modeled, the marginated sickle cells are found to induce substantial fluctuations in wall shear stress even with no direct contact with the wall -- these would only be exacerbated by cell-wall adhesion. In fact, the aforementioned experimental studies on shear stress-induced endothelial expression were all performed in the absence of adhesive events. Hence, our results suggest that the aberrant physical interactions between sickle cells and endothelial cells alone are sufficient to cause endothelial dysfunction.}

The present work does not incorporate adhesive interactions between diseased cells and vessel walls. While these play a significant role in the acute vaso-occlusive crisis \cite{Kaul3356,Kaul1994,Frenette2002,Lei11326,Lei2015}, which in turn, may cause endothelial dysfunction, our aim here was to address the question of whether non-adhesive effects may also lead to a pro-inflammatory state of the endothelium via mechanotransductive mechanisms. In fact, the aforementioned experimental studies on shear stress-induced endothelial expression were all performed in the absence of adhesive events. 
%Hence, this work demonstrates that purely physical (non-adhesive) interactions between endothelial cells and sickle cells are sufficient to cause endothelial inflammation, which provides innovative and vital insight into SCD pathophysiology that extends beyond, and is independent of, vaso-occlusion. 
Hence, this work suggests that purely physical (non-adhesive) interactions between endothelial cells and sickle cells may be sufficient to cause endothelial inflammation, which provides innovative and vital insight into SCD pathophysiology that extends beyond, and is independent of, vaso-occlusion. In the case where adhesive events become important, the phenomena revealed here would still be of prime importance, as the strong sickle cell margination demonstrated here brings the cells into the vicinity of the walls, which is of course a prerequisite for adhesion to them.

\section{CONCLUSION} \label{sec:conclusion}
%  \input{conclusion}
%\begin{itemize}
%\item Proposed a hypothesis for the mechanism for endothelial damage associated with sickle cell disease (SCD);
%\item Simulation of flowing suspensions of healthy and sickle RBCs:
%\subitem - stiff RBCs marginate, while healthy RBCs accumulate both near the centerline and at the edge of CFL; 
%\subitem - curvature shows a minor effect on the segregation behavior.
%\item Orbital dynamics of RBCs in flowing suspensions:
%\subitem - healthy RBCs tend to approach a rolling motion in the higher shear rate region;
%\subitem - sickle RBCs (prolate capsules) tend to approximate a rolling motion inside CFL;
%\subitem - dynamics of healthy and sickle RBCs are consistent with experimental observations;
%\subitem - the interactions between sickle RBCs (prolate capsules) and the walls are quantified.
%\item Understanding the segregation behavior in suspensions:
%\subitem - wall-induced migration rate and post-collisional displacement after shear-induced pair collisions;
%\subitem - the capsules inside CFL are likely to remain stable in it;
%\subitem - sickle RBCs in the bulk are potentially still evolving toward the walls;
%\subitem - curvature seems to show a minor effect.
%\item The margination of sickle RBCs and their near-wall dynamics may help substantiate the aforementioned hypothesis.
%\end{itemize}
%
In this study, we investigated different flowing suspensions of deformable capsules subjected to pressure-driven flow in a planar slit using boundary integral simulations, assuming that the flow is in the Stokes regime.
%\st{ A base case is a homogeneous suspension of flexible biconcave discoidal capsules representing purely healthy RBCs. Additionally, three binary suspensions are studied. In one case, the two components in the suspension differ only in membrane rigidity with the same biconcave discoidal rest shape to illustrate the isolated effect of rigidity contrast.} 
The focus of this study is a binary suspension of healthy RBCs and sickle cells, which describes the microvascular blood flow in sickle cell disease.
%\st{ The sickle cells are modeled as stiff curved prolate capsules, displaying substantial contrasts with healthy RBCs in rigidity, shape, and size. Another binary suspension containing healthy RBCs and stiff straight prolate capsules is also considered for direct comparison with the case with sickle cells to show the effect of curvature of the stiff component. In each binary suspension, the stiff component is dilute, taking up 10\% of the total number of capsules. The mechanics of the capsule membranes are described using a model incorporating both shear and bending elasticities.} 
The parameter regime is based on the range of experimentally determined values for RBCs and blood flow in the microcirculation.
%\st{, except that the suspending fluid and the fluids enclosed by all capsules are assumed to have the same viscosity, i.e., the viscosity ratio $\lambda = 1$, for the purpose of keeping the computational cost manageable for the simulations}.

We first characterized the cross-stream distributions and orientational dynamics of different types of capsules in each suspension (Section~\ref{sec:cell_distribution_and_dynamics}), and revealed a number of key features.

(i) In a homogeneous suspension of healthy (flexible) RBCs, a cell-free layer is formed next to the channel walls. The spontaneous shape of RBCs plays a nontrivial role in the orientational dynamics of single cells in the suspension, but has a negligible effect on the number density distribution.
%\st{Healthy RBCs accumulate both around the centerplane of the channel and near the wall right beyond the cell-free layer, showing two corresponding peaks in the number density distribution profile. In the near-wall region where local shear rate is high, healthy RBCs approximate a rolling orbit with the major axis nearly aligned with the vorticity ($z$) direction, which is in agreement with the observations in prior studies }\cite{Goldsmith351,Bitbol1986,Lanotte13289,Cordasco:2013hb,Mendez2018}\st{ for the dynamics of single RBCs at high shear rate with a more physiological viscosity ratio $\lambda = 5$. No qualitative changes are observed for the distribution profile and dynamics of healthy (flexible) RBCs in binary suspensions.}
In a binary suspension of flexible and stiff RBCs with the same rest shape but large contrast in membrane rigidity, the dilute stiff RBCs are largely drained from the center of the channel and display substantial margination toward the walls.
%\st{, showing a high near-wall peak in the number density distribution profile}. 
This suggests that rigidity contrast by itself is sufficient to induce the segregation behavior in a binary suspension. The marginated stiff RBCs assume an approximate in-plane tumbling orbit. 

(ii) In a binary suspension of healthy RBCs and sickle cells, the dilute stiff sickle cells are almost completely drained from the bulk of the suspension, and strongly aggregate inside the cell-free layer. The marginated sickle cells tend to roll in flow with the major axis aligned with the vorticity ($z$) direction. Curvature plays a minor role in the segregation behavior and orientational motion of the stiff capsules.   

Furthermore, we quantified the physical effect of each suspension on the walls (Section~\ref{sec:wall_shear_stress}). In particular, the additional wall shear stress induced by the capsules was computed for each suspension. We found that, compared to the small fluctuations in wall shear stress for the case with purely healthy RBCs, large local peaks in wall shear stress are induced by all binary suspensions, a hydrodynamic effect of the marginated cells on the walls. Detailed quantification showed that the shape and size of the stiff component have a secondary role, even though for the parameter space considered in this study, straight prolate capsules seem to have a greater overall effect on the wall shear stress than do stiff RBCs and sickle cells.

Overall, all these results represent an effort to gain an improved understanding of the behavior of flowing suspensions of cells or capsules in complex scenarios, particularly the blood flow in SCD which is the focus of the current study. More importantly, this work is of essential interest from the medical point of view in that the results may help explain the mechanism for complications associated with SCD, such as endothelial inflammation, that are still poorly understood. In particular, this work addresses how circulating stiffened and misshapen RBC subpopulations may directly contribute to vasculopathy in SCD, not via vaso-occlusion or cell adhesion as canonically depicted, but via biophysical effects due to the margination of diseased cells. 

One limitation of this work is that the properties of cells of each type were identical and unchanging -- physiologically, sickle cells vary considerably in cell morphology and membrane stiffness, which should be addressed in future studies. Other parameters of the suspensions, such as the total volume fraction and the number fraction of each component in a binary system, can also be altered in extended studies to investigate the effects of these parameters on the suspension dynamics. Furthermore, it has been observed in experiments \cite{Loiseau2015} that the flow of a suspension of sickle cells around an acute corner of a triangular pillar or a bifurcation leads to an enhanced deposition and aggregation of cells. Therefore, complex geometries of the flow domain other than a planar slit should be considered in future simulations. Finally, future experimental studies are needed to elucidate to what extent the cross-stream distribution of RBCs in SCD would be adversely altered in the microcirculation, and if so, how this process may contribute to SCD vasculopathy. Such work could directly lead to a new paradigm of biophysical therapeutic strategies by mitigating the potential margination of aberrant sickle RBCs in SCD.

\begin{acknowledgments}
\XZrevise{This work was supported by NSF Grant No.~CBET-1436082 and NIH Grant No.~R21MD011590-01A1. The authors gratefully acknowledge helpful discussions with Yumiko Sakurai from the Lam Lab. This work used the Extreme Science and Engineering Discovery Environment (XSEDE) \cite{xsede} SDSC Dell Cluster with Intel Haswell Processors (Comet) through allocations TG-CTS190001 and TG-MCB190100.}\end{acknowledgments}

% Bibliography
%\bibliographystyle{achemso}%{osa}
% references
%\clearpage
%\bibliography{RBCsuspension.bib}

%%% main.bbl 
%apsrev4-2.bst 2019-01-14 (MD) hand-edited version of apsrev4-1.bst
%Control: key (0)
%Control: author (8) initials jnrlst
%Control: editor formatted (1) identically to author
%Control: production of article title (0) allowed
%Control: page (0) single
%Control: year (1) truncated
%Control: production of eprint (0) enabled
%

\end{document}